\documentclass[12pt,a4paper]{article} 
\usepackage{style}
\usepackage{amsmath}
\usepackage{amssymb}
\usepackage{amsthm}
\usepackage{psfrag}
\usepackage{graphicx}
\usepackage{hyperref}
\usepackage{feynmp}
\usepackage{color}
\usepackage{bm}
\usepackage{longtable,array}
\definecolor{editcol}{rgb}{0.70,0.05,0.20}
\newif\ifshowedits \showeditstrue

\makeatletter
\def\LT@makecaption#1#2#3{\LT@mcol\LT@cols c{\hbox to\z@{\hss\parbox[t]\LTcapwidth{\reset@font\small
  \sbox\@tempboxa{#1{#2. }#3}\ifdim\wd\@tempboxa>\hsize #1{#2. }#3\else\hbox to\hsize{\hfil\box\@tempboxa\hfil}\fi
  \endgraf\vskip\baselineskip}\hss}}}
\makeatother

\newcommand{\be}{\begin{equation}}
\newcommand{\ee}{\end{equation}}
\newcommand{\beqa}{\begin{eqnarray}}
\newcommand{\eeqa}{\end{eqnarray}}

\newcommand\m{\mu}
\newcommand\p{{\bf p}}

\renewcommand\a{\alpha}

\newcommand{\HH}{{\cal H}}

\newcommand\x{{\bf x}}
\renewcommand\k{{\bf k}}
\newcommand\q{{\bf q}}

\newcommand{\khat}{\hat{k}}
\newcommand{\qhat}{\hat{q}}
\newcommand{\Mpch}{h^{-1}\text{Mpc}}
\newcommand{\hMpc}{h\text{Mpc}^{-1}}
\newcommand{\kmax}{k_{\rm max}}
\newcommand{\knl}{k_{\rm NL}}
\newcommand{\kp}{{k_\parallel}}

\def\d{\partial}
\newcommand{\bseq}{\begin{subequations}}
\newcommand{\eseq}{\end{subequations}}

\renewcommand{\ln}{\mathop{\rm ln}\nolimits}

\renewcommand{\L}{\Lambda}

\renewcommand{\k}{{\bf k}}
\newcommand{\z}{\hat{\bf z}}

\relax

\title{Don't miss the forest for the trees: the Lyman alpha forest power spectrum in 
effective field theory }

\author[1]{Mikhail M. Ivanov\note{ivanov99@mit.edu}}
\affiliation[]{Center for Theoretical Physics, Massachusetts Institute of Technology, \\
Cambridge, MA 02139, USA}

\abstract{We derive an effective field theory (EFT) for 
cosmological Lyman alpha forest fluctuations
valid for the power spectrum at the one-loop order. 
The ``bottom-up'' EFT expansion at the level of the transmitted flux
is identical to the line-of-sight 
dependent bias model first derived by Desjacques et al.
We confirm this result by a ``top-down'' derivation based on the 
exponential map of the optical depth field. 
Specifically, we show that the combination of the exponential map
and conditions of renormalizability 
generates the same EFT expansion as the ``bottom-up'' approach.  
In passing, we point out inconsistencies of the tree-level perturbative 
expansion of the exponential map without counterterms.
To facilitate practical applications, 
we generalize the FFTLog method for efficient calculations of one-loop integrals
from new line-of-sight dependent operators. Finally, we compare the one-loop 
EFT model against data from the Sherwood hydrodynamic simulation. 
The theory fits the data with sub percent accuracy up 
to $k= $ 3 $h$Mpc$^{-1}$ at $z= 2.8$ for both 3D and 1D correlations.
Our model can be readily used for cosmological full-shape analyses of  
Lyman alpha forest data. 
}

\begin{document}

\begin{flushright}
MIT-CTP/5607
\end{flushright}

\maketitle

\section{Introduction}

The Lyman alpha forest 
is the key source of information on the matter power 
spectrum at small scales ($\sim$ Mpc and smaller)
and high redshifts, \mbox{$2\lesssim z\lesssim 5$}~\cite{Hernquist:1995uma,Miralda-Escude:1995evh,1997ApJ...488..532C,Croft:1997jf,McDonald:1999dt,Croft:2000hs}.
It has been extensively used to constrain the $\Lambda$CDM model~\cite{Zaldarriaga:2000rg,Zaldarriaga:2000mz,Zaldarriaga:2001xs,SDSS:2004aee,SDSS:2004kjl,Seljak:2006bg,2011JCAP...09..001S,Irsic:2017sop,Chabanier:2018rga},
as well as neutrino masses~\cite{Seljak:2006bg,2012MNRAS.420.2551B,Palanque-Delabrouille:2014jca,Pedersen:2019ieb,Garny:2020rom},
primordial black hole dark matter~\cite{Afshordi:2003zb},
dark-matter baryon interactions~\cite{Dvorkin:2013cea},
and other non-minimal cosmological models~\cite{Garny:2018byk,Fuss:2022zyt,Goldstein:2023gnw}.
The forest is also a unique 
probe of warm dark matter~\cite{Viel:2005qj,Seljak:2006qw,Boyarsky:2008xj,Viel:2013fqw,Baur:2015jsy,Garzilli:2015iwa,Baur:2017stq,Boyarsky:2018tvu,Palanque-Delabrouille:2019iyz,Garzilli:2019qki} 
and fuzzy dark matter models~\cite{Hui:2016ltb,Irsic:2017yje,Rogers:2020ltq}
that is not affected by ambiguities of the galaxy formation physics. 
In addition, the large-scale Lyman alpha forest is a valuable source of 
high-redshift baryon acoustic oscillation (BAO) information, see e.g.~\cite{Slosar:2013fi,BOSS:2013igd,Aubourg:2014yra,duMasdesBourboux:2020pck}. In addition to the BAO, 
the broadband shape of the large-scale 3D Lyman alpha 
power spectrum may also be a powerful probe of cosmology~\cite{Slosar:2011mq,Cuceu:2021hlk,Cuceu:2022wbd,Gordon:2023dua}.

The Lyman alpha forest has been one of the key targets 
for past and ongoing large-scale structure surveys such as 
BOSS~\cite{BOSS:2015ids,Chabanier:2018rga}, 
eBOSS~\cite{duMasdesBourboux:2020pck}, 
XQ-100~\cite{Irsic:2017sop}, MIKE/HIRES~\cite{Viel:2013fqw},
and DESI~\cite{Aghamousa:2016zmz,DESI:2023pir,Gordon:2023dua}. 
Entering the regime of high-precision Lyman alpha cosmology 
with DESI
requires a robust theoretical understanding of the forest.
The standard approach to the forest modeling has been  
hydrodynamical simulations, see e.g.~\cite{Borde:2014xsa,Rossi:2014wsa,Bird:2014pia,Bolton:2016bfs,Arinyo-i-Prats:2015vqa,Anderson:2018zkm,Smith:2021hqg,Tillman:2023pby}. 
In this work we explore an alternative method and 
study to what extent the Lyman alpha fluctuations can be 
modeled perturbatively~\cite{Greig:2012zw,Garny:2018byk,Garny:2020rom,Givans:2020sez,Chen:2021rnb,Givans:2022qgb}. 
On the one hand, perturbative approaches break down at short scales, where the power
spectrum is fully nonlinear, which ultimately limits their utility. 
On the other hand, they provide a high level of accuracy and flexibility on 
mildly non-linear scales.
Indeed, as long as the EFT formally applies,\footnote{The conditions of the applicability of the EFT will be specified shortly.} it
gives a systematic program of consecutive approximations 
that can be executed to arbitrary accuracy.
The high flexibility of the EFT is especially important for the efficient  
exploration of beyond-$\L$CDM models. 
For example, it is currently unfeasible to run a full
simulation-based
Monte Carlo analysis of a large grid of 
$\L$CDM extensions 
similar to the analysis by the Planck collaboration~\cite{Aghanim:2018eyx}.\footnote{Although recently there has been 
significant progress in cosmological analyses of Lyman alpha forest 
beyond $\Lambda$CDM based on approximating, interpolating~\cite{Palanque-Delabrouille:2014jca,Murgia:2018now,Hooper:2022byl} and 
emulating~\cite{Rogers:2018smb,Bird:2018efe,Pedersen:2020kaw} a grid of simulations.}
Such an analysis {is}, however, perfectly possible with 
perturbation theory-based pipelines analogous to the ones
recently applied to galaxy clustering data, see e.g.~\cite{Ivanov:2019pdj,DAmico:2019fhj,Ivanov:2019hqk,Chudaykin:2020ghx,Ivanov:2020ril,Philcox:2021kcw,Ivanov:2021zmi,White:2021yvw,Chen:2021wdi,Chen:2022jzq,Chudaykin:2022nru,Ivanov:2023qzb}.
In addition, perturbative descriptions are based on 
different assumptions than simulations. In principle, they provide alternative 
first-principle 
models that are agnostic about the physics of
intergalactic medium.\footnote{See~\cite{Seljak:2012tp,Cieplak:2015kra}
for relationship between the underlying physics of the Lyman alpha forest 
and perturbation theory parameters.}
Therefore, they represent a valuable addition to the Lyman alpha 
cosmology toolbox. 

The aim of this work is to develop an effective field theory (EFT) description
for the Lyman alpha forest, analogous to that of other large-scale
structure tracers~\cite{Baumann:2010tm,Carrasco:2012cv,Desjacques:2016bnm,Cabass:2022avo,Ivanov:2022mrd}.
The main philosophy of the EFT is to describe dynamics on large scales
using only relevant symmetries.
These ideas have been explored before
both in the context of the Lyman alpha forest~\cite{Garny:2018byk,Givans:2020sez,Chen:2021rnb} 
and for general line-of-sight dependent bias tracers~\cite{Desjacques:2018pfv}. 
In particular,
our work is a direct extension of Refs.~\cite{Desjacques:2018pfv,Chen:2021rnb}.
Desjacques et al.~\cite{Desjacques:2018pfv} 
derived a general perturbative model for a tracer of matter that depends on the 
line-of-sight selection effects. 
Although this description has been developed primarily in the context of 
galaxies~\cite{2009MNRAS.399.1074H}, one can argue that it applies, practically without modifications, to the Lyman alpha forest. 
This is quite natural, as the EFT bias expansion~\cite{Desjacques:2016bnm} 
is based on symmetries, and the 
galaxies in the presence of selection effects enjoy the same key symmetries
as the Lyman alpha forest: the $SO(2)$ rotations around the line-of-sight 
and the equivalence principle.

Chen et al.~\cite{Chen:2021rnb} pursued an alternative route and built a perturbative 
expansion for the Lyman alpha forest flux $F$ starting from the exponential map of the 
effective optical depth field $\tau$, $F=\exp(-\tau)$. 
This field is often assumed to be a selection-independent 
tracer of dark matter, whose density is conserved 
during transformations from real to redshift space,
see e.g.~\cite{Seljak:2012tp,Cieplak:2015kra}.
Chen et al. then showed how the combination of dynamics 
and the exponential map produces terms in the
anisotropic bias expansion for the Lyman alpha forest derived in~\cite{Desjacques:2016bnm}. 
In this regard, the main theoretical goal of this work 
is to establish a direct connection between the two approaches. 
We will explicitly show that the combination of the 
exponential map and the EFT renormalizability principle
explicitly reproduces the complete set 
of EFT operators. 
In order to emphasize the role of renormalization, 
we also carry out an alternative derivation in which we ignore counterterms, 
i.e. treat the perturbative expansion
\textit{\`a la} standard perturbation theory 
(SPT)~\cite{Bernardeau:2001qr}. In that case
the exponential map produces
an incomplete set of line-of-sight dependent bias operators at the tree-level,
with strong constraints on their bias coefficients.
We call this naive SPT-like expansion ``the tree-level Tau model.''
We show explicitly that this model is mathematically inconsistent. 
Once the model is properly renormalized, its tree-level constraints get violated by loops. 
We argue that within this particular derivation the 
loops should also generate the rest of the EFT operators allowed by symmetries. 
Our calculations thus suggest that the ``renormalized'' SPT-like approach is  
equivalent to the full EFT\footnote{At least at the level of the one loop power spectrum. } once the loop corrections are taken into account.

It is also important to note 
that Refs.~\cite{Zheng:2010jf,Wyithe:2011mt,Arinyo-i-Prats:2015vqa}
found that the optical depth is actually a selection-dependent 
tracer, i.e. in contrast to idealized galaxies\footnote{From now on when discussing galaxies, we will always assume an idealized situation 
when the line-of-sight selection effects~\cite{2009MNRAS.399.1074H,Desjacques:2018pfv} are negligible, although this assumption may not always be warranted, see e.g.~\cite{Martens:2018uqj,Obuljen:2020ypy}. }
its fluctuations trace the line-of-sight velocity gradients.
This further strengthens the motivation for a bottom-up approach such as the EFT.

On a more practical side, we 
extend a fast logarithmic Fourier transform (FFTLog) algorithm of {Refs.}~\cite{Simonovic:2017mhp,Chudaykin:2020aoj} 
to quickly compute the full EFT one-loop power spectrum of the 
3D Lyman alpha forest.  
Having done this calculation, we compare our EFT power spectrum model 
to the Lyman alpha forest data from the Sherwood simulations~\cite{Bolton:2016bfs}. 
We find this model to be sub percent level accurate 
up to $\kmax= 3~\hMpc$ for $z=2.8$, and up to $\kmax= 5~\hMpc$ for $z=3.2$. 
We have also detected  
strong deviations from the predictions of the tree-level Tau model. 
This may be considered 
``experimental'' evidence in favor of the full EFT expansion 
and validity of the renormalization program. 
Finally, we show how the EFT approach works out at the level of the one-dimensional 
flux power spectrum. 
The methods that we have presented are ready to be applied to data. 
This paves the way for systematic and efficient EFT-based cosmological 
analyses of the Lyman alpha forest.

Our paper is structured as follows.
In Sec.~\ref{eq:basics} we recap the Lyman alpha forest physics and 
discuss its aspects relevant for the EFT: scales and power counting. 
Section~\ref{sec:eft} outlines the EFT 
model first derived by \cite{Desjacques:2016bnm}
and discusses its features relevant for the Lyman alpha. 
There we also discuss the FFTLog implementation 
of our one-loop calculation. 
Section~\ref{sec:expmap} establishes the relationship
between the EFT model and the exponential map. There we 
also introduce the tree-level Tau model, which we use 
as a case study to underline the importance of loop 
corrections and renormalization. 
In Sec.~\ref{sec:sims} we compare the full one-loop EFT 
model with Sherwood simulations. 
Sec.~\ref{sec:1d} is devoted to the one-dimensional
flux power spectrum modeling. Finally, we outline our main 
results and draw conclusions in Sec.~\ref{sec:disc}.
Some technical material, including the one-loop renormalization conditions of the EFT model, is collected in 
appendices.

\section{Preliminaries}
\label{eq:basics}

\subsection{Basics of Lyman alpha forest}

Our Universe is filled 
with optically thin 
neutral hydrogen (HI) clouds at redshifts $2 \lesssim z\lesssim 5$. 
Background quasars 
emit UV radiation that propagates towards us through these clouds
along given lines of sight. 
The atoms in the cloud absorb the 
background radiation 
if the radiation's rest-frame frequency happens to match that of a transition between
hydrogen levels. 
Notably, if the relevant wavelength is 121.6 nm, 
a hydrogen atom undergoes a Lyman alpha transition.
Since the quasar spectrum is continuous in a wide frequency range, and 
there are many clouds along the line of sight at different redshifts, 
the observed quasar spectrum features 
a dense ``comb'' of absorption lines, called the Lyman alpha forest.
The fraction of transmitted flux for each quasar spectrum 
is not homogeneous, and its fluctuations trace cosmological 
mass fluctuations along the line of sight. 
Specifically, under the assumption of photoionization-recombination equilibrium, 
the optical depth 
of neutral hydrogen clouds is proportional {to} the 
Lyman alpha absorption cross section
and the neutral hydrogen density $n_{HI}$ along the line of sight. 
The latter scales with the fractional overdensity of gas (baryons) $\delta_b$,
gas temperature $T$, and the ionizing
background amplitude $J$ as 
\be
\label{eq:nHI}
n_{HI}\propto  \frac{(1+\delta_b)^2}{T^{0.7} J}\,.
\ee
Assumptions of the adiabatic expansion and 
the photoionization equilibrium lead to the tight relation between 
temperature and density, 
$T=T_0(1+\delta_b)^{\gamma-1}$, where $\gamma\simeq 1.6$.
Assuming that $T_0$, $J$ and $\gamma$ are constants (i.e. do not have 
spatial fluctuations), we arrive at the 
well-known conclusion (see e.g.~\cite{Rauch:1998xn,Hui:2016ltb,2013fgu..book.....L}) that the fluctuations of $n_{HI}$
observed through the forest trace the underlying 
fluctuations of gas density, which in turn, must 
reflect the matter over-density $\delta$.\footnote{Recall that in standard linear cosmological perturbation theory $\delta_b=\delta$ sufficiently deep into the matter
domination regime. Note that this condition is not always correct beyond $\Lambda$CDM, see e.g.~\cite{Blas:2012vn,Audren:2014hza}.} In this simplified model, 
known as the fluctuating Gunn-Peterson approximation~\cite{1965ApJ...142.1633G,Croft:1997jf}, 
the optical depth $\tau$ is proportional to $n_{HI}$,
while the transmitted flux we measure is simply given by $F=\exp(-\tau)$.

One way to study the Lyman alpha fluctuations is to model 
the optical depth field. 
In our work, however, we employ a different approach. Since
we are interested in fluctuations of the flux, we focus directly 
on this observable. Instead of modeling 
explicitly underlying physics that affects the flux, we 
will build a perturbative (gradient) expansion 
that is based on symmetries of the problem and 
a minimal set of additional assumptions. In particular, 
we will not explicitly employ relations such as Eq.~\eqref{eq:nHI},
but simply assume that the large-scale flux contrast is 
a certain unknown function of only a few variables (degrees of freedom): 
the tidal field, velocity gradients, and stochastic noise.
This function is then represented as a general Taylor expansion that 
contains all possible dependencies allowed by symmetries.
This approach in general is known as the EFT of 
large-scale structure~\cite{Baumann:2010tm,Carrasco:2012cv,Cabass:2022avo,Ivanov:2022mrd}.

In the EFT of large-scale structure one builds
a perturbative expansion of relevant observables in terms 
of the matter overdensity field, which is linear on large scales. 
This linear field $\delta^{(1)}$ is the seed function
of all perturbative EFT calculations.
Its statistical
properties
are fully fixed in terms of the linear matter power spectrum,
\be
\langle \delta^{(1)}(\k)\delta^{(1)}(\k')\rangle  = (2\pi)^3\delta_D^{(3)}(\k+\k')P_{\rm lin}(k,z)\,.
\ee
In what follows we will often drop the explicit redshift-dependence 
and assume that all quantities are evaluated at a given redshift
of the Lyman alpha forest, which we take to be $z=2.8$ to match
the simulation data that we use here. This redshift is similar to the 
redshifts of Lyman alpha forest data from BOSS~\cite{Chabanier:2018rga}, 
eBOSS~\cite{eBOSS:2020yzd} and DESI~\cite{Aghamousa:2016zmz}.

\subsection{Relevant scales}

The EFT is a gradient expansion based on the assumption of scale separation. 
Let us discuss how well this assumption 
is satisfied by the Lyman alpha forest fluctuations.

The first relevant scale in the Lyman alpha forest physics is the 
non-linear scale where the density field becomes fully non-linear. 
In momentum space, it can be estimated as a wavenumber 
for which the amplitude of the dimensionless linear 
matter power spectrum becomes unity, 
\be
\frac{k^3_{\rm NL}}{2\pi^2}P_{\rm lin}(k_{\rm NL},z)=1\,,~~~\Longrightarrow\quad k_{\rm NL}\simeq 5~\hMpc\,,\quad \text{for}\quad z\simeq 3\,.
\ee
A perturbative description of the Lyman alpha forest 
is possible only if $k\ll k_{\rm NL}$. 
The second relevant parameter is the smoothing scale. 
The forest is smoothed by two effects\footnote{It is worth mentioning that the actual data 
features certain observational effects that effectively act 
as an additional source of smoothing, e.g. an imperfect resolution~\cite{Palanque-Delabrouille:2014jca}. These effects can also be absorbed, to some extent, into the EFT expansion. } : gas pressure (3D effect)~\cite{Gnedin:1997td}
and thermal broadening (1D effect)~\cite{Hui:1997dp,Theuns:1999mz,Zaldarriaga:2000mz}. 
These smoothing 
effects are naturally incorporated within the EFT as part of 
the gradient expansion.
The gas perturbations are smoothed by pressure on the Jeans 
scale $k_J\sim 18~\hMpc$~\cite{Gnedin:1997td,Garny:2020rom}. The absorption lines are
also subject to thermal broadening, whose characteristic 
scale $k_S\sim 11~\hMpc$ in the 1D flux power spectrum
is comparable to that of the Jeans smoothing~\cite{Garny:2020rom}.\footnote{We stress that both scales are not known exactly. The values quoted in~\cite{Garny:2020rom} 
are benchmark values that are sufficient for our order-of-magnitude 
estimates. }
Both smoothing length scales are typically 
shorter than the non-linear scale at the lower end of redshifts
relevant for the forest, $z\simeq 3$. This gives us the 
following hierarchy of scales in the EFT for Lyman alpha:
\be 
k\ll k_{\rm NL}\lesssim k_J,~k_s\,.
\ee

High redshift Lyman alpha forest 
is affected by potentially significant spatial
fluctuations of the ionizing background,
and temperature fluctuations due to inhomogeneous (patchy) reionization.\footnote{In our notation this can be thought of as a spatial variation of $T_0$ in Eq.~\eqref{eq:nHI}.}
The first effect modulates the power spectrum shape
on a broad range of scales, up to 100 $\Mpch$~\cite{Pontzen:2014ena,GontchoAGontcho:2014xbj}.
These modulations are, however, quite smooth, which suggests
that they can be captured perturbatively with an appropriately modified
bias expansion.
As far as the patchy reionization 
is concerned, Ref.~\cite{DAloisio:2015exk}
suggests that the gas temperature fluctuates on 
a scale of $\sim 25~\Mpch$ at $z\sim 5$. These results are consistent with 
simulations of Ref.~\cite{Montero-Camacho:2019ucp}, which additionally 
imply that patchy reionization may impact 
the 3D Lyman alpha power spectrum at wavenumbers as large as~$0.2~\hMpc$
at a few percent level at low redshifts. 
This suggests that the typical scales associated with UV background and 
temperature fluctuations are significantly
longer than the non-linear scale, and hence their
systematic description within a gradient expansion approach, 
such as EFT, may present a theoretical
challenge\footnote{See however Ref.~\cite{Cabass:2018hum} for an approach to resum the large gradients  
due to the radiation transfer effects.}.
In this work we build the 
EFT for Lyman alpha forest starting with an idealized scenario where the 
photoionization and temperature 
fluctuations are absent.
We stress though that our description is still realistic 
at low redshifts ($z\sim 3$) and small scales ($k \gtrsim 0.1~\hMpc$), where
the effects in question
are suppressed~\cite{GontchoAGontcho:2014xbj,Montero-Camacho:2019ucp}.

Within our approximation, the most relevant scale for the 
EFT expansion is the non-linear scale, just like in the usual
EFT of LSS. In what follows we will construct 
a perturbative expansion of the Lyman alpha forest 
fluctuations around linear theory, which will be 
implicitly controlled by a small parameter $k/k_{\rm NL}$.
Let us sketch the structure of this expansion.

\subsection{Estimates of perturbative terms in the scaling Universe}

\begin{figure}
    \centering
    \includegraphics[width=0.7\textwidth]{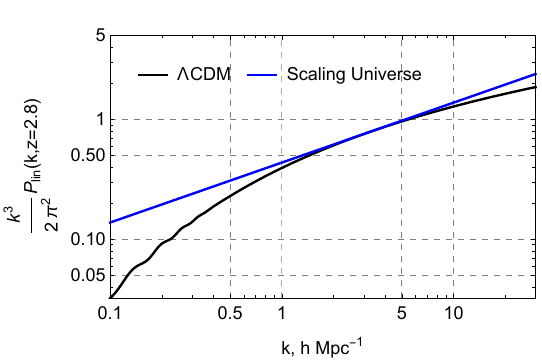}
    \caption{The dimensionless linear matter power spectrum in $\Lambda$CDM 
    and a scaling Universe from Eq.~\eqref{eq:su}. Here we assume $z=2.8$,
  a redshift of the Sherwood simulation snapshot that we use in this work.}
    \label{fig:scalingU}
\end{figure}

Let us estimate the size of various terms in perturbation theory.
It is convenient to approximate the actual $\Lambda$CDM power spectrum with a power-law $P_{\rm lin}\propto k^n$. For the range of scales relevant for our analysis,
and for $z\approx 3$,  
$k\approx 3~\hMpc$, the linear matter power spectrum can be well approximated as
\be
\label{eq:su}
P_{\rm lin}(k)=2\pi^2\frac{k^n}{k^{n+3}_{\rm NL}}\,,\quad \text{with}\quad 
n\approx -2.5\,,\quad 
k_{\rm NL}\approx 5.3~\hMpc\,,
\ee
see Fig.~\ref{fig:scalingU}, where the scaling universe curve is normalized to the 
linear power spectrum at $k=3~\hMpc$.
From the dimensional analysis, the $L$'th-loop correction to this 
power spectrum must scale as~\cite{Bernardeau:2001qr,Pajer:2013jj}
\be
 \frac{k^3}{2\pi^2} P_{L-\text{loop}}(k)=\frac{\alpha_L(2\pi^2)}{k^{3}_{\rm NL}}\left(\frac{k}{k_{\rm NL}}\right)^{(n+3)(L+1)}\,,
\ee
where $\alpha_L$ is an order one number. 
These terms will be referred to as ``mode-coupling corrections''
in what follows.
The higher derivative (counterterm) corrections scale as $k^2P_{\rm lin}\propto k^{n+2}$.
Thus, the total power spectrum in the EFT reads, schematically
\be
\label{eq:scale1l}
\begin{split}
P_{\rm NL}(k)&=2\pi^2\frac{k^n}{k^{n+3}_{\rm NL}}\left(\underbrace{1}_{\text{linear}}+
\underbrace{\alpha_1\left(\frac{k}{k_{\rm NL}}\right)^{n+3}}_{\text{1-loop}}
+\underbrace{\alpha_2\left(\frac{k}{k_{\rm NL}}\right)^{2(n+3)}}_{\text{2-loop}}
+ \underbrace{\alpha_{k^2}\left(\frac{k}{k_{\rm NL}}\right)^2}_{\text{higher deriv.}}+ ...\right)\\
&=2\pi^2\frac{k^{-2.5}}{k^{0.5}_{\rm NL}}\left(\underbrace{1}_{\text{linear}}+
\underbrace{\alpha_1\left(\frac{k}{k_{\rm NL}}\right)^{0.5}}_{\text{1-loop}}
+\underbrace{\alpha_2\left(\frac{k}{k_{\rm NL}}\right)^{1}}_{\text{2-loop}}
+ \underbrace{\alpha_{k^2}\left(\frac{k}{k_{\rm NL}}\right)^2}_{\text{higher deriv.}}+ ...\right)\,,
\end{split} 
\ee
where we used $n=-2.5$ from Eq.~\eqref{eq:su} in the second line.
Note that we explicitly included only the mode-coupling and $k^2$-corrections above.
We see that in our toy model scaling Universe the one-loop corrections 
due to mode coupling are the leading contributions in the regime $k\ll k_{\rm NL}$.
The $k^2$ corrections become important only at the three loop order.\footnote{See~\cite{Blas:2013aba,Konstandin:2019bay} for the three loop order matter calculations.}

Our discussion so far has considered non-linearity 
in the context of the matter density field. 
In the EFT, additional non-linearities
appear due to biasing and redshift-space distortions.
These effects are conceptually similar to the dark matter 
non-linearities, and hence our scaling universe estimates 
are expected to
apply to the more realistic case without significant modifications.\footnote{It is important to note that the velocity field is more non-linear than the density one, see e.g.~\cite{Pueblas:2008uv,Schmittfull:2020trd}, i.e. 
the wavenumber associated with the velocity non-linear scale is smaller.} 
One aspect worth mentioning is the presence of 
constant stochastic shot noise contribution, 
which is, however, 
negligibly small in the scaling Universe where it scales as $P_{\rm shot}\sim k_{\rm NL}^{-3}$. 

From Eq.~\eqref{eq:scale1l} we see that in contrast to galaxies (see e.g.~\cite{Cabass:2022wjy,Cabass:2022ymb}), the leading order corrections 
to linear theory in the Lyman alpha forest case 
come from the one-loop mode coupling 
effects.\footnote{Here we distinguish between the one-loop mode-coupling 
integrals and higher-derivative counterterms, both of which are parts of the 
one-loop power spectrum.} 
Recall that for the scales relevant to galaxy 
clustering the effective power law is less steep, 
$n\approx -1.5$, so that the $k^2-$type higher derivative terms become important 
already at the one-loop level.
Our $\Lambda$CDM Universe is, of course, more complicated 
than the toy model we consider here. 
Nevertheless, the scaling universe model gives us a good sense of 
the importance of various terms in the perturbation expansion.

\section{Effective field theory for Lyman alpha fluctuations}
\label{sec:eft}

\subsection{Overview of galaxies in redshift space}

It is instructive to start our discussion of the bias model for 
the Lyman alpha with a recap of the bias model for galaxies. 
In real space, and on large scales, the 
galaxy overdensity field $\delta_g$ can be 
expressed through a perturbative expansion over 
the velocity gradients and tidal fields~\cite{Senatore:2014eva,Assassi:2014fva,Mirbabayi:2014zca,Lazeyras:2017hxw,Desjacques:2016bnm}. 
This expansion 
is, in general, non-local in time, but at the cubic in density 
order, relevant for the one-loop power spectrum calculation, 
it can be expressed through local in time operators~\cite{Desjacques:2016bnm}.
Specifically, at this order, we have the following expression 
for the deterministic part of the galaxy bias relation, 
\be
\label{eq:bias_gal}
\begin{split}
\delta_g(\x) = b^g_1 \delta(\x) + \frac{b^g_2}{2}(\delta(\x)^2 -\langle \delta(\x)^2\rangle ) + b^g_{\mathcal{G}_2}\mathcal{G}_2(\x) + b^g_{\Gamma_3}\Gamma_3(\x) + b^g_{\nabla^2\delta} R_*^2\nabla^2 \delta\,,
\end{split} 
\ee
where $\delta$ is the (nonlinear) matter overdensity field, and we also defined 
\be
\begin{split}
& \mathcal{G}_2(\x) = (\d_i \d_j \Phi)^2 - (\d^2 \Phi)^2\,,\\
& \Gamma_3(\x) =\mathcal{G}_2[\Phi] - \mathcal{G}_2[\Phi_v]\,, 
\end{split} 
\ee
where we introduced density and velocity potentials $\Phi$ and $\Phi_v$,
respectively, which satisfy
\be
\d^2 \Phi = \delta\,,\quad \d_i \Phi_v = v^i\,, 
\ee
where $v^i$ is the peculiar velocity field.
The bias coefficients 
$b^g_1,b^g_2,b^g_{\mathcal{G}_2},b^g_{\Gamma_3},b^g_{\nabla^2\delta}$
are free parameters that characterize a given galaxy selection;
$R_*$ is the typical length scale associated with galaxies,
which we inserted to make $b^g_{\nabla^2\delta}$
dimensionless. Typically, this scale is assumed to be of the order 
of the Lagrangian radius of the host halos~\cite{Desjacques:2016bnm,Lazeyras:2019dcx}, 
but it can be as large as 100~$\Mpch$ 
if the radiation transfer 
effects are included, see e.g.~\cite{Sanderbeck:2018lwc,Cabass:2018hum}.

There are several aspects of the galaxy bias model in 
Eq.~\eqref{eq:bias_gal} that are worth emphasizing. 
It is the most general local in time 
perturbative expansion in the matter density in real space,
modulo terms that do not contribute at the 
level of the one-loop galaxy power spectrum, e.g.~$\delta^3$.
This expansion 
involves all possible operators allowed by symmetries 
of the problem, the spherical symmetry 
and the equivalence principle.
In our context this means the relevant operators are 
scalars under $SO(3)$ rotations, and there is no velocity bias
on large scales, i.e. galaxies undergo the same acceleration 
as dark matter.
Using the general symmetry-motivated expansions like~\eqref{eq:bias_gal}
is the key principle of effective field theory.  
In this philosophy, 
the distribution of galaxies on large scales 
is expressed in terms of underlying 
basic long-wavelength 
degrees of freedom, which in our context are matter density, velocity gradients,
and tidal fields, plus stochastic terms that will be discussed later.
The tree-level 
bias coefficients that appear in~\eqref{eq:bias_gal}
are, strictly speaking, not well defined since they receive formally 
infinite loop corrections. 
In the EFT jargon these are called ``bare'' parameters, or Wilson coefficients. Only their finite ``renormalized'' parts
make physical sense and can be matched to the data
that describes physical observables.

Since the galaxies are observed in redshift space,
one needs to transform the galaxy bias expansion~\eqref{eq:bias_gal}
using the following velocity-dependent mapping 
(see e.g.~\cite{Bernardeau:2001qr}),\footnote{Using this expansion, we
assume a flat sky approximation that is accurate for small scales 
relevant for our work.} 
\be
\label{eq:rsdmap}
\delta^{(s)}_g(\k) = \delta_g(\k) +\int d^3x~e^{-i\k\x}\left(
e^{-ik_z v_{gz}(\x)/(aH)}-1\right)(1+\delta_g(\x))\,,
\ee
where we have switched to the 
Fourier-space representation, and used $a$ and $H$
to denote the metric scale factor and the Hubble parameter, 
respectively, $\HH=aH$ is the conformal Hubble parameter, 
whilst the subscript $z$ stands for the projection
onto the line-of-sight, which we will describe with the 
unit vector $\z$. $v_{g}$ above is the tracer's velocity, 
which is equal to the matter peculiar velocity $v$
at the zeroth order in the derivative expansion
due to the equivalence principle.

In what follows it will be convenient 
to use the following notation for the normalized 
velocity gradient of matter along the line-of-sight, 
\be
\label{eq:etadef}
\eta \equiv \frac{\d_z v_z}{aH} \,.
\ee
Note that in linear theory in Fourier space we have 
\be
\eta \approx -f \mu^2 \delta^{(1)}_\k \,, \quad \mu = \frac{(\z\cdot \k)}{k}\,,
\ee
where we introduced 
$f\equiv d\ln D_+/d\ln a$ ($D_+$ is the growth factor).

Assuming that the velocity field is perturbative, 
one Taylor expands the exponent in~\eqref{eq:rsdmap}
and finds, at linear order, 
\be
\label{eq:rsdlin}
\delta_g^{(s)}= b_1^g \delta -\eta \,.
\ee
Crucially, the second term, which depends on the line-of-sight, 
does not have any free parameter in front of it. This embodies the absence of selection effects, and follows from the 
conservation of the galaxy number density under the coordinate 
transformation from the galaxy rest frame to observed redshift space.

Going to higher order in the Taylor expansion of~\eqref{eq:rsdmap}, 
one obtains corrections to the observed galaxy 
overdensity due to redshift-space distortions (RSD).
Note that this expansion generates extra UV sensitivity,
which must be removed by appropriate 
counterterms~\cite{Senatore:2014vja,Lewandowski:2015ziq},
which start at order $k^2\delta$.
Apart from these higher-derivative contributions, 
RSD mapping~\eqref{eq:rsdmap} does not 
generate new free parameters for the deterministic part of the 
galaxy power spectrum in redshift space. The redshift space galaxy overdensity reads\footnote{We use the notation $\int_\q\equiv \int\frac{d^3q}{(2\pi)^3}$.}
\be 
\label{eq:Zexp}
\begin{split}
\delta_g^{(s)}(\k)=&Z_1(\k)\delta^{(1)}_\k + \int_{\q_1}
\int_{\q_2} 
(2\pi)^3\delta_D(\k-\q_{12})Z_2(\q_1,\q_2)\delta^{(1)}_{\q_1}
\delta^{(1)}_{\q_2}+\\
&+\int_{\q_1} \int_{\q_2}\int_{\q_3}
(2\pi)^3\delta_D(\k-\q_{123})Z_3(\q_1,\q_2,\q_3)\delta^{(1)}_{\q_1}
\delta^{(1)}_{\q_2} \delta^{(1)}_{\q_3}+...\,,
\end{split}
\ee
where the dots denote higher loop corrections and higher-derivative counterterms, 
$\delta^{(1)}_\k$ is the linear matter density field, 
while the redshift-space kernels $Z_{1,2,3}$
are given by
\bseq 
\label{eq:Zs}
\begin{align}
\label{eq:z1}
&Z_1(\k)  = b^g_1+f\mu^2\,,\\
&Z_2(\k_1,\k_2)  =\frac{b^g_2}{2}+b^g_{\mathcal{G}_2}\left(\frac{(\k_1\cdot \k_2)^2}{k_1^2k_2^2}-1\right)
+b^g_1 F_2(\k_1,\k_2)+f\mu^2 G_2(\k_1,\k_2)\notag\\
&\qquad\qquad\quad~~+\frac{f\mu k}{2}\left(\frac{\mu_1}{k_1}(b^g_1+f\mu_2^2)+
\frac{\mu_2}{k_2}(b^g_1+f\mu_1^2)
\right)
\,,\\
&Z_3(\k_1,\k_2,\k_3)  =2b^g_{\Gamma_3}\left[\frac{(\k_1\cdot
     (\k_2+\k_3))^2}{k_1^2(\k_2+\k_3)^2}-1\right]
\big[F_2(\k_2,\k_3)-G_2(\k_2,\k_3)\big]
\notag
\\  
&\quad
+b^g_1 F_3(\k_1,\k_2,\k_3)+f\mu^2 G_3(\k_1,\k_2,\k_3)+\frac{(f\m k)^2}{2}(b^g_1+f \mu_1^2)\frac{\m_2}{k_2}\frac{\m_3}{k_3}
\notag
\\
&\quad
+f\mu k\frac{\mu_3}{k_3}\left[b^g_1 F_2(\k_1,\k_2) + f \mu^2_{12} G_2(\k_1,\k_2)\right]
+f\m k (b^g_1+f \mu^2_1)\frac{\m_{23}}{k_{23}}G_2(\k_2,\k_3)
\notag
\\
&\quad+b^g_2 F_2(\k_1,\k_2)+2b^{{g}}_{\mathcal{G}_2}\left[\frac{(\k_1\cdot (\k_2+\k_3))^2}{k_1^2(\k_2+\k_3)^2}-1\right]F_2(\k_2,\k_3)
+\frac{b^g_2f\mu k}{2}\frac{\mu_1}{k_1}
\notag
\\
&\quad+b^g_{\mathcal{G}_2}f\mu k\frac{\m_1}{k_1}\left[\frac{(\k_2\cdot \k_3)^2}{k_2^2k_3^2}-1\right]
\,,
\end{align} 
\eseq
where the density and velocity SPT kernels $F_n$ and $G_n$ 
can be found, e.g. in \cite{Bernardeau:2001qr}. Note that 
$Z_3$ must be symmetrized over its arguments.
We also introduced cosines between the line-of-sight and 
momentum vectors, 
\be
\mu_i \equiv \frac{(\z\cdot \k_i)}{k_i} \,,
\ee
and use $\k$ to denote the sum of all arguments, i.e. for $Z_3$ we have 
$\k\equiv \k_1+\k_2+\k_3$.

Note that the density field variance $\langle \delta(\x)^2\rangle$
is used in the Eq.~\eqref{eq:bias_gal} in order to enforce 
the vanishing of the overdensity v.e.v.,
$\langle \delta \rangle = 0$, 
which a properly defined fluctuating field 
must satisfy. 
The $\k=0$ correction to~\eqref{eq:Zs} produced by the
density variance is not explicitly shown above, 
but it is needed to ensure that the line-of-sight 
dependent bias operators are not 
generated by loop corrections for galaxies.

\subsection{General bias expansion for the Lyman alpha forest}

The Lyman alpha forest is different from galaxies 
in several important aspects. 
First, unlike galaxies, 
the forest has a preferred direction - the line-of-sight. 
Indeed, in contrast to galaxies, the flux fluctuations intrinsically exist 
only in redshift space. 
This means that new operators are allowed in the Lyman alpha 
forest bias expansion. 
Physically, this happens because 
the absorption probability for the Lyman-$\alpha$
line depends on the tidal field along the line of sight.  
Mathematically, this property 
may be traced to the nonlinear mapping between flux 
and the optical depth, $F=\exp{(-\tau)}$.
If one assumes that the $\tau$ field is a matter tracer
whose number density is conserved, just like for galaxies,
there would be no selection effects, and the bias expansion would take the form~\eqref{eq:rsdlin} at the lowest order.
The nonlinear mapping, however, breaks the conservation of tracer number density~\cite{Seljak:2012tp} and forces us to introduce line-of-sight selection
effects.
At linear order, and in the rest frame 
of a neutral hydrogen cloud, we should write 
\be
\label{eq:dF_lin0}
\frac{F-\bar F}{\bar F}\equiv \delta_F = b_1 \delta -f\, b_\eta \hat{z}^i\hat{z}^j \d_i\d_j \Phi \,,
\ee
where $\bar F\equiv \langle F\rangle$ is the mean transmitted flux.
In linear theory the tidal field is related to the velocity gradient $\eta$ defined in Eq.~\eqref{eq:etadef}, so that one 
can rewrite the above expression as
\be
\label{eq:dF_lin}
\delta_F = b_1 \delta + b_\eta \eta \,.
\ee
The new selection-dependent parameter $b_\eta$
will be referred to as ``velocity gradient bias'' in what follows~\cite{Cieplak:2015kra}.
The linear model~\eqref{eq:dF_lin} gives the well-known tree-level result\footnote{We ignore here the RSD mapping, which shifts $b_\eta$ by $-1$. This shift is unobservable, and in what follows we use the shifted, observed $b_\eta$.}
\be
\label{eq:tree}
P_{\rm tree}(k,\mu)=(b_1-b_\eta f\mu^2)^2 P_{\rm lin}(k)\,,
\ee
which is a simple generalization of the Kaiser formula for galaxies~\cite{Kaiser:1987qv}.
We stress that Eq.~\eqref{eq:dF_lin0} is not the most 
general expression. As in the context of galaxies, a consistent 
bias expansion should depend on the history of the relevant 
operators along their fluid trajectories $\x_{\rm fl}(\x,\tau)$  ($\x$ is the last point
of the trajectory)~\cite{Carrasco:2013sva,Senatore:2014eva,Mirbabayi:2014zca,Desjacques:2016bnm,Ivanov:2022mrd},\footnote{In analogy with the smoothing scale of the Lyman 
alpha fluctuations~\cite{Gnedin:1997td}, one may say that the Lyman alpha bias parameters  
in general are expected to depend on the whole reionization history of the Universe. }  
\be 
\delta_F(\x,\tau)\supset \int^\tau T(\tau,\tau')  \mathcal{O}(\tau',\x_{\rm fl}(\x,\tau'))\,,
\ee
where $\tau$ is 
conformal time.
Using the matter equations of motion, the dependence on the fluid 
trajectory and time can be Taylor expanded and summed up into a 
finite number of 
linearly independent terms, which eventually reproduce the 
\textit{local-in-time} bias expansion similar to Eq.~\eqref{eq:dF_lin}.
All in all, it is important to keep in mind that the 
bias expansion is intrinsically non-local in time, but for the purposes 
of EFT calculations 
this non-locality is captured by the 
Eulerian bias operators $\Pi^{[n]}_{ij}$~\cite{Mirbabayi:2014zca,Desjacques:2016bnm},
which produce the  
local-in-time expansion.

It is straightforward to generalize the bias model~\eqref{eq:dF_lin}
to operators that are higher order in the linear density field.  
Specifically, Ref.~\cite{Desjacques:2018pfv} showed 
that a general selection-dependent biased tracer 
has the following deterministic bias model up to cubic order:
\be
\label{eq:F_full}
\begin{split}
\delta_F = & b_1\delta + \frac{b_2}{2}\delta^2  + b_{\mathcal{G}_2}\mathcal{G}_2 + b_{\Gamma_3}\Gamma_3 \\
&+ b_\eta \eta + b_{(KK)_\parallel} K_{ij} K_{jl} \hat{z}^i\hat{z}^l+ b_{\delta\eta}\eta \delta + b_{\eta^2}\eta^2 +b_{\Pi^{[2]}_\parallel} \Pi^{[2]}_{ij}\hat{z}^i\hat{z}^j \\
& + b_{\delta\Pi^{[2]}_\parallel} \delta\Pi^{[2]}_{ij}\hat{z}^i\hat{z}^j
+ b_{\eta\Pi^{[2]}_\parallel} \eta\Pi^{[2]}_{ij}\hat{z}^i\hat{z}^j
+ b_{(K\Pi^{[2]})_\parallel} K_{ij}\Pi^{[2]}_{jl}\hat{z}^i\hat{z}^l
+ b_{\Pi^{[3]}_\parallel} \Pi^{[3]}_{ij}\hat{z}^i\hat{z}^j\,,
\end{split}
\ee
where we kept only the terms that contribute non-trivially 
to the 
one-loop power spectrum, and introduced the new 
operators
\be
\begin{split}
& K_{ij} = \frac{\d_i\d_j\delta}{\Delta} -\frac{1}{3}\delta_{ij}\delta \\
& \Pi^{[1]}_{ij} =\d_i\d_j \Phi = \frac{\d_i\d_j\delta}{\Delta}\,,\\
& \Pi^{[n]}_{ij} = \frac{1}{(n-1)!}
\left[(aHf)^{-1}\frac{D}{D\tau}\Pi^{[n-1]}_{ij}-{(n-1)\,}\Pi^{[n-1]}_{ij}\right], \quad n>1
\,,
\end{split}
\ee
where we used the convective derivative 
$\frac{D}{D\tau}=\d_\tau +v^i \d_i$.
{As in Eq.~\eqref{eq:bias_gal}, all composite operators in Eq.~\eqref{eq:F_full} are mean-subtracted, e.g.\ $\delta^2\to\delta^2-\langle\delta^2\rangle$, so that $\langle\delta_F\rangle=0$. This is important because the RSD mapping conserves the $\k=0$ mode: without the subtraction the mapping produces a tadpole contribution $\propto \langle\mathcal{O}\rangle\eta$ that shifts $b_\eta$ by $-\sigma^2[\frac{b_2}{2}-\frac{f}{3}b_{\delta\eta}+\frac{f^2}{5}b_{\eta^2}+\frac29b_{(KK)_\parallel}+\frac13b_{\Pi^{[2]}_\parallel}]$, where $\sigma^2=\langle\delta^2\rangle$ and the coefficients are the rest-frame ones of Eq.~\eqref{eq:F_full}. In the full EFT this shift is absorbed by the free parameter $b_\eta$, but it matters for the Tau model of Section~\ref{sec:expmap}.}
The explicit expressions for the tensors $\Pi^{[2]}_{ij}$
and $\Pi^{[3]}_{ij}$ in perturbation theory can be 
found in Ref.~\cite{Desjacques:2018pfv}.
In total, we have 9 new selection 
bias coefficients in addition to the 4 selection free ones.\footnote{Our discussion so far does not include higher derivative counterterms 
and stochastic bias contributions, which will generate 
additional free parameters.}

The expression~\eqref{eq:F_full} is written in 
the rest-frame of the tracer. In order to describe 
the actual observations, this has to be transferred to the 
observer's frame using the standard RSD mapping~\eqref{eq:rsdmap}~\cite{Desjacques:2016bnm,Desjacques:2018pfv}.
Note that the physical meaning of the RSD mapping here is somewhat different from that in the galaxies' case.
Neglecting selection effects, the mapping simply acts on the 
real-space
bias expansion that captures the actual number density 
in tracer's rest frame. If the probability to detect a galaxy 
depends on line-of-sight properties~\cite{2009MNRAS.399.1074H,Obuljen:2020ypy},
the line-of-sight selection effects should be treated as part of the 
bias expansion that describes the \textit{apparent}  
number density of the tracer in its rest frame.
The apparent density can be defined as the observed galaxy distribution 
formally inversely 
transformed from redshift space to real space. 
Obviously, it is different from the \textit{actual}, 
rest-frame (real-space) galaxy number density, which, in principle, can 
be reconstructed with better selection criteria or better measurements~\cite{2009MNRAS.399.1074H}.
Contrasting the actual number density and the observed one in redshift-space, 
we may say that the RSD mapping does not conserve the galaxy number
density in the presence of selection effects.
In the Lyman alpha case, 
the line-of-sight dependent bias expansion also captures the 
apparent number density. This number density is by definition subject to
the RSD mapping that conserves it.
The conceptual difference with galaxies is that 
the Lyman alpha fluctuations intrinsically live along lines of sight, 
so that a notion of the actual rest frame density does not 
make sense in this case. From the technical point of view,
however, the treatment of the mapping is similar:  one has to apply the RSD mapping~\eqref{eq:rsdmap}
to the selection-dependent bias expansion~\eqref{eq:F_full}.

When expanded, the RSD mapping will produce additional 
velocity-dependent  
contributions into~\eqref{eq:F_full}, most of which 
will be degenerate with the selection terms that are already present.
This will lead to an unobservable re-definition of the 
selection bias coefficients. 
There will be, however, a few extra operators that are not
present in~\eqref{eq:F_full}.
These terms correspond to pure projection effects,
i.e. they explicitly depend on locally unobservable tracer's velocity.
The coefficients in front of these terms are protected by 
the equivalence principle and do not get renormalized
by non-linear effects. With these terms included, we have the net 
expression~\cite{Desjacques:2018pfv}:
\be 
\label{eq:F_full2}
\begin{split}
\delta^{(s)}_F = \delta_F[\text{from Eq.~(\ref{eq:F_full})}]  
 {-} \frac{v_z}{aH}\hat{z}^i\d_i
 \left[b_1\delta +b_\eta \eta + b_{\Pi^{[2]}_\parallel }\Pi^{[2]}_{kl}\hat{z}^k\hat{z}^l
 \right]\,.
\end{split}
\ee
{The bias coefficients in Eqs.~\eqref{eq:F_full2}--\eqref{eq:K3full} are the observed ones. They differ from the rest-frame coefficients of Eq.~\eqref{eq:F_full}, denoted here by a superscript ``rf'', by the contributions of the RSD mapping: $b_\eta=b_\eta^{\rm rf}-1$, $b_{\delta\eta}=b^{\rm rf}_{\delta\eta}-b_1$, $b_{\eta^2}=b^{\rm rf}_{\eta^2}-b_\eta$ and $b_{\eta\Pi^{[2]}_\parallel}=b^{\rm rf}_{\eta\Pi^{[2]}_\parallel}-b_{\Pi^{[2]}_\parallel}$, while the other coefficients are unchanged.}
The final mode-coupling 
three-dimensional one-loop power spectrum that comes from the 
expansion~\eqref{eq:F_full2}
is given by
\be
\label{eq:deltaP1L}
\begin{split}
\Delta P^{\text{1-loop}}(k,\mu)\equiv & 2\int_\q K_2^2(\q,\k-\q)
P_{\text{lin}}(|\k-\q|)P_{\text{lin}}(q) \\
& + 6 K_1(\k)P_{\text{lin}}(k)\int_\q K_3(\k,-\q,\q)P_{\text{lin}}(q)\,,
\end{split}
\ee
where we introduced the new selection-dependent redshift space 
kernels, 
\be
\label{eq:K2full}
\begin{split}
K_1(\k) = & b_1-b_\eta f\mu^2\,,\\
K_2(\k_1,\k_2)=&\frac{b_2}{2}+b_{\mathcal{G}_2}\left(\frac{(\k_1\cdot \k_2)^2}{k_1^2 k_2^2}-1\right)+b_1F_2(\k_1,\k_2)-b_\eta f\mu^2 G_2(\k_1,\k_2)\\
& - fb_{\delta \eta}\frac{\mu_2^2+\mu_1^2}{2} +b_{\eta^2}f^2\mu_1^2\mu_2^2\\
& +b_1f\frac{\mu_1\mu_2}{2}\left(\frac{k_2}{k_1} + \frac{k_1}{k_2}\right)
-b_\eta f^2\frac{\mu_1\mu_2}{2}\left(\frac{k_2}{k_1}\mu_2^2 + \frac{k_1}{k_2}\mu_1^2\right)\\
& + b_{(KK)_\parallel}\left(\mu_1\mu_2 \frac{(\k_1\cdot \k_2)}{k_1k_2}
-\frac{\mu_1^2+\mu_2^2}{3}+\frac{1}{9}
\right)\\
& + b_{\Pi^{{[2]}}_\parallel}\left(\mu_1\mu_2 \frac{(\k_1\cdot \k_2)}{k_1k_2}+\frac{5}{7}\mu^2 \left(1-\frac{(\k_1\cdot \k_2)^2}{k_1^2 k_2^2}\right)\right)\,,
\end{split} 
\ee
\be 
\label{eq:K3full}
\begin{split}
& \int_{\q} K_3(\k,\q,-\q) 
P_{\rm lin}(q)
\\
=\,&
b_1
\int_{\q} F_3(\k,\q,-\q)P_{\text{lin}}(q)
-
f b_\eta\mu^2
\int_{\q} G_3(\q,-\q,\k)P_{\text{lin}}(q)
\\
&\:+ 
\int_{\q}
\left[
1 - \left(\khat\cdot\qhat\right)^2
\right]
P_{\rm lin}(q)
\\
&\:\times
\Bigg\{
\frac{4}{21}
(
5b_{\mathcal{G}_2}
+ 
2b_{\Gamma_3})
\left[\left(\frac{(\k-\q)\cdot\q}{|\k-\q|q}\right)^2 - 1\right]
- 
\frac{2}{21} f b_{\delta\eta}
\left[
   \frac{3(k_\parallel-q_\parallel)^2}{|\k-\q|^2}
   +
   \frac{5q_\parallel^2}{q^2}
\right]
\\
&\quad\:
+\frac47 f^2 b_{\eta^2}
\frac{q_\parallel^2}{q^2}
\frac{(k_\parallel-q_\parallel)^2}{|\k-\q|^2}
+
\frac{20}{21} b_{(KK)_\parallel}
\left[
    \frac{(\k\cdot\q-q^2)(k_\parallel-q_\parallel)q_\parallel}{|\k-\q|^2q^2} 
    - \frac13\frac{(k_\parallel-q_\parallel)^2}{|\k-\q|^2} - \frac13 \frac{q_\parallel^2}{q^2} + \frac19
\right]
\\
&\quad\:+\frac{10}{21}
b_{\Pi_\parallel^{[2]}}
\frac{(\k\cdot\q-q^2)}{|\k-\q|^2}
\frac{(k_\parallel-q_\parallel)^2}{q^2}
+
\frac{10}{21}
\left[
b_{\delta\Pi_\parallel^{[2]}} 
-\frac13 b_{(K\Pi^{[2]})_\parallel}
- 
f b_{\eta\Pi_\parallel^{[2]}} 
\frac{q_\parallel^2}{q^2}
\right]
\frac{(k_\parallel-q_\parallel)^2}{|\k-\q|^2}
\\
&\quad\:+
\frac{10}{21}
b_{(K\Pi^{[2]})_\parallel}
\frac{(\q\cdot\k-q^2)}{q|\k-\q|}
\frac{q_\parallel(k_\parallel-q_\parallel)}{q|\k-\q|}
+
\frac{10}{21}
f b_{\Pi^{[2]}_\parallel}
\frac{q_\parallel(k_\parallel-q_\parallel)^3}{q^2|\k-\q|^2}
\\
&\quad\:
+
(b_{\Pi_\parallel^{[3]}}+2b_{\Pi_\parallel^{[2]}})
\Bigg[
\frac{13}{21}
\frac{\k\cdot\q-q^2}{|\k-\q|^2}
\frac{q_\parallel(k_\parallel-q_\parallel)}{q^2} 
-
\frac{5\mu^2}{9}
\left[\left(\frac{(\k-\q)\cdot\q}{|\k-\q|q}\right)^2 - \frac{1}{3}\right]
\Bigg]
\\
&\quad\:+
\frac{2}{21}f b_1 
\left[
5
\frac{q_\parallel(k_\parallel-q_\parallel)}{q^2}
+
3
\frac{q_\parallel(k_\parallel-q_\parallel)}{|\k-\q|^2}
\right]
-
\frac27 
f^2 
b_\eta
\frac{q_\parallel(k_\parallel-q_\parallel)}{q^2|\k-\q|^2}
\left[
(k_\parallel-q_\parallel)^2
+
q_\parallel^2
\right]
\Bigg\}\,.
\end{split}
\ee
Note that for convenience, 
we presented the $K_3$ integrand instead of the kernel, 
as in this case the expression greatly simplifies.
{Eq.~\eqref{eq:K3full} reproduces the full one-loop $P_{13}$ for all bias structures up to scaleless terms, which vanish in dimensional regularization and hence in our FFTLog evaluation of Section~\ref{sec:fftlog}. If the integral is evaluated with a hard cutoff instead, the otherwise unobserved scaleless pieces must be restored. These are, first, the IR term required by the equivalence principle, $-K_1(\k)\int_\q(2f(\k\cdot\q)k_\parallel q_\parallel+f^2k_\parallel^2q_\parallel^2)P_{\rm lin}(q)/(6q^4)$, which must be added to the right-hand side of Eq.~\eqref{eq:K3full}, and, second, UV constants that renormalize $b_1$ and $b_\eta$, proportional to the variance $\sigma^2_\Lambda$ of the linear field below the cutoff $\Lambda$ (Eq.~\eqref{eq:sLambda}; Appendix~\ref{app:renorm}).

Up to UV constants, the cubic operators produce only four independent $P_{13}$ shapes, whilst the quadratic operators do not produce any new shapes. This follows from exact identities between the angle-averaged 
$P_{13}$-type integrands $\mathcal{I}_{\mathcal O}(k,q,\mu)\equiv3\langle K_3(\k,\q,-\q)\rangle_{\hat q}\big|_{b_{\mathcal O}}$ of the corresponding bias parameters 
$b_{\mathcal O}$,
which hold for all $q$:
\be
\label{eq:degen}
\begin{split}
&\mathcal{I}_{\Pi^{[3]}_\parallel}-\tfrac{35}{24}\mathcal{I}_{\Gamma_3}-\tfrac{109}{90}\mathcal{I}_{\delta\Pi^{[2]}_\parallel}+\tfrac{101}{30}\mathcal{I}_{(K\Pi^{[2]})_\parallel}=\tfrac{181}{450}+\tfrac{112}{75}\mu^2\,,\\
&\mathcal{I}_{\mathcal{G}_2}-\tfrac52\mathcal{I}_{\Gamma_3}=0\,,\qquad
\mathcal{I}_{(KK)_\parallel}-2\,\mathcal{I}_{(K\Pi^{[2]})_\parallel}=\tfrac{226}{945}\left(1-3\mu^2\right)\,,\\
&\mathcal{I}_{\delta\eta}+\tfrac{3f}{5}\mathcal{I}_{\delta\Pi^{[2]}_\parallel}=-\tfrac{f}{35}\left(11+12\mu^2\right)\,,\qquad
\mathcal{I}_{\eta^2}+\tfrac{6f}{5}\mathcal{I}_{\eta\Pi^{[2]}_\parallel}=-\tfrac{2}{25}f^2\mu^2\,.
\end{split}
\ee
The right-hand sides of the above equations do not depend on $q$, so they reduce to $(a+b\mu^2)\sigma^2_\Lambda$ upon loop integration with the linear power spectrum. These terms vanish in dimensional regularization and otherwise only renormalize $b_1$ and $b_\eta$.

The above relation between $\mathcal{I}_{\mathcal{G}_2}$ and $\mathcal{I}_{\Gamma_3}$ stemming from the SPT-evolved quadratic tidal operator $\mathcal{G}_2$ and ${\Gamma_3}$
is well known in the context of galaxy bias.
All other relations in the above group except the first one are its Lyman-$\alpha$ analogs.
In contrast, the first relation is an identity
involving only the cubic kernels. This identity can be used 
to eliminate one of the cubic operators. Using this, in what follows we set $b_{\Gamma_3}=0$,
which can be used without loss of generality
as long as one models the power spectrum
to one-loop order.

\subsection{Higher derivative counterterms and stochastic 
contributions}

The line-of-sight dependent higher derivative and stochastic 
contributions 
relevant for the one-loop power spectrum of a generic 
selection-dependent tracer 
are also known~\cite{Desjacques:2018pfv},
\be
\label{eq:hd}
\begin{split}
\delta_F^{(s)}\Big|_{\text{stoch+h.d.}} = \epsilon 
+ {R_*^2\left[b_{\nabla^2\delta}\nabla^2 \delta+b_\eta\left(\beta_{\nabla^2 \textbf{v}}\nabla^2\eta + \beta_{\d^2_{||} \textbf{v}}\d^2_{||}\eta\right)\right]}\,,
\end{split} 
\ee 
where $\epsilon$ is the stochastic field uncorrelated with $\delta$. At linear order Eq.~\eqref{eq:hd} gives 
\[\delta_F^{(s)}\Big|_{\text{h.d.}} =
  -k^2(\gamma_0+\gamma_2\mu^2+\gamma_4\mu^4)\delta
\] in Fourier space, with $\gamma_0=b_{\nabla^2\delta}R_*^2$, $\gamma_2=-fb_\eta\beta_{\nabla^2\textbf{v}}R_*^2$ and $\gamma_4=-fb_\eta\beta_{\d^2_{||}\textbf{v}}R_*^2$.
Note that we do not explicitly add 
the redshift-space EFT counterterms $\sim k^2\mu^{2n}\delta$~\cite{Senatore:2014vja} in Eq.~\eqref{eq:hd} as they are
fully degenerate with the higher-derivative operators 
that we already have. Indeed, at linear order the equivalence principle, rotations around the line of sight and the $z\to-z$ parity allow exactly the structures $k^{2m}\mu^{2j}\delta$ with $j\leq m+1$. At order $k^2$ these are exactly $k^2\delta$, $k^2\mu^2\delta$ and $k^2\mu^4\delta$, which Eq.~\eqref{eq:hd} already spans. 

At the same time, our scaling Universe analysis 
suggests that the higher-derivative contributions 
scale as three-loop order contributions, and hence may be ignored
at the one-loop order we are interested in. 
On the other hand, the  
precision of the Sherwood simulation
data is sufficient to nominally detect these terms even 
when their Wilson coefficients are very small, 
of the order $10^{-3}~[\Mpch]^2$.
The inclusion of these corrections 
noticeably improves the fit at $k\sim 3~\hMpc$,
which is why we prefer to keep them.
The theoretical 
rationale behind this choice is that these corrections 
can be treated as a proxy 
for higher loop contributions,
which can be partially absorbed into the counterterms~\eqref{eq:hd}.
Thus, their inclusion may indeed improve the fit 
even if two-loop and three-loop contributions are
not explicitly included.

As far as the stochastic corrections are concerned, their
expression up to $(k/\knl)^2$ order is given by
\be 
\label{eq:stochP}
\langle \epsilon(\k)\epsilon(\k') \rangle  =(2\pi)^3\delta_D^{(3)}(\k+\k')\left(
P_{\text{shot}}+a_0\frac{k^2}{\knl^2}+a_2\frac{k^2\mu^2}{\knl^2}{+a_4\frac{k^2\mu^4}{\knl^2}}\right)\,,
\ee
where $P_{\rm shot}$, $a_{0,2,4}$ 
are dimensional
stochastic counterterms. 
$P_{\rm shot}$, $a_{0,2}$ terms are standard redshift-space stochasticity terms derived e.g.
in~\cite{Perko:2016puo,Desjacques:2018pfv}
The $a_4$ term is a new term absent for galaxies, whose rest-frame statistics are isotropic. For a line-of-sight dependent tracer it is both allowed by symmetry and generated by loops.
Its origin can be understood as follows.
$\delta_F$ depends on the stochastic 
part of the dark matter field
through the line-of-sight operators 
$\eta,\Pi^{[n]}_{\parallel}$ in the 
Eulerian bias expansion, which receive 
stochastic contributions from the dark matter
stress tensor in the EFT. If $\textbf{v}_\epsilon$ is the stochastic dark matter velocity, we get 
a contribution of the form
$\delta_F\supset b_\eta \eta_\epsilon=b_\eta\d_z v_{z,\epsilon}/(aH)$, which generates a Fourier space
contribution 
\be\label{eq:stoch_tau}
\delta_F\propto \hat{z}^i\hat{z}^jk_ik_j \frac{\theta_{\epsilon}}{k^2} \propto \hat{z}^i\hat{z}^jk_ik_j \frac{k^lk^m\tilde\tau_{lm}}{k^2}\,,
\ee
where $\theta_{\epsilon}$ is the stochastic
dark matter
velocity divergence stemming from the dark matter stress tensor $\tilde\tau_{lm}$ integrated over time with
the velocity Green's function. Symmetries allow 
$\tilde\tau_{lm}$
to have an anisotropic correlation with 
the Lyman-$\alpha$ stochasticity field $\epsilon$ in the rest frame, 
\[
  \langle \epsilon \tilde\tau_{lm} \rangle'=n\delta^{lm}+n'\hat{z}^l\hat{z}^m\,,
\]
which together with Eq.~\eqref{eq:stoch_tau} implies
\be 
 \langle \epsilon \delta_F^\epsilon  \rangle'\supset n' \hat{z}^i\hat{z}^jk_ik_j \frac{k_lk_m\hat{z}^l\hat{z}^m}{k^2}=n'\mu^4 k^2\,,
\ee 
where $\delta_F^\epsilon$ is the stochastic 
part of the flux contrast inherited from
the underlying matter non-linearity, see Eq.~\eqref{eq:stoch_tau}.
The isotropic noise piece $n$
produces the $k^2\mu^2$ contribution which we already have. 
We generalize this argument and explain the origin of the $k^2\mu^4$ stochastic term in detail
in Appendix~\ref{app:a4}.

The shot noise corrections are expected to be small
and can be 
ignored in practice for the 3D correlators, but are important for the 1D 
power spectrum, see Section~\ref{sec:1d} for more detail. 

\subsection{IR resummation}

Baryon acoustic oscillations 
are affected by large IR effects, 
which may be resummed along the lines of Refs.~\cite{Senatore:2014via,Vlah:2015zda,Baldauf:2015xfa,Blas:2015qsi,Blas:2016sfa,Ivanov:2018gjr,Vasudevan:2019ewf}.
These effects are fully controlled by the  
displacement and velocity fields, which are both locally unobservable. 
In the context of the Eulerian fluid description, the structure of these IR sensitive terms is dictated by the 
equivalence principle, which demands that the IR-enhanced 
contributions cancel in the formal limit $\{q_i\}\to 0$, where $\{q_i\}$ denote 
a set of 
loop momenta. The cancellation is inexact if
some $q_i$'s are greater than the BAO scale $r_{\rm BAO}\sim 110~{\Mpch}$.
In that case (IR) resummation is needed. 
Since IR resummation is controlled by the IR divergent terms 
whose form is completely fixed by the equivalence 
principle, it would be unaffected by the line-of-sight selection bias
operators, which are finite in the IR limit by construction. This implies that 
IR resummation 
for a generic line-of-sight dependent tracer 
would be identical to that of redshift space galaxies~\cite{Ivanov:2018gjr},
as was first pointed out in~\cite{Desjacques:2018pfv}.
Indeed, one can explicitly check that the new 
line-of-sight dependent operators do not 
produce any new IR singularities 
in the soft limit, see e.g.~\eqref{eq:K3full}. 

Even though IR resummation can be straightforwardly included,
we defer its explicit implementation
for future work. First, the main goal of this paper is to study 
the performance of the EFT on small scales, $k\sim 3~\hMpc$, where 
the BAO feature is completely washed out. 
Second, the hydrodynamical simulations
that we use here do not have enough large-scale resolution for a significant detection of 
the BAO in
the data. Given these reasons, we believe that the non-linear effects 
on the BAO deserve a separate dedicated analysis that goes beyond the scope of this paper.

\subsection{Calculation of loop integrals with FFTLog}
\label{sec:fftlog}

The one-loop expression~\eqref{eq:deltaP1L} can be transformed to the form
suitable for the evaluation with the FFTLog method~\cite{Simonovic:2017mhp} (see also~\cite{McEwen:2016fjn,Fang:2016wcf}). 
The basic idea of this method is to represent the linear matter power spectrum,
which is an input in all loop calculations, as a sum of the power-law functions.
Then the loop integrals for each power law can be done analytically, 
and a loop calculation reduces to a matrix multiplication
problem.

In redshift space, following Ref.~\cite{Chudaykin:2020aoj} (see also~\cite{Chen:2020fxs}), 
one has to expand the loop integrands over a basis 
of master operators involving contractions of the line of sight vector $\z$
and loop momenta. 
One technical difficulty w.r.t. the case of galaxies 
is that one needs such master {expressions} up to eight momenta instead of four 
in the galaxy case. 
This is quite straightforward albeit the expressions are 
rather lengthy. The eventual expressions for the relevant master 
integrals after taking all the necessary contractions of the
basis operators are given by
\be
\label{eq:masterint}
\begin{split}
&\int_\q\frac{(\z\cdot \q)^5}{q^{2\nu_1}|\q-\k|^{2\nu_2}}=k^{3-2\nu_{12}}\cdot
k^5\mu(A_5+\mu^2 B_5 + \mu^4 C_5)\,,\\
&\int_\q\frac{(\z\cdot \q)^6}{q^{2\nu_1}|\q-\k|^{2\nu_2}}=k^{3-2\nu_{12}}\cdot
k^6(A_6+\mu^2 B_6 + \mu^4 C_6+\mu^6 D_6)\,,\\
&\int_\q\frac{(\z\cdot \q)^7}{q^{2\nu_1}|\q-\k|^{2\nu_2}}=k^{3-2\nu_{12}}\cdot
k^7\mu(A_7+\mu^2 B_7 + \mu^4 C_7+\mu^6 D_7)\,,\\
&\int_\q\frac{(\z\cdot \q)^8}{q^{2\nu_1}|\q-\k|^{2\nu_2}}=k^{3-2\nu_{12}}\cdot
k^8(A_8+\mu^2 B_8 + \mu^4 C_8+\mu^6 D_8 + \mu^8 E_8)\,,
\end{split} 
\ee
where $\nu_{12}\equiv \nu_1+\nu_2$ and the master functions $A_n,B_n,C_n,D_n,E_n$
for $n=5,6,7,8$
are given in Appendix~\ref{app:master}. The rest of the master integrals 
are presented in Appendix A of~\cite{Chudaykin:2020aoj}. 
With the help of master integrals like~\eqref{eq:masterint},
the total mode-coupling power spectrum may be written as
\be
\Delta P^{\rm 1-loop}=\sum_{n=0}^4 \mu^{2n} P^{(22)}_n + (b_1-b_\eta f\mu^2)
\sum_{n=0}^2 \mu^{2n} P^{(13)}_n\,,
\ee
where $ P^{(22)}_n$ and $P^{(13)}_n$ depend on the biases that appear in 
Eq.~\eqref{eq:F_full2}. 

An important aspect of the FFTLog method is the choice of the so-called
FFTLog bias parameter $\nu$ which appears 
in the FFTLog power law approximation as $P_{\rm lin}\propto k^\nu$, and hence
it controls the convergence of the loop integrals. Most of the 22-type 
integrals that we encounter converge for $-1<\nu<1/2$. The lower end of this range 
corresponds to spurious IR contributions that cancel in the power spectrum
due to the equivalence principle~\cite{Blas:2013bpa,Blas:2015qsi}.
If this IR contribution is taken into account, one can choose an FFTLog
bias in the range $-3 <\nu < -1$. 
Certain bias integrals, like the ones proportional to 
$b_2^2$, converge only for $-3<\nu<-3/2$. 
For $\nu>-3/2$ they have formally UV divergent constant parts
that behave as~\cite{Assassi:2014fva,Simonovic:2017mhp} 
\be 
\label{eq:const}
I_{\delta^2\delta^2}(0)\equiv \frac{1}{2}\int_{\q}  P^2_{\rm lin}(q)\,.
\ee 
In the case of 
galaxies the standard practice is to 
subtract these terms from the final answer as they simply renormalize the 
constant shot noise contribution. For the Lyman alpha, however, the intrinsic
shot noise contribution produced by tracer discreteness is 
negligibly small, while the constant piece~\eqref{eq:const}
is quite large for the physical $\Lambda$CDM power spectrum,
\be 
\label{eq:const2}
I_{\delta^2\delta^2}(0)\Big|_{z=2.8}\approx 28~[\Mpch]^3\,.
\ee
Note that this integral is saturated in the IR, so it would have the same
value even if we cut it off at $\Lambda\sim k_{\rm NL}$. Hence, 
Eq.~\eqref{eq:const2} represents a calculable, physical 
constant power spectrum contribution that is deterministic 
by origin, and must be present in the model. Let us estimate its magnitude.
When discussing the Tau model below, we will get a natural
estimate $b_2\sim  0.3$, so that
\be
\label{eq:pshotest}
P_{\rm shot}\sim \frac{b_2^2}{2} \int_{\q,|q|\leq k_{\rm NL}}  P^2_{\rm lin}(q)\approx b_2^2 \times I_{\delta^2\delta^2}(0)\approx 2.5~[\Mpch]^3\,.
\ee
This estimate is two orders of magnitude 
larger than the naive scaling Universe 
result $P_{\rm shot}\sim k_{\rm NL}^{-3}\sim 10^{-2}~[\Mpch]^3$, 
and also many orders of {magnitude} larger than the  
discreteness shot noise. 

The constant 
terms like~\eqref{eq:pshotest}
are fully calculable. They are 
produced mostly by perturbative modes and 
they do not require introducing new parameters as counterterms, as their 
UV-sensitivity is very mild for $\Lambda$CDM. 
The deterministic constant pieces like~\eqref{eq:pshotest}
should be contrasted with the stochastic constant shot
noise contributions due to discreteness effects. 
The latter are extremely small for the
Lyman alpha forest due to high column densities, and hence can be
neglected for the precision we are interested in. 
Note that unlike the stochastic constant pieces, 
the deterministic constant pieces do not require free parameters -- their 
contributions are fully fixed in terms of non-trivial 
bias operators that survive in the limit $K_2(\q,-\q)$, like $b_2$. {Explicitly, the constant is $S\int_\q P^2_{\rm lin}(q)$ with $S=2\langle K_2(\q,-\q)^2\rangle_{\hat q}$, which involves all operators that survive at zero total momentum, including the line-of-sight dependent ones (Appendix~\ref{app:renorm}).}
In order to retain these terms, we choose the FFTLog bias for $P_{22}$-type
integrals to be $\nu = -1.7${.}

As far as the 13-type corrections are concerned, there is no
single choice of $\nu$ that would make them converge both in the IR and UV. 
Following the discussion above, we use $\nu=-1.7$ that 
leads to the convergence in the UV, but misses 
the spurious IR-sensitive contributions. Together with our choice $\nu=-1.7$ for 
the $P_{22}$-type integrals, all spurious IR corrections are consistently
absent in our loop integrals.

The convergent UV contributions
of one-loop integrals
are proportional to the displacement variance,
\be 
\label{eq:velvar}
\sigma_v^2 =\frac{1}{3}\int_\q \frac{P_{\rm lin}(q)}{q^2}\Big|_{\Lambda\text{CDM}, z=2.8}
=4.0~[\Mpch]^2\,.
\ee
With our choice of the FFTLog bias, this correction is computed over the 
entire range of scales. The physical 
calculable part, however, should only involve modes with $k\lesssim k_{\rm NL}$.
This means that our choice of the FFTLog bias will introduce an error of 
\be 
\frac{1}{6\pi^2}\int_{k_{\rm NL}}^\infty  dq~q^2\frac{P_{\rm lin}(q)}{q^2}\approx 4\cdot 10^{-3}~[\Mpch]^2~\,.
\ee
Even if this error is very small in absolute terms, we will find that the 
Sherwood data is actually sensitive to these corrections. This is another 
reason to keep the higher-derivative counterterms in the fit: they compensate 
for errors in loop calculations. More precisely, the contribution of modes with $q\gg k$ to the 13-type correction can be written as $2K_1(\k)P_{\rm lin}(k)\,k^2\,T(\mu)\int_\q P_{\rm lin}(q)/q^2$, where $T(\mu)$ is a polynomial of degree four in $\mu$.
The error estimated above is therefore exactly absorbed by a shift of the counterterms of Eq.~\eqref{eq:hd}, which we discuss in detail in Section~\ref{sec:sims} and Appendix~\ref{app:renorm}.

\section{Top-down derivation from the exponential map}
\label{sec:expmap}

The symmetry-based EFT expansions~\eqref{eq:F_full} and~\eqref{eq:F_full2}
are complete and general. 
It is interesting to see directly how these expressions
arise from the non-linear transformation of the optical depth field.  
The optical depth itself is assumed to be a selection-free
biased tracer of the underlying matter density, just like 
the galaxies. In that case all selection-dependent contributions 
in the flux bias model should stem from the exponential map.
The main objective of this section is to show how this happens. 
Since we aim here at reproducing the known results~\eqref{eq:F_full2},
the content of this Section would be somewhat academic in nature.
We will argue that it is still important to the general understanding 
of the perturbative Lyman alpha modeling. 

The question of how the exponential map 
generates selection-dependent bias 
contributions was previously studied by Ref.~\cite{Chen:2021rnb}.
Specifically, this work showed how the exponential 
map and the SPT kernels 
generate certain terms in the EFT expansion.
In this {section} we build on results of~\cite{Chen:2021rnb}  
and study the relationship between the exponential map and 
the EFT bias expansion in more detail. 
The main novelty of our work is that we show explicitly 
that the mapping \textit{and} the requirement of renormalization 
generate a full set of the line-of-sight dependent EFT operators~\eqref{eq:F_full2}.
If we formally drop the requirement of renormalization,
the exponential map will produce only
a particular incomplete set of EFT operators 
with fixed Wilson coefficients, at the tree level.
We call the corresponding perturbative 
model ``the tree-level Tau-model.''
Naively, this model is extremely 
predictive, as it requires only one extra parameter
(velocity gradient bias, or the mean optical depth) in addition to the standard set of 
selection-free bias parameters.
We show, however, 
that this model receives infinitely large loop contributions, 
which renders it inconsistent. 
Once the loop corrections are appropriately renormalized, 
their finite pieces generate sizable corrections 
to all bias parameters, which will break 
the tree-level constraints imposed by the exponential map.
All in all, this means that even if we ignore  
counterterms in the perturbative expansion 
of the Lyman alpha forest flux at the beginning,
the loop corrections will generate them and make 
the exponential map expansion equivalent to the full EFT.

\subsection{Renormalized exponential map expansion}

We start with a {top-down} derivation involving the renormalization of the
flux contrast field. It is a {non-linear} transformation 
of the optical depth field $\tau$, which we assume to be a selection-independent tracer of matter. The exponential map reads
\be
\label{eq:tau}
F=e^{-\tau_0(1+\epsilon\delta_\tau)}\,,
\ee
where $\tau_0$ is the mean optical depth.
{In the} spirit of {the} EFT, this is a ``bare'' parameter that
has to be properly renormalized. We will address this issue soon.
$\epsilon$ is our order counting parameter that keeps track of the
order of our perturbative solution. It is to  be set to 1 in the final result. 
Note that $\tau$ is a function of the {redshift-space} coordinate~\cite{McDonald:1999dt}, 
i.e. the full RSD mapping~\eqref{eq:rsdmap}
has already been applied to it. Taylor expanding Eq.~\eqref{eq:rsdmap}
up to cubic order we obtain,\footnote{We use the notation 
$[g]_\k=\int d^3x ~e^{-i\k\x}g(\x)~$
for an arbitrary function $g(\x)$.}
\be 
\label{eq:rsdtau}
\begin{split}
\delta^{(s)}_\tau(\k)= & \delta^{(r)}_\tau(\k) - \frac{ik_zv_{\tau,z}(\k)}{\HH}
-\epsilon\frac{ik_z}{\HH}[\delta^{(r)}_\tau v_{\tau,z}]_\k
+\epsilon\frac{i^2k_z^2}{2\HH^2}[v^2_{\tau,z}]_\k \\
& +\epsilon^2\frac{i^2k_z^2}{2\HH^2}[v^2_{\tau,z}\delta^{(r)}_\tau]_\k
-\epsilon^2\frac{i^3k_z^3}{6\HH^3}[v^{{3}}_{\tau,z}]_\k\,,
\end{split}
\ee
where $\delta_\tau^{(r)}$ is the optical depth fluctuation field in real space, 
and we have used our bookkeeping parameter $\epsilon$ to trace 
perturbative order.
In what follows we will drop the $(s)$ superscript and use 
$\delta_\tau$ for the optical depth contrast in redshift space.
As a next step, we Taylor expand the expression~\eqref{eq:tau},
\be
\label{eq:Ftayl}
F= e^{-\tau_0}\left(1-\epsilon\tau_0 \delta_\tau +\frac{(\epsilon\tau_0\delta_\tau)^2}{2}-\frac{(\epsilon\tau_0\delta_\tau)^3}{6}+O(\epsilon^4)\right)\,{.}
\ee
We are interested in one-loop order so the series goes up to $O(\epsilon^4)$.
At this order, assuming $\langle \delta_\tau\rangle=0$, the mean flux is given by 
\be
\langle F \rangle  = e^{-\tau_0}\left(1 + \epsilon^2 \frac{\tau_0^2}{2}
\langle \delta^2_\tau\rangle 
+...\right)\,.
\ee
Here we assume that 
$\langle\delta_\tau\rangle=0$, i.e.\ $\tau_0=\langle\tau\rangle$ in redshift space. This is equivalent to using mean-subtracted composite operators in the bias expansion of $\delta^{(r)}_\tau$. Otherwise the mapping generates a spurious shift $\Delta b_\eta=\frac{\tau_0}{2}b_2^\tau\sigma^2_\Lambda$ at one loop (with $\sigma^2_\Lambda$ of Eq.~\eqref{eq:sLambda}), which leads to a redefinition $\tau_0\to\langle\tau\rangle$.
The mean flux is a function of the ``bare coupling'' $\tau_0$
that receives corrections order by order in perturbation theory.
These corrections are unobservable since they are
absorbed into $\tau_0$ in order to produce 
the physically observable mean flux $\langle F \rangle $ after renormalization.
The situation here is similar to the renormalization of the ``bare''
mean number density of galaxies in the context of the perturbative bias expansion~\cite{Desjacques:2016bnm}.
The flux fluctuations{, $\delta_F=F/\langle F\rangle-1$,} are given by
\be
\label{eq:dF_map}
\begin{split}
\delta_F=-\tau_0\epsilon \delta_\tau{\left(1-\frac{\epsilon^2\tau_0^2}{2}\langle\delta^2_\tau\rangle\right)} +
\frac{\tau_0^2}{2}\epsilon^2 (\delta^2_\tau - \langle \delta^2_\tau\rangle)
-\frac{(\epsilon\tau_0\delta_\tau)^3}{6}+{O}(\epsilon^4)\,.
\end{split} 
\ee
Setting $\epsilon=1$,
and going into Fourier space we get,
 \be
\begin{split}
\delta_F(\k)=-\tau_0{\left(1-\frac{\tau_0^2}{2}\langle\delta^2_\tau\rangle\right)}\delta_\tau(\k) +
\frac{\tau_0^2}{2} [\delta^2_\tau]_\k  
-\frac{(\tau_0)^3}{6}[\delta^3_\tau]_\k\,,
\end{split} 
\ee
where $\delta_\tau$ is given in Eq.~\eqref{eq:rsdtau}.
Note that so far we have not used the selection-independent 
bias model for $\delta^{(r)}_\tau$.
Similar to Eq.~\eqref{eq:rsdtau}, the Taylor expansion above 
produces operators evaluated at the same point in space that are highly UV sensitive. 
In the EFT approach, we need 
to make sure that $\delta_F$ receives contributions 
only from modes up to $k_{\rm NL}$ that are under perturbative control.
That means all functions in the r.h.s. of Eq.~\eqref{eq:Ftayl}, e.g.
$\delta_\tau$ as well as 
the composite local operators [$\delta_\tau^2$] and $[\delta_\tau^3]$ need to be {renormalized}.

Let us start with the linear term $[\delta_\tau]$, see Eq.~\eqref{eq:rsdtau}. 
First, one can plug the perturbative bias model for the real space $\delta_\tau$, 
which, according to our assumptions, does not have line-of-sight 
selection effects, see Eq.~\eqref{eq:bias_gal}. This model
is closed under renormalization so we do not need to add new counterterms.
The velocity terms in Eq.~\eqref{eq:rsdtau} are renormalized 
in the standard way, see Ref.~\cite{Senatore:2014vja}. Hence, this term also
does not require additional counterterms. 
Let us move to the operators [$\delta_\tau^2$] and $[\delta_\tau^3]$. 
Their renormalized versions must include all possible counterterms built out of the long-wavelength fields
and allowed by symmetries. 
In our case these are the equivalence principle and $SO(2)$ rotations around the line-of-sight. The equivalence 
principle means the counterterms can only be
functions of $\d_i\d_j\Phi$ and $\d_i v_j$ and time derivatives along the flow. 
Up to cubic order, these operators have already been introduced in Eq.~\eqref{eq:F_full}.
Thus, we write
\be
\begin{split}
&  [\delta^2_\tau]_\k\Big|_{\rm ren}= [\delta^2_\tau]_\k + b'_1\delta + b'_\eta \hat{z}^i \hat{z}^j \d_i \d_j \Phi + \sum b'_{\mathcal{O}}\mathcal{O} \,,\\
& [\delta^3_\tau]_\k\Big|_{\rm ren}= [\delta^3_\tau]_\k + b^{''}_1\delta + b^{''}_\eta \hat{z}^i \hat{z}^j \d_i \d_j \Phi + \sum b^{''}_{\mathcal{O}}\mathcal{O} \,,
\end{split}
\ee
where $\sum b^{',''}_{\mathcal{O}}\mathcal{O}$ stand for the quadratic and cubic operators 
in Eq.~\eqref{eq:F_full}.
This way we generate all the terms in the expansion that we already had before
in the ``bottom-up'' case.
The bias coefficients $b^{'}_\mathcal{O}$ and $b^{''}_\mathcal{O}$, along with 
the biases present in $[\delta^n_\tau]_\k$,
will sum up into the bias coefficients of Eq.~\eqref{eq:F_full}.
Note that velocity-dependent contributions dictated by the equivalence principle 
are contained inside the expressions $\delta^2_\tau$ and $\delta^3_\tau$. 
They are 
unaffected by the counterterms, 
as the EFT cannot generate terms that depend on the velocity 
itself, only on its gradients. 
All in all, this way we have arrived at the same EFT expansion as Eq.~\eqref{eq:F_full}.
We see that even if we start directly  
from the exponential map, the smoothing and renormalization 
will generate all the necessary selection-dependent bias operators.

To sum up, the renormalization procedure requires that 
coarse-grained composite operators stemming from the exponential map
contain all possible line-of-sight dependent bias operators as counterterms,
\be 
\label{eq:rencount}
[\delta^n_\tau]_\k\Big|_{\rm ren}= [\delta^n_\tau]_\k +\text{counterterms}\,.
\ee
This is similar to usual renormalization conditions for local operators in 
quantum field theory. 
If we were to ignore the counterterms, the perturbative expansion 
generated by the exponential map would only involve 
simple contractions of $\delta^n_\tau$ whose bias coefficients 
would be completely fixed by the mapping. It is instructive to 
study the structure of this expansion and contrast it with the full EFT. 
Since the counterterms 
are typically considered along {with} the loop corrections, with
some abuse of language we call such an expansion ``the tree-level Tau model.''
The goal of this discussion is to give a pedagogical example 
of how the counterterms in Eq.~\eqref{eq:rencount} will reappear
from loops even if we ignore them to begin with.

\subsection{Tree-level Tau model}
\label{sec:tautree}

Let us study in more detail the ``unrenormalized'' part of the exponential map expansion. 
A similar derivation was recently done by~\cite{Chen:2021rnb}. 
We reproduce parts of this
derivation here and extend the results of~\cite{Chen:2021rnb} to non-linear selection-dependent 
bias terms. 
It is interesting to {re-derive them} 
in the context of the standard perturbation theory (SPT) without renormalization~\cite{Bernardeau:2001qr}.
Our starting point would be the exponential map~\eqref{eq:tau}
that we will Taylor expand and arrive at Eq.~\eqref{eq:dF_map}
for the fluctuations.
The difference now is that we assume that fluctuations of 
$\tau$ are perturbative in the matter density, 
and they satisfy the standard bias expansion similar
to~\eqref{eq:bias_gal},\footnote{In the SPT context this assumption 
is wrong because it is precisely coarse-graining and renormalization
that guarantee the adequacy of the perturbative regime.} 
\[
\delta_\tau = \delta^{(1)}_\tau+\epsilon\delta^{(2)}_\tau+\epsilon^2\delta^{(3)}_\tau\,,
\]
where $\delta^{(n)}_\tau\sim Z_n [\delta^{(1)}]^n$. 
Plugging this into \eqref{eq:dF_map} we get:
\be
\label{eq:expansiontau}
\begin{split}
\delta_F =&-\tau_0 \epsilon \delta^{(1)}_\tau 
-\tau_0\epsilon^2\delta^{(2)}_\tau -\tau_0\epsilon^3\delta^{(3)}_\tau +\\
&+\frac{\tau_0^2\epsilon^2}{2} ((\delta^{(1)}_\tau )^2+2\epsilon\delta^{(2)}_\tau \delta^{(1)}_\tau  - \langle \delta^2_\tau \rangle)-\frac{(\epsilon\tau_0\delta^{(1)}_\tau )^3}{6}{+\frac{\epsilon^3\tau_0^3}{2}\langle\delta^2_\tau\rangle\delta^{(1)}_\tau}+O(\epsilon^4)\,.
\end{split} 
\ee
Using the explicit 
expression for the standard redshift-space 
kernels~\eqref{eq:Zexp}, the expansion above 
can be rewritten in Fourier space 
in the standard perturbation theory form
as
\be 
\label{eq:KTau}
\begin{split}
{\delta_F}(\k)=&\tilde{K}_1(\k)\delta^{(1)}_\k + \int_{\q_1}
\int_{\q_2} 
(2\pi)^3\delta_D(\k-\q_{12})\tilde{K}_2(\q_1,\q_2)\delta^{(1)}_{\q_1}
\delta^{(1)}_{\q_2}+\\
&+\int_{\q_1} \int_{\q_2}\int_{\q_3}
(2\pi)^3\delta_D(\k-\q_{123})\tilde{K}_3(\q_1,\q_2,\q_3)\delta^{(1)}_{\q_1}
\delta^{(1)}_{\q_2} \delta^{(1)}_{\q_3}+...
\end{split}
\ee
where
\be 
\label{eq:Ktilde}
\begin{split}
& \tilde{K}_1(\k) = (-\tau_0 b^{\tau}_1)+ ( -\tau_0 ) f\mu^2 \,,\\
& \tilde{K}_2(\k_1,\k_2)  = \frac{(-\tau_0b^{\tau}_2)}{2}+(-\tau_0b^{\tau}_{\mathcal{G}_2})\left(\frac{(\k_1\cdot \k_2)^2}{k_1^2 k_2^2}-1\right)+(-\tau_0 b^{\tau}_1)F_2(\k_1,\k_2)-\tau_0 f\mu^2 G_2(\k_1,\k_2)\\
&-\tau_0\frac{f\mu k}{2}\left(\frac{\mu_1}{k_1}(b^{\tau}_1+f\mu_2^2)+\frac{\mu_2}{k_2}(b^{\tau}_1+f\mu_1^2)\right) 
+ \frac{\tau_0^2}{2}((b^{\tau}_1)^2 
+ b^{\tau}_1f(\mu_1^2+\mu_2^2)+f^2\mu_1^2\mu_2^2)\,,\\
&\tilde{K}_3(\k_1,\k_2,\k_3) =-\tau_0 Z_3({\k_1,\k_2,\k_3}) +\tau_0^2 [Z_2({\k_1,\k_2})Z_1({\k_3})]_{\rm symm.} 
- \frac{\tau_0^3}{3!}Z_1({\k_1})Z_1({\k_2})Z_1({\k_3})\,.
\end{split} 
\ee
Comparing the linear kernel $\tilde K_1$ with the EFT expression~\eqref{eq:K2full} we identify
\be
\begin{split}
& b_1  = -\tau_0 b^\tau_1 \,, \\ 
& b_\eta = \tau_0 \,.
\end{split}
\ee
As anticipated, the exponential map generates the velocity gradient bias
already at the linear level~\cite{Seljak:2012tp}.
Let us focus now on the quadratic kernel $\tilde K_2$.
Rewriting {the} expression in {the} second line of the $\tilde K_2$ 
expansion above as
\be
\begin{split}
&-\tau_0\frac{f\mu k}{2}\left(\frac{\mu_1}{k_1}(b^{\tau}_1+f\mu_2^2)
+\frac{\mu_2}{k_2}(b^{\tau}_1+f\mu_1^2)\right) + \frac{\tau_0^2}{2}(
b^{\tau}_1f(\mu_1^2+\mu_2^2)+f^2\mu_1^2\mu_2^2)\\
&= - f(\tau_0 b^{\tau}_1-\tau_0^2 b^{\tau}_1)\frac{\mu_2^2+\mu_1^2}{2} 
+(-\tau_0+\frac{\tau_0^2}{2})f^2\mu_1^2\mu_2^2 \\
&+(-\tau_0b^{\tau}_1)f\frac{\mu_1\mu_2}{2}\left(\frac{k_2}{k_1} + \frac{k_1}{k_2}\right)
-\tau_0 f^2\frac{\mu_1\mu_2}{2}\left(\frac{k_2}{k_1}\mu_2^2 + \frac{k_1}{k_2}\mu_1^2\right)\,,\\ 
\end{split}
\ee
and matching the entire kernel with \eqref{eq:K2full} we obtain:
\be
\begin{split}
& b_{2} = -\tau_0 b^{\tau}_{2}+\tau_0^2 (b^{\tau}_1)^2\,, \quad
 b_{\mathcal{G}_2} = -\tau_0 b^{\tau}_{\mathcal{G}_2}\,, \quad
 b_{\delta \eta} = \tau_0 b^{\tau}_1-\tau_0^2 b^{\tau}_1\,,\\
& b_{\eta^2} = -\tau_0 +\frac{\tau_0^2}{2}\,,
\quad ~~~~~~~~ b_{(KK)_\parallel}=0\,,
\quad ~~~~~ b_{\Pi^{[2]}_\parallel}=0\,.
\end{split}
\ee
Assuming that the bias parameters of the optical depth field 
are order one numbers, just like those of galaxies, $b_1^\tau\sim b_2^\tau \sim 1$,
we get an estimate $b_1 \sim b_2\sim \tau_0\approx 0.3$~\cite{Arinyo-i-Prats:2015vqa}.

As far as the $\tilde K_3$ 
{is} concerned, we see that in addition to the standard term $\sim Z_3$,
there are new 
contributions proportional to $Z_1^3$ and $Z_2Z_1$ kernels,
which are generated by the exponential map. 
However, one can easily check that only the $Z_2Z_1$ 
term contributes non-trivially to the one-loop power spectrum. 
More specifically, only the term containing $G_2$
has a finite contribution that {consistently} reproduces
the cubic order terms that must stem from the non-trivial
$\delta \eta$, $\eta^2$ and $\eta$ operators. 
The rest of the additional contributions are proportional to 
the UV-diverging mass variance 
\be 
\label{eq:sLambda}
\sigma^2_\Lambda = \int_0^\Lambda \frac{dq}{2\pi^2}~q^2P_{\text{lin}}(q)\,,
\ee
where $\L$ is a UV cutoff. 
We will discuss the renormalization of these 13-type terms shortly. 
At the kinematic configurations 
relevant for the one-loop power spectrum we thus have
\be 
\label{eq:tildeK3}
\begin{split}
& \int_{\q} \tilde K_3(\k,\q,-\q) 
P_{\text{lin}}(q)
=
-\tau_0b^\tau_1
\int_{\q} F_3(\k,\q,-\q)P_{\text{lin}}(q)
-
f \tau_0\mu^2
\int_{\q} G_3(\q,-\q,\k)P_{\text{lin}}(q)
\\
&\:+ 
\int_{\q}
\left[
1 - \left(\hat\k\cdot\hat\q\right)^2
\right]
P_{\text{lin}}(q)
\\
&\:\times
\Bigg\{
\frac{4}{21}
(
5b_{\mathcal{G}_2}
+ 
2b_{\Gamma_3}
)
\left[\left(\frac{(\k-\q)\cdot\q}{|\k-\q|q}\right)^2 - 1\right]
- 
\frac{2}{21} f{\left(\tau_0b_1^\tau-\tau_0^2b_1^\tau\right)}
\left[
   \frac{3(k_\parallel-q_\parallel)^2}{|\k-\q|^2}
   +
   \frac{5q_\parallel^2}{q^2}
\right]
\\
&\quad\:
+\frac47 f^2 {\left(-\tau_0+\frac{\tau_0^2}{2}\right)}
\frac{q_\parallel^2}{q^2}
\frac{(k_\parallel-q_\parallel)^2}{|\k-\q|^2}
\\
&\quad\:+
\frac{2}{21}f (-\tau_0b^\tau_1) 
\left[
5
\frac{q_\parallel(k_\parallel-q_\parallel)}{q^2}
+
3
\frac{q_\parallel(k_\parallel-q_\parallel)}{|\k-\q|^2}
\right]
-
\frac27 
f^2 
\tau_0
\frac{q_\parallel(k_\parallel-q_\parallel)}{q^2|\k-\q|^2}
\left[
(k_\parallel-q_\parallel)^2
+
q_\parallel^2
\right]
\Bigg\}\,.
\end{split}
\ee
Comparing this with the full EFT expression, we observe that the Tau model 
predicts that all EFT cubic 
biases are zero except $b_{\Gamma_3}=-\tau_0b_{\Gamma_3}^\tau$.

We call Eq.~\eqref{eq:Ktilde} the tree-level Tau model. 
This model involves only one extra parameter, $\tau_0$. It also suggests
that most of the one-loop selection-dependent 
EFT operators vanish,
\be 
\label{eq:constr0}
\begin{split}
b_{(KK)_\parallel}=b_{\Pi^{[2]}_\parallel}=b_{\Pi^{[3]}_\parallel}
=b_{\delta\Pi^{[2]}_\parallel}=b_{\eta\Pi^{[2]}_\parallel}
=b_{(K\Pi^{[2]})_\parallel}=0\,,
\end{split}
\ee
and that all non-trivial quadratic
selection-dependent biases in the EFT expansion 
can be expressed through linear biases as\footnote{The vanishing of $b_{(KK)_\parallel}$ and $b_{\Pi^{[2]}_\parallel}$ happens in the basis of Eq.~\eqref{eq:F_full}, which uses $\eta$ as the linear line-of-sight operator. In this basis it holds to all orders in the perturbative expansion of the exponential map, before renormalization. If the tidal operator $-f\,\Pi^{[1]}_{ij}\hat z^i\hat z^j$, which coincides with $\eta$ only at linear order, is used instead, one finds $b_{\Pi^{[2]}_\parallel}=-b_{(KK)_\parallel}=\frac25f\tau_0$.}
\be
\label{eq:constr}
\begin{split}
& b_{\delta \eta} = -b_1(1-b_\eta)\,,\quad 
b_{\eta^2} = -b_\eta +\frac{b_\eta^2}{2}\,.
\end{split}
\ee
Our discussion so far has not included 
the loop corrections, which we discuss in detail now.

\subsection{The effect of loop corrections}
\label{sec:loops}

The inconsistencies of the tree-level Tau model
become apparent once we take loops into account. In particular, they will generate 
formally 
infinite corrections to the tree-level bias parameters of 
the linear density and projected velocity gradient fields. 
These biases get renormalized by the following $P_{13}$-type contribution:
\be
\label{eq:renormP13}
\begin{split}
&\delta P_{13} = (b_1-b_\eta f \mu^2)P_{\text{lin}}(k)
\int_\q P_{\text{lin}}(q)\,{4\tau_0^2 Z_2(\k,\q)Z_1(\q)} \,,
\end{split}
\ee
which produces the leading order in $k$ UV contributions. {Here we have omitted the term $-\tau_0Z_3$ of $\tilde K_3$, which renormalizes the bias of the optical depth field in the standard way, see below. The cubic term of Eq.~\eqref{eq:dF_map} contains the contraction $-\frac{\tau_0^3}{2}\langle\delta^2_\tau\rangle\delta_\tau$, which is cancelled exactly by the term $\propto\tau_0^3$ that stems from the normalization by the mean flux. As a result, the renormalization of $b_1$ and $b_\eta$ in the Tau model receives no contributions of order $\tau_0^3$ (Appendix~\ref{app:p13uv}). The UV limit of Eq.~\eqref{eq:renormP13} is proportional to the short-scale mass variance~\eqref{eq:sLambda}. Expanding the integrand for $q\gg k$ (Appendix~\ref{app:p13uv}), we obtain}
\be 
\label{eq:renormP13UV}
\delta P_{13}\Big|_{\rm UV}=2(b_1-b_\eta f\mu^2) P_{\rm lin}(k)\,{\tau_0^2(\mathcal{A}+\mathcal{B} f\mu^2)}\,\sigma_\Lambda^2\,,
\ee
where the {order-one} constants {$\mathcal{A}$ and $\mathcal{B}$ are given in Eq.~\eqref{eq:AB}}. {The $P_{13}$-type term $-\tau_0Z_3$ renormalizes the linear bias of the optical depth, $b_1^{\tau,R}=b_1^\tau+(\frac{34}{21}b_2^\tau+\frac12b_3^\tau-\frac43b^\tau_{\delta\mathcal{G}_2})\sigma^2_\Lambda$, where $b^\tau_3$ and $b^\tau_{\delta\mathcal{G}_2}$ multiply the standard cubic operators $\delta^3/6$ and $\delta\mathcal{G}_2$, which are omitted in Eq.~\eqref{eq:bias_gal} because they only renormalize $b_1^\tau$. Since $\tau$ is a conserved tracer, this term does not affect $b_\eta$, provided $\langle\delta_\tau\rangle=0$.}

It is convenient now to recover our bookkeeping 
perturbation theory parameter $\epsilon$.
Adding the linear result, we 
obtain the following expression for 
the tree-level plus the UV part of the one-loop power spectrum in the Tau model,
\be
\begin{split}
P_{\rm tree}(k)+P_{\rm 1-loop}(k)\Big|_{\rm UV}=&(b_1-b_\eta f\mu^2)^2 P_{\rm lin}(k) + 2(b_1-b_\eta f\mu^2) P_{\rm lin}\,{\tau_0^2(\mathcal{A}+\mathcal{B} f\mu^2)}\epsilon^2\sigma_\Lambda^2 \\
= & P_{\rm lin}(k) (b_1+{\tau_0^2\mathcal{A}}\sigma_\Lambda^2\epsilon^2-(\tau_0{-\tau_0^2\mathcal{B}}\epsilon^2\sigma_\Lambda^2)f\mu^2)^2 
+ O(\epsilon^4)\,,
\end{split}
\ee
where $O(\epsilon^4)$ are formally two-loop corrections that can be ignored
at the precision level we are interested in. 
We have seen that the leading order loop corrections in the Tau model 
produce infinite contributions to the linear bias parameters, including 
the velocity gradient bias. 
This merely 
implies that the SPT
expansion for the Lyman alpha forest is inconsistent,
and one has to use the full EFT model that includes all operators 
allowed by symmetries. 

In the EFT approach one must treat $b_1$ and $b_\eta$
as formally infinite ``bare'' parameters that absorb the leading order UV behavior 
from the loops. 
Then their finite, physically observable, 
renormalized 
values of $b_1$ and $b_\eta$ are given by the sum of the ``bare'' parameters and the 
UV-sensitive loop
contribution,
\be 
\label{eq:bren}
b_\eta^{\rm ren}=\tau_0 {-\tau_0^2\mathcal{B}}\sigma_\Lambda^2\,,\quad b^{\rm ren}_1=-\tau_0 b^{\tau{,R}}_1+{\tau_0^2\mathcal{A}}\sigma_\Lambda^2\,.
\ee
Let us estimate the finite parts of $b_\eta^{\rm ren}$ and $b_1^{\rm ren}$ now. 
To that end we can use $\Lambda \sim k_{\rm NL}$,
so that $\sigma^2_{\rm NL}\sim 1$ by definition. 
In the absence 
of fine tuning, 
they should be of the order of 
\be
\label{eq:betaren}
{b_\eta^{\rm ren}-\tau_0\sim -\tau_0^2\mathcal{B}\,\sigma^2_{\rm NL}\,,\quad b_1^{\rm ren}+\tau_0b_1^{\tau,R}\sim \tau_0^2\mathcal{A}\,\sigma^2_{\rm NL}} \,.
\ee
Hence, even if we take the tree-level Tau model at 
face value, the
loops would generate {a significant} breaking\footnote{The breaking is suppressed by $\tau_0^2\approx0.1$, but the coefficients $\mathcal{A}$ and $\mathcal{B}$ are not small. Inverting the tree-level relations of Section~\ref{sec:tautree} at the best-fit values of Table~\ref{tab:kmax08} ($\tau_0=b_\eta$), we find $\mathcal{A}\approx2.2$ and $\mathcal{B}\approx2.3$, i.e.\ $b_1^{\rm ren}+\tau_0b_1^{\tau,R}\approx0.23\,\sigma^2_\Lambda$ and $b_\eta^{\rm ren}-\tau_0\approx-0.24\,\sigma^2_\Lambda$. For the linear power spectrum of the Sherwood simulation $\sigma^2_\Lambda \simeq 1.0$ at $\Lambda= 3~\hMpc$, so that these corrections amount to about $110\%$ of $b_1$ and $75\%$ of $b_\eta$.} 
of the relationship between the 
EFT operators following from the tree-level Tau model. 
Specifically, Eq.~\eqref{eq:betaren} implies that $b_\eta^{\rm ren}$
and $b_1^{\rm ren}$
differ from $\tau_0$ and $-\tau_0 b_1^{\tau,R}$ by amounts comparable to these parameters, and hence 
the tree-level exponential mapping constraints~\eqref{eq:constr}
do not hold once the loops are taken into account. 
This also suggests that the other constraint~\eqref{eq:constr0}, i.e. the vanishing 
of specific selection-dependent EFT operators, 
would not hold true as well because of the loops. 
This is quite natural to expect, as 
there is no symmetry that would guarantee the vanishing 
of these operators in the presence of loops. In the EFT jargon one says that 
operators like $\Pi^{[2]}_\parallel$ are generated by loops even if they are 
not present in the exponential mapping at the tree-level.

\section{Comparison with simulations}
\label{sec:sims}

\subsection{Fitting details}

In this Section we compare the one-loop EFT power spectrum model for the 
Lyman alpha flux power spectrum against simulation data. 
To that end we use the state-of-the-art suite of Sherwood 
simulations~\cite{Bolton:2016bfs}. 
These are large-scale, high resolution hydrodynamic
simulations that reproduce the
evolution of intergalactic medium
with a large number of particles (up to $1.7\times 10^{10}$).
The fiducial cosmology of these simulations 
is a flat $\Lambda$CDM with massless neutrinos 
and $\Omega_m=0.308$, $\Omega_b=0.0482$, $\sigma_8=0.829$, $n_s=0.961$,
$h=0.678$. The details of the Sherwood simulations can be found in Refs.~\cite{Bolton:2016bfs,Givans:2022qgb}. 

In this work, we fit the 3D Lyman alpha power spectrum at $z=2.8$,
extracted from the simulation box \texttt{L160\_N2048}, which 
followed 2048$^3$ gas and cold dark matter particles
in a {cube} with side length L=160$~\Mpch$.
The simulations use a homogeneous ionizing background model,
and assume that the gas is in ionization equilibrium 
and optically thin~\cite{Bolton:2016bfs}.
The final 3D power spectrum measurements are publicly available\footnote{
\href{https://github.com/andreufont/sherwood\_p3d}{\textcolor{blue}{https://github.com/andreufont/sherwood\_p3d}}}.
They are presented as functions of the Fourier mode wavenumber
$k$ and the cosine $\mu$ of the angle between the corresponding 
Fourier vector and the line of sight. In practice, the $\mu$
space is sampled by 16 bins, while the $k$ space is log-sampled
in the range [$k_F,k_{\rm Ny}$], where $k_F=0.039~\hMpc$ is the fundamental mode,
and $k_{\rm Ny}=20~\hMpc$ is the Nyquist frequency.

Following~\cite{Givans:2022qgb}, we fit the power spectrum by Monte-Carlo
Markov Chain sampling the following pseudo-$\chi^2$ function, 
\be
\label{eq:llk}
\chi^2 = \sum_{i}\frac{[P^{\rm data}_i - P^{\rm model}(k_i,\mu_i) ]^2
}{2(P^{\rm data}_i)^2/N_i}\,,
\ee
where $P^{\rm data}_i$ are measurement power spectrum bins,
and $N_i$ is the number of modes in the bin. Note that in Eq.~\eqref{eq:llk}
we have explicitly assumed a Gaussian diagonal covariance 
for the power spectrum. While this assumption is definitely  
correct on large scales, it has not been systematically validated 
on small scales, relevant for our analysis. 
Studies of the galaxy power spectrum covariance 
in Refs.~\cite{Wadekar:2019rdu,Wadekar:2020hax,Philcox:2020zyp} showed 
that the Gaussian approximation is very accurate on 
mildly non-linear scales because in this regime the effective 
covariance is dominated by the theoretical error due to the marginalization
over nuisance parameters~\cite{Baldauf:2016sjb,Chudaykin:2019ock,Chudaykin:2020hbf}.
In what follows we assume that the same 
argument is true for the Lyman alpha power 
spectrum, and we proceed with the Gaussian 
diagonal covariance.
We warn, however, that the interpretation of our results is, 
strictly speaking, contingent upon our covariance matrix assumptions.

Note that unlike~\cite{Givans:2022qgb}, we do {not} add the \textit{ad hoc}
noise floor correction to the covariance, as our goal here is to fit 
the data given its actual precision. This allows us to 
illustrate the main advantage of the EFT
over phenomenological approaches: its high precision
on the scales where the EFT is applicable.

The model vector $P^{\rm model}(k_i,\mu_i)$ is a one-loop EFT model 
that includes the necessary counterterms, 
\be
\begin{split}
P^{\rm 1-loop}(k,\mu) &= (b_1-b_\eta f\mu^2)^2P_{\text{lin}}(k) \\
& +2\int_\q K_2^2(\q,\k-\q)
P_{\text{lin}}(|\k-\q|)P_{\text{lin}}(q)  \\
&+ 6 K_1(\k)P_{\text{lin}}(k)\int_\q K_3(\k,-\q,\q)P_{\text{lin}}(q)\\
&-2(c_0+c_2\mu^2+c_4\mu^4)k^2 P_{\text{lin}}(k)\,.
\end{split}
\ee
The $k^2P_{\rm lin}$ corrections are 
formally beyond the one-loop in our power counting and 
added here 
as proxy for the two-loop contributions.
In principle, the consistent form of these counterterms that is closed under renormalization is $-2K_1(\k)(\gamma_0+\gamma_2\mu^2+\gamma_4\mu^4)k^2P_{\rm lin}(k)$, with the $\gamma_n$ of Eq.~\eqref{eq:hd}. 
We nevertheless use the above phenomenological form for simplicity.  We have checked that adding an extra parameter $\propto k^2\mu^6P_{\rm lin}$
has no impact on the fit. 
In addition, we found that the effect of the stochastic 
counterterms (see Eq.~\eqref{eq:stochP}) is negligible for our data,
and hence we set all of them, including the stochastic shot noise $P_{\rm shot}$, to zero.
At the same time the deterministic constant of Eq.~\eqref{eq:pshotest} is retained in the $P_{22}$ integrals. This is consistent with our theoretical arguments.

At face value, the one-loop EFT model depends on 16 free parameters: 2 linear biases, 
11 non-linear biases, and 3 higher derivative operators. 
One cubic parameter may be eliminated by
virtue of identity~\eqref{eq:degen}, which allows us to set
$b_{\Gamma_3}=0$.\footnote{The situation here is similar but somewhat different from that 
of galaxies. There the known degeneracy between $b_{\mathcal{G}_2}$ and $b_{\Gamma_3}$
at the power spectrum level~\cite{Ivanov:2019pdj,Nishimichi:2020tvu,Ivanov:2021kcd,Philcox:2021kcw,Ivanov:2023qzb} is not exact, and may be broken, in principle,
in the limit of an infinite signal-to-noise
of the data. The Lyman-$\alpha$ degeneracy~\eqref{eq:degen}, in contrast, is an exact kernel degeneracy which cannot be broken
even in the limit of an infinite signal-to-noise. 
}
For the remaining parameters, we use the following priors:
\be
\label{eq:fitparams}
\begin{split}
& b_1\in [-2,2]\,,\quad  b_\eta\in [-2,2]\,,\quad b_2\sim \mathcal{N}(0,2^2)\,,\quad b_{\mathcal{G}_2}\sim \mathcal{N}(0,2^2)\,,\\
& b_{(KK)_\parallel} \sim \mathcal{N}(0,2^2)\,,\quad 
b_{\Pi^{[2]}_\parallel} \sim \mathcal{N}(0,2^2)\,,\quad 
b_{\Pi^{[3]}_\parallel} \sim \mathcal{N}(0,2^2)\,,\quad  
\\
& b_{\delta \eta} \sim \mathcal{N}(0,2^2)\,,\quad 
b_{\eta^2} \sim \mathcal{N}(0,2^2)\,,\quad 
\frac{c_{0,2,4}}{[\Mpch]^2} \sim \mathcal{N}(0,0.1^2)\,,
\\
& b_{(K\Pi^{[2]})_\parallel} \sim \mathcal{N}(0,2^2)\,,\quad 
b_{\delta\Pi^{[2]}_\parallel} \sim \mathcal{N}(0,2^2)\,,\quad 
b_{\eta\Pi^{[2]}_\parallel} \sim \mathcal{N}(0,2^2)\,,
\end{split} 
\ee
where $\mathcal{N}(\mu,\sigma^2)$ stands for a Gaussian distribution 
with mean $\mu$ and r.m.s. $\sigma$.
Let us discuss in detail the motivation for this choice of priors. 
First, the priors for $b_1$ and $b_\eta$ are chosen to be 
flat and uninformative. Second, for all the quadratic and cubic 
selection-dependent bias parameters we choose large uninformative Gaussian 
priors with zero mean and r.m.s. equal to 2. In general, the EFT naturalness
arguments dictate that they should be $O(1)$ numbers, i.e. we expect their absolute values to roughly be in the range $[1,10]$. 
We note, however, 
that some additional suppression may come from the mapping, i.e. 
the flux bias parameters are $O(1)$ bias parameters of the $\tau$ field that have got multiplied by $\tau_0\simeq 0.3$. 
Given this reason, we reduced the r.m.s. of 
our prior to 2, instead of the conservative choice of 10 that 
could be made without any \textit{a priori} knowledge. 

As far as the counterterms are concerned, 
as we discussed earlier, we include them mainly 
in order to absorb some two-loop corrections that 
are not explicitly computed.  Indeed, we have found
that they actually improve the fit, which justifies their 
presence \textit{a posteriori}. 
The typical size of 
these corrections expected in the EFT is $\knl^{-2}\sim 0.05~[\Mpch]^2$.
We additionally multiply this by 2 in order to be conservative.
Indeed, one might expect the velocity field to be more nonlinear 
than the density one, and hence the $k^2$-contributions may be 
enhanced, as  is the case for galaxies~\cite{Ivanov:2019pdj,Ivanov:2021fbu}.
Jumping ahead, let us note that we will actually find 
the higher derivative corrections whose values are consistent with $\knl^{-2}$,
with  error bars that are much tighter than the priors. 

Only the 15 nuisance parameters~\eqref{eq:fitparams}
are assumed to vary in our MCMC chains.
We leave the fitting of cosmological 
parameters for future work.
We sample the likelihood analytically 
marginalized over all parameters that enter the likelihood quadratically, 
i.e. 
\[
{\{c_0,c_2,c_4,b_{\Pi^{[3]}_\parallel},b_{(K\Pi^{[2]})_\parallel},b_{\delta\Pi^{[2]}_\parallel},b_{\eta\Pi^{[2]}_\parallel}\}}\,,
\]
as in Refs.~\cite{Philcox:2020zyp,Chudaykin:2020hbf}.
The approximate 
posterior distributions and best-fit values for these parameters 
are later recovered from the MCMC chains for the marginalized likelihood.
The rest of the EFT parameters in Eq.~\eqref{eq:fitparams}
are explicitly varied in our MCMC chains. 

The MCMC chains are run with the \texttt{Montepython} sampler~\cite{Audren:2012wb,Brinckmann:2018cvx}
and analyzed with \texttt{getdist}~\cite{Lewis:2019xzd}.

\subsection{Results}

\begin{figure}
    \centering
    \includegraphics[width=\textwidth]{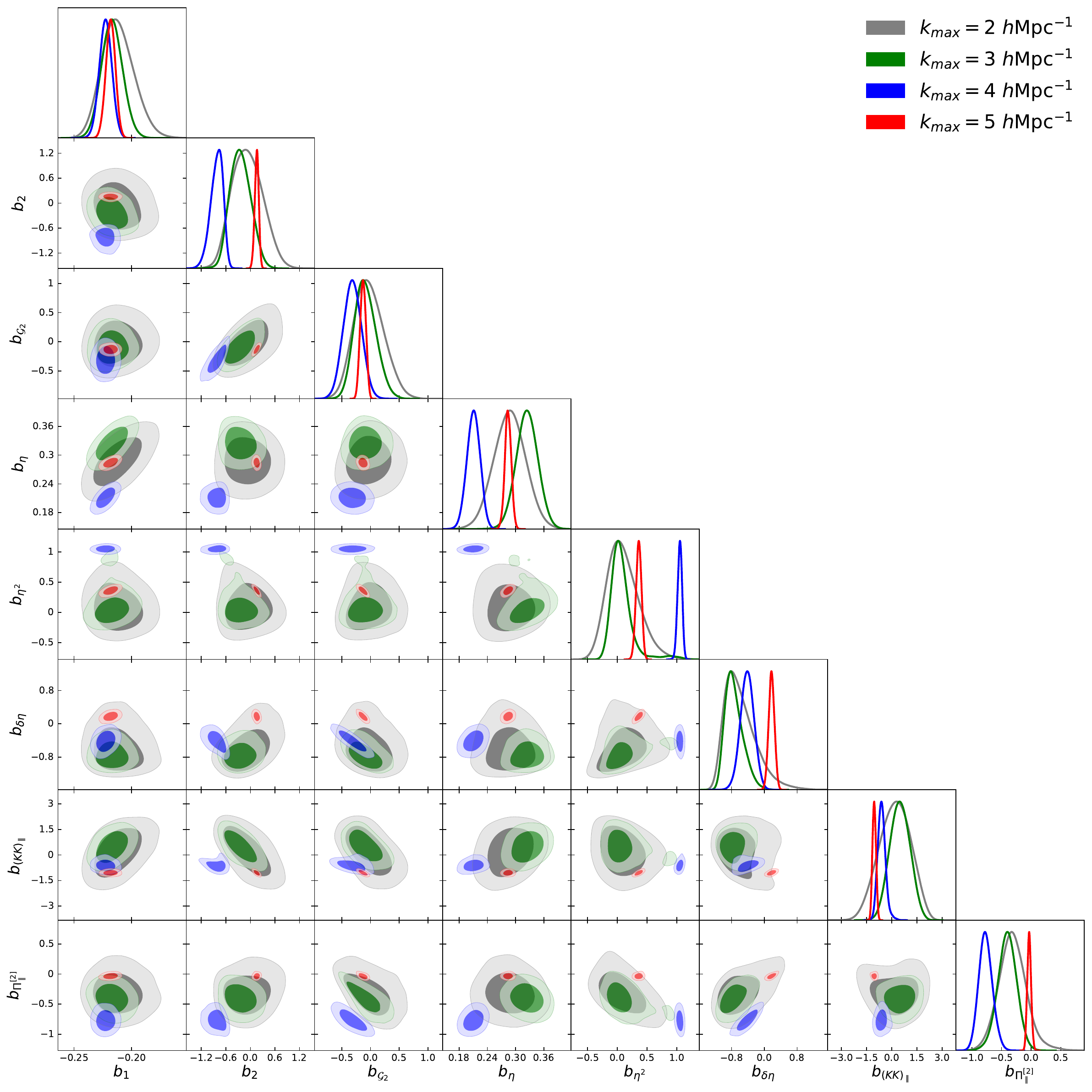}
    \caption{Triangle plot and marginalized projections for bias parameters
    of the EFT model for the Lyman alpha flux power spectrum of the Sherwood 
    simulations at $z=2.8$. We show results for four choices of $\kmax$: 2, 3, 4, and 5 $\hMpc$ (gray, green, blue, red, respectively).}
    \label{fig:bias_eft}
\end{figure}

\begin{table}[!htb]
\centering
\begin{tabular}{|l|c|c|c|c|}
 \hline
Param & best-fit & mean$\pm\sigma$ & 95\% lower & 95\% upper \\ \hline
$b_{1 }$ &$-0.2168$ & $-0.2167_{-0.0096}^{+0.0093}$ & $-0.2356$ & $-0.1976$ \\
$b_\eta$ &$0.324$ & $0.325_{-0.024}^{+0.023}$ & $0.278$ & $0.373$ \\
$b_{2 }$ &$-0.35$ & $-0.24_{-0.28}^{+0.24}$ & $-0.72$ & $0.27$ \\
$b_{\mathcal{G}_2 }$  &$-0.14$ & $-0.087_{-0.21}^{+0.18}$ & $-0.46$ & $0.30$ \\
$b_{\eta^2}$  &$0.041$ & $0.072_{-0.19}^{+0.1}$ & $-0.28$ & $0.45$ \\
$b_{\delta\eta}$  &$-0.85$ & $-0.74_{-0.26}^{+0.16}$ & $-1.13$ & $-0.29$ \\
$b_{(KK)_\parallel}$ &$0.56$ & $0.48_{-0.6}^{+0.6}$ & $-0.63$ & $1.63$ \\
$b_{\Pi^{[2]}_\parallel}$  &$-0.41$ & $-0.41_{-0.16}^{+0.16}$ & $-0.72$ & $-0.08$ \\
\hline
$10^3 c_0/[\Mpch]^2$ &$4.90 $ & $4.90_{-1.0}^{+1.0}$ & $2.80$ & $7.00$ \\
$10^3 c_2/[\Mpch]^2$ &$1.31$ & $1.31_{-0.62}^{+0.62}$ & $0.07$ & $2.54$ \\
$10^3 c_4/[\Mpch]^2$ &$-3.71$ & $-3.71_{-0.75}^{+0.75}$ & $-5.21$ & $-2.21$ \\
$b_{\Pi^{[3]}_\parallel}$ &$0.99$ & $0.99_{-0.05}^{+0.05}$ & $0.89$ & $1.08$ \\
$b_{\delta\Pi^{[2]}_\parallel}$ &$-1.03$ & $-1.03_{-0.05}^{+0.05}$ & $-1.13$ & $-0.93$ \\
$b_{(K\Pi^{[2]})_\parallel}$ &$-2.53$ & $-2.53_{-0.09}^{+0.09}$ & $-2.69$ & $-2.35$ \\
$b_{\eta\Pi^{[2]}_\parallel}$ &$1.67$ & $1.67_{-0.1}^{+0.1}$ & $1.47$ & $1.87$ \\
 \hline
 \end{tabular} 
 \caption{One-dimensional marginalized constraints on 
 nuisance parameters of the one-loop EFT model fit to the Lyman alpha 
 forest flux power spectrum of the Sherwood simulation
at $z=2.8$. The upper group of parameters were directly sampled in our MCMC chains. 
 The lower parameters are analytically marginalized over 
 in the likelihood. Their posteriors are recovered from the chains~\textit{a posteriori}.
 \label{tab:kmax08}}
\end{table}

In Fig.~\ref{fig:bias_eft}
we present 1d and 2d marginalized posterior distributions 
for the
bias parameters of the one-loop EFT model that were directly sampled 
in our chains. We show results for four choices of $k_{\rm max}=2,3,4$
and $5$ $\hMpc$. We see that the posteriors for $k_{\rm max}=2~\hMpc$ and 
$k_{\rm max}=3~\hMpc$ are fully consistent with each other. 
The contour for the $k_{\rm max}=4~\hMpc$ case is, however, 
significantly shifted w.r.t. the $2$ and $3$ $\hMpc$ posteriors.
Although the $k_{\rm max}=5~\hMpc$ contour shifts back inside the 
$k_{\rm max}=2~\hMpc$ posterior, it is still inconsistent 
with the $k_{\rm max}=3~\hMpc$ contour. 
These dramatic shifts and reduction of the posterior volume
suggest that the fit is biased for 
$k_{\rm max}=4~\hMpc$ and $k_{\rm max}=5~\hMpc$ due to two-loop 
corrections that are not included in our model. 
As a {frequentist} confirmation of our scale cuts, we also see 
a significant deterioration of the best-fit $\chi^2$ statistic for $k_{\rm max}>3~\hMpc$.
This suggests {choosing} the $k_{\rm max}=3~\hMpc$ case as a baseline. 
Note that a similar behavior of the posteriors when sliding $\kmax$
was previously observed in the case of 
redshift-space galaxy mocks~\cite{Nishimichi:2020tvu},
where the same criterion of the ``stability'' and consistency 
of posteriors w.r.t. variations of $\kmax$ was applied.

The best-fit values, 68\% and 95\% confidence limits
for the fit parameters at $k_{\rm max}=3~\hMpc$ are presented in Table~\ref{tab:kmax08}. First, we see that the linear bias parameters $b_1$ and $b_\eta$
are measured to $\sim 5\%$ precision. The second important observation
is that many Lyman alpha bias parameters are 
detected at high significance: 
$b_{\delta \eta}, b_{\Pi^{[2]}_\parallel}$, $b_{\Pi^{[3]}_\parallel}$,
$b_{(K\Pi^{[2]})_\parallel}$, $b_{\delta\Pi^{[2]}_\parallel}$,
$b_{\eta\Pi^{[2]}_\parallel}$.
Detections of these operators in the data clearly indicate an inconsistency 
of the tree-level Tau model, and the power of the EFT approach. 
The best-fit value of $b_{\delta \eta}$
is 4$\sigma$ away from the prediction of the tree-level Tau model $\tilde  b_{\delta\eta}=-b_1(1-b_\eta)\approx 0.15$,  
evaluated with Eq.~\eqref{eq:constr} using the best-fit value of $b_\eta$. {Strictly speaking, Eq.~\eqref{eq:constr} relates bare parameters, which differ from the renormalized ones by loop corrections of order $\tau_0^2\sigma^2_\Lambda$ (Section~\ref{sec:loops}), so this comparison is only indicative.} 
The apparent tension between the tree-level 
Tau model and the data is ``experimental'' 
evidence for the inconsistency of the tree-level calculations for the Lyman alpha 
forest. 
As far as higher derivative counterterms are concerned, we have a significant 
detection of $c_0$ and $c_4$.

\begin{figure}
    \centering
    \includegraphics[width=0.49\textwidth]{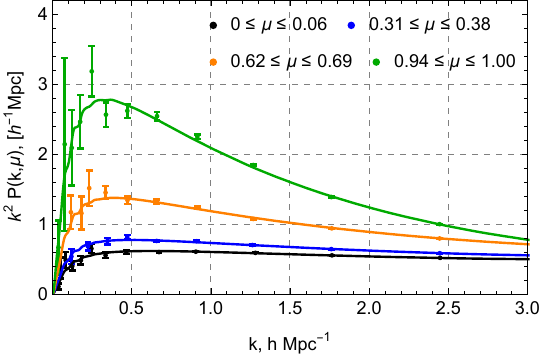}
    \includegraphics[width=0.49\textwidth]{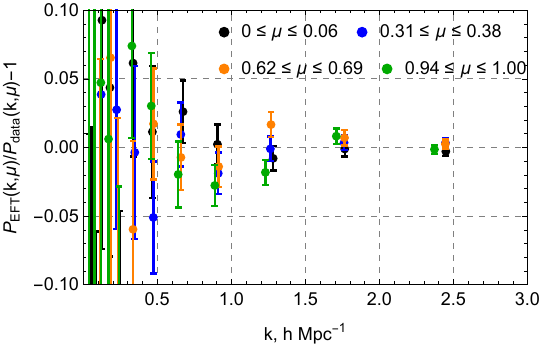}
    \caption{Best-fit EFT models against the simulated power spectra (left panel),
    and the residuals between the model and the data (right panel).}
    \label{fig:bestfit}
\end{figure}

\begin{figure}
    \centering    
    \includegraphics[width=0.49\textwidth]{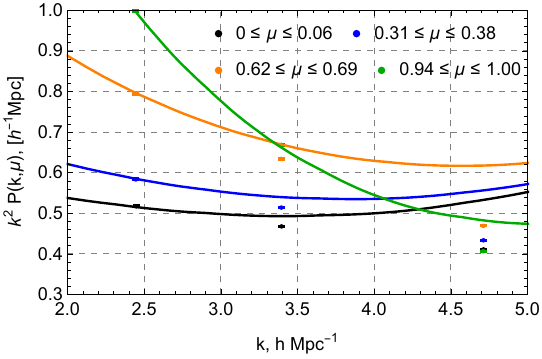}
    \includegraphics[width=0.49\textwidth]{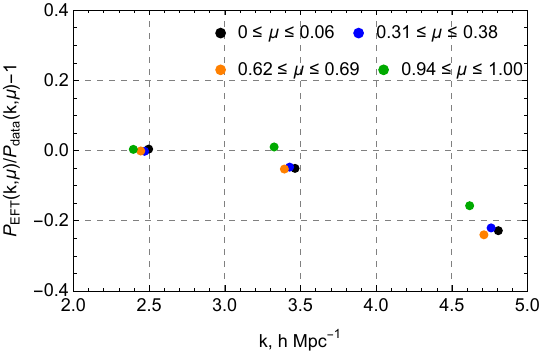}
      \caption{Best-fit EFT models fit from $\kmax=3~\hMpc$ against the simulated power spectra up to $k=5~\hMpc$ (left panel),
    and the residuals between them (right panel).}
    \label{fig:bf3}
\end{figure}

\begin{figure}
    \centering    
    \includegraphics[width=0.49\textwidth]{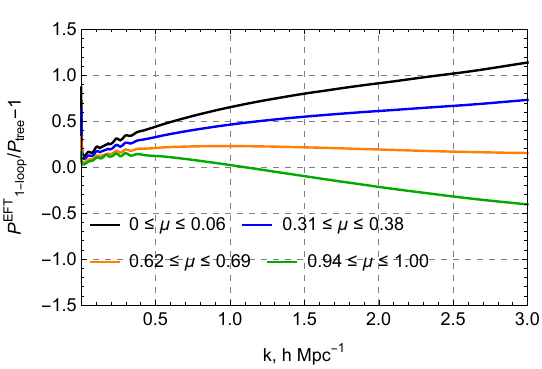}
    \includegraphics[width=0.49\textwidth]{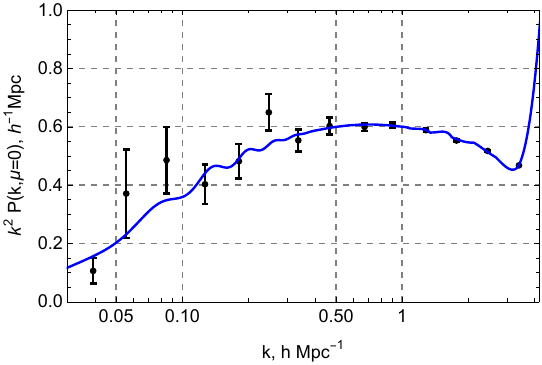}
    \caption{Left panel: the magnitude of one-loop corrections relative to the linear theory answer. Right panel: the fit for the transverse modes.}
    \label{fig:bf2}
\end{figure}

The best-fit models for four angular bins are shown against 
the data in Fig.~\ref{fig:bestfit}.
The nominal $\chi^2$ statistics across the 161 data points is 188, 
which indicates a satisfactory fit for 15 free parameters. 
It is important to keep in mind, however, that our EFT parameters are quite degenerate 
and hence the counting of degrees of freedom is not straightforward in our
case. In addition, our Gaussian covariance assumptions are likely 
not very accurate at the scales of interest. This is important 
to keep in mind when interpreting the $\chi^2$ statistics.

In order to illustrate the robustness of our fit, we extrapolate the 
$\kmax=3~\hMpc$ best fit model up to $k=5~\hMpc$ in Fig.~\ref{fig:bf3}.
Remarkably, the model calibrated at $\kmax=3~\hMpc$ describes the data quite well, within 10\%, even at $k=3.5~\hMpc$, beyond the regime of the validity of the original fit.
This can be interpreted as evidence of absence of overfitting. We note, however, 
that the 10\% residuals between the theory and the data are greater than the nominal error bars, which is why the fit at 
$\kmax=4~\hMpc$ was found earlier to be biased. The model breaks down gradually, 
reaching $20\%$ at  
$k=5~\hMpc$, which is consistent with the effect of the higher order corrections.

An important consistency check is that the one-loop corrections 
are smaller than the tree level result. 
To verify that this is the case, in Fig.~\ref{fig:bf2} 
we plot the one-loop contributions divided by the tree level model~\eqref{eq:tree}. 
We see that the magnitude of the perturbative corrections does not 
exceed the linear theory power spectrum except for the $\mu\simeq 0$ bin.
The size of the one-loop corrections is comparable to 
the tree-level result for $k\simeq 3~\hMpc$, which suggests 
that higher loop corrections may not be negligible. 
Given that no biases are observed 
at the level of the parameter estimation,  
it is likely that the effect of the two-loop
corrections was partly absorbed by the one-loop 
nuisance parameters and counterterms. 
A separate dedicated analysis is 
needed in order to quantify the effect of 
the two loop corrections more accurately. 
We note{,} however, that even if two-loop corrections 
are not negligible 
at $k=3~\hMpc$ in the strict sense, 
the EFT can still be applied on these scales
as a phenomenological model that is capable of
fitting the data with sub-percent accuracy.

Finally, the right panel of Fig.~\ref{fig:bf2} illustrates 
that for the transverse ($\mu\approx 0$) modes the EFT provides a good fit 
even for $\kmax=4~\hMpc$.
This suggests that the breakdown of the one-loop
model happens because of the velocity 
field that dominates the signal along the line 
of sight. 
Indeed, in general, the velocity field is more nonlinear than the density 
field for dark matter and galaxies~\cite{Pueblas:2008uv,Mercolli:2013bsa,Chudaykin:2020hbf,Ivanov:2021fbu,Ivanov:2021zmi}.
Fig.~\ref{fig:bf2}
shows, however, that the modes along the line of sight and transverse to the line of sight 
are roughly equally nonlinear for the Lyman alpha forest at $z=2.8$. 
Overall, the quality of the $z=2.8$ data
does not allow us to make a definitive conclusion about the role
of the velocity non-linearities. More insight may be obtained 
from the analysis of the $z=3.2$ data, which we discuss below.

We present the analysis of 
the Lyman alpha forest power spectrum at $z=3.2$
in Appendix~\ref{app:z3p2}.
At this redshift, the non-linear wavenumber
becomes significantly larger, $\knl = 10~\hMpc$, suggesting that the EFT expansion
should converge better and one can push the 
one-loop fit to smaller scales where the error bars are smaller.
We found that this is indeed the case. The EFT provides a good fit to the data 
up to $\kmax=5~\hMpc$. 
The optimal values of the Lyman alpha bias parameters are larger at $z=3.2$,
and they are detected with higher
significance, which is consistent with the expectations that 
tracers become more biased with redshift for uniform selection criteria. 
In particular, we have a significant 
detection of the RSD counterterms, whose impact increases towards $\mu=1$, 
e.g. the $c_4$ counterterm is detected at more than $20\sigma$.  
This supports
the above argument that the velocity field 
is primarily responsible for the 
breakdown of perturbation theory at small scales.
On the practical side, this suggests that a more optimal strategy for parameter 
inference would be 
to use a $\mu$-dependent $\kmax$ cutoff. 
We will explore this option 
in future work.

\section{One dimensional power spectrum}
\label{sec:1d}

We have seen that the EFT can accurately describe the full 3D power
spectrum of the Lyman alpha forest. Let us show now how the EFT technique 
can be used to calculate the one-dimensional (1D) power spectrum 
of the Lyman alpha forest. 
Imagine that the flux fluctuations have an underlying power spectrum $P_{\rm 3D}(k,k_\parallel)$ in 3D,
where $k_\parallel$ is the wavevector projection along the line-of-sight. 
Then
the power spectrum of these fluctuations seen only along the line-of-sight is given by~\cite{1989MNRAS.238..293L}
\be
\label{eq:p1d}
P_{\rm {1D}}(k_\parallel)= \frac{1}{2\pi}\int_{k_\parallel}^\infty dk~k P_{\rm 3D}(k,k_\parallel)\,.
\ee 
A conceptual difficulty of this observable is that
the 1D power spectrum is an integral over the 3D power spectrum that involves
UV modes that are not under robust analytic control. 
This is precisely the issue that the EFT is aimed to resolve. 
In fact, Eq.~\eqref{eq:p1d} has the form of a loop integral, so we can 
treat this UV sensitivity along the lines of the 
one-loop EFT renormalization for the matter power spectrum~\cite{Ivanov:2022mrd}. 
To that end we assume that the integral in Eq.~\eqref{eq:p1d} is cut off
at a scale $\Lambda \lesssim k_{\rm NL}$. 
A first relevant observation is that in perturbation theory 
$P_{\rm 3D}(k,k_\parallel)$ has a simple polynomial dependence on $k_\parallel=k\cdot \mu$.
For instance, at one loop order we have 
\be
 P_{\rm 3D}(k,k_\parallel)=\sum_{n=0}^{4}\left(\frac{k_\parallel}{k}\right)^{2n}P_{2n}(k)\,.
\ee
A calculation of $P_{\rm 3D}(k,k_\parallel)$ at a given order in 
perturbation theory 
is valid only up to a certain scale that we call $k_{\rm trust}$.
Then Eq.~\eqref{eq:p1d} can be equivalently recast as 
\be
\label{eq:p1dv2}
P_{\rm {1D}}(k_\parallel)= \frac{1}{2\pi}\int_{k_\parallel}^{k_{\rm trust}} dk~k P_{\rm 3D}(k,k_\parallel)+\frac{1}{2\pi}\int_{k_{\rm trust}}^{\Lambda} dk~k P_{\rm 3D}(k,k_\parallel)\,.
\ee 
The rightmost integral is given by simple polynomials of $k_\parallel$
times some function of the cutoff, 
\be 
\label{eq:UVdiv}
\begin{split}
& \frac{1}{2\pi}\int_{k_{\rm trust}}^{\Lambda} dk~k P_{\rm 3D}(k,k_\parallel)= 
\sum_{n=0}^{4}k^{2n}_\parallel {C}_n^\Lambda\,,\quad
 {C}_n^\Lambda\equiv \frac{1}{2\pi}\int_{k_{\rm trust}}^{\Lambda}dk~k^{1-2n}P_{2n}(k)\,.
\end{split}
\ee 
These UV-sensitive contributions are naturally absorbed into the EFT 
stochastic counterterms, which have the scale dependence that 
exactly matches Eq.~\eqref{eq:UVdiv}. Indeed, Eq.~\eqref{eq:stochP} implies that
\be
\label{eq:p3dst}
 P^{\rm stoch}_{\rm 3D}(k,k_\parallel)=P_{\rm shot}+a_0\frac{k^2}{\knl^2}+a_2\frac{k^2_\parallel}{\knl^2}{+a_4\frac{k^4_\parallel}{k^2\knl^2}}+\cdots\,,
\ee
where ``$\cdots$'' stands for terms higher order in $k/\knl$. 
$P_{\rm shot},a_0,a_2 ,a_4$ above are $\Lambda$-dependent Wilson coefficients that 
cancel the UV sensitivity from the integrals \eqref{eq:UVdiv}.  
Plugging Eq.~\eqref{eq:p3dst} into Eq.~\eqref{eq:p1dv2}
we get:
\be
\label{eq:p1dst}
\begin{split}
& \frac{1}{2\pi}\int_{k_\parallel}^{k_{\rm trust}} dk~k P^{\rm stoch}_{\rm 3D}(k,k_\parallel)= 
{P_{\rm shot}}\frac{k_{\rm trust}^2-k_\parallel^2}{4\pi} + 
{a_0}\frac{k_{\rm trust}^4-k_\parallel^4}{8\pi \knl^2} \\ +
&\quad {a_2}k_\parallel^2\frac{k_{\rm trust}^2-k_\parallel^2}{4\pi \knl^2} 
{+a_4\frac{k_\parallel^4}{2\pi\knl^2}\ln\frac{k_{\rm trust}}{k_\parallel}} \\
&=\frac{2P_{\rm shot}k_{\rm trust}^2+a_0k_{\rm trust}^4/\knl^2}{8\pi }  + k_\parallel^2\left(
a_2\frac{k_{\rm trust}^2}{4\pi\knl^2}-\frac{P_{\rm shot}}{4\pi}\right)\\
&\quad+\frac{k_\parallel^4}{4\pi\knl^2}\left(
-\frac{a_0}{2}-{a_2}{+2a_4\ln\frac{k_{\rm trust}}{\knl}}\right){-\frac{a_4k_\parallel^4}{2\pi\knl^2}\ln\frac{k_\parallel}{\knl}}\,,\\
&\equiv C'_0 + k_\parallel^2 C'_1 + k_\parallel^4 C'_2{+C'_{\rm L}\,k_\parallel^4\ln\frac{k_\parallel}{\knl}}\,.
\end{split} 
\ee
Here $C'_{\rm L}\equiv-a_4/(2\pi\knl^2)$. The $\ln k_{\rm trust}$ dependence of the $a_4$ term enters $C'_2$, where it is compensated by the running of $\bar C_2$ below, while the term $C'_{\rm L}k_\parallel^4\ln(k_\parallel/\knl)$ is non-analytic in $k_\parallel$ and cannot be absorbed by a polynomial counterterm. Its coefficient is fixed by the 3D parameter $a_4$. Due to the non-analytic dependence on the matching scale $k_{\rm trust}$, the combination 
$ \bar{C}_2\equiv -\left(
\frac{a_0}{2}+{a_2}\right)/(4\pi\knl^2)$ entering the coefficient $C'_2$
picks up a classical running with scale. Indeed, requiring the $C'_2$ in Eq.~\eqref{eq:p1dst} to be 
independent of $k_{\rm trust}$, 
$\frac{dC'_2}{d\ln k_{\rm trust}}=0$, implies
the beta-function
\be 
\frac{d\bar{C}_2}{d\ln k_{\rm trust}}=-\frac{a_4}{2\pi k_{\rm NL}^2}\,.
\ee 
Adding Eq.~\eqref{eq:p1dst} to \eqref{eq:UVdiv} we obtain
a finite cutoff-independent expression, 
\be
\begin{split}
&\frac{1}{2\pi}\int_{k_\parallel}^{k_{\rm trust}} dk~k {P^{\rm stoch}_{\rm 3D}}+\frac{1}{2\pi}\int_{k_{\rm trust}}^{\Lambda} dk~k P_{\rm 3D}\\
&=\left(C_0^\Lambda+C'_0\right)+k_\parallel^2\left(C_1^\Lambda+C'_1\right) + k_\parallel^4\left(C_2^\Lambda+C'_2\right){+C'_{\rm L}\,k_\parallel^4\ln\frac{k_\parallel}{\knl}}+\cdots\\
&=C_0^{\rm finite} + C_1^{\rm finite}k_\parallel^2 + C_2^{\rm finite}k_\parallel^4{+C'_{\rm L}\,k_\parallel^4\ln\frac{k_\parallel}{\knl}}+\cdots\,,
\end{split} 
\ee
where ${C}_n^{\rm finite}$ are finite parts of the EFT coefficients that have to be 
matched to data. 

In what follows we neglect the $a_4$ term, as we do with the stochastic contributions in the 3D fits, i.e. we set $C'_{\rm L}=0$.
The rationale behind this is that 
this coefficient is completely fixed by the 
3D stochasticity, which has a negligible 
contribution in our simulation fits. At the same time, ${C}_n^{\rm finite}$ receive
corrections from the UV parts of the 1D integral and thus become sizeable 
even if the 3D stochasticity can be ignored.

To sum up, in the EFT the 1D power spectrum calculation
amounts to doing the integral over the deterministic part of the 
3D spectrum up to $k_{\rm trust}$,
which is natural to identify with $\kmax$ from our 3D fits,
and supplementing it with simple power-law functions of  
$k_\parallel^2$ whose free parameters ${C}_n^{\rm finite}$
need to be fit from the data,
\be 
P^{\rm EFT}_{\rm {1D}}(k_\parallel)= 
\frac{1}{2\pi}\int_{k_\parallel}^{\kmax} dk~k [P_{\rm 3D}-P_{\rm stoch}]+C_0^{\rm finite} + C_1^{\rm finite}k_\parallel^2 + C_2^{\rm finite}k_\parallel^4~\,.
\ee
The part that contains ``$-P_{\rm stoch}$'' can be re-absorbed into ${C_{0,1,2}^{\rm finite}}$, yielding 
\be 
\label{eq:p1deft}
P^{\rm EFT}_{\rm {1D}}(k_\parallel)= 
\frac{1}{2\pi}\int_{k_\parallel}^{\kmax} dk~k P_{\rm 3D}(k,k_\parallel)+\mathcal{C}_0 + \mathcal{C}_1k_\parallel^2 + \mathcal{C}_2k_\parallel^4~\,.
\ee
Note that counterterms similar to $\mathcal{C}_n$ were 
first discussed in the context of regularization of the 1D Lyman alpha 
power spectrum integrals in \cite{Garny:2018byk}. In particular, 
our $\mathcal{C}_0$ matches the counterterm $\bar{I}_0$ from \cite{Garny:2018byk}.

\begin{figure}
    \centering    
    \includegraphics[width=0.49\textwidth]{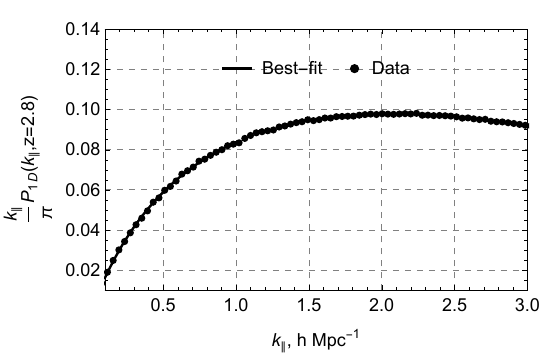}
    \includegraphics[width=0.49\textwidth]{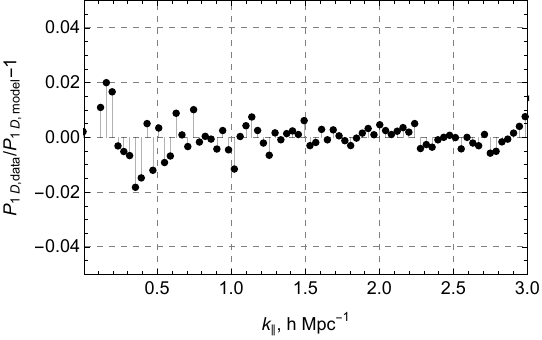}
    \caption{Left panel: dimensionless 1D flux power spectrum of the Sherwood simulations
    against the best-fit theory model from Eq.~\eqref{eq:p1deft}, 
    where all the EFT parameters
    except $\mathcal{C}_{n}$'s are kept fixed to their 3D best-fit values. Right panel: residuals between 
    the theory and the data.}
    \label{fig:1D}
\end{figure}

Let us estimate calculable parts of the  
1D counterterms $\mathcal{C}_n$ in our Universe for $z=2.8$. To that end 
we can approximate them assuming $P_{\rm 3D}$ from linear theory up to $\kmax=3~\hMpc$ 
and $\knl=5~\hMpc$, 
\be
\label{eq:Cest}
 \mathcal{C}_n \simeq 
 \Bigg\{
    \begin{array}{lr}
& \int^{\knl}_{\kmax}\frac{dk}{2\pi}~k b_1^2 P_{\rm lin}(k)= 0.02~[\Mpch], \quad  n=0\,,\\
& \int^{\knl}_{\kmax}\frac{dk}{2\pi}~k^{-1} (-2b_1b_\eta {f}) P_{\rm lin}(k)= 4\cdot 10^{-3}~[\Mpch]^{3}, \quad  n=1\,,\\        
& \int^{\knl}_{\kmax}\frac{dk}{2\pi}~k^{-3} b^2_\eta P_{\rm lin}(k)= 3\cdot 10^{-4}~[\Mpch]^5, \quad  n=2\,,
    \end{array}
\ee
where we used the best-fit values for $b_1$ and $b_\eta$ from Tab.~\ref{tab:kmax08}.
We see that the power law contributions in the r.h.s. of Eq.~\eqref{eq:p1deft}
are suppressed for small $k_\parallel$, 
so they represent a convergent gradient expansion,
which justifies dropping higher order terms in Eq.~\eqref{eq:p1deft}.
In a power law Universe these terms 
naturally scale as powers of $k_\parallel/\knl$. Using Eq.~\eqref{eq:su} we get the following 
order-of-magnitude
estimates for the 1D power spectrum
\be
\begin{split}
&\frac{k_\parallel}{2\pi}P_{\rm 1D}(k_\parallel)\sim \kp\int_\kp^{\knl} dk~
\Bigg[\underbrace{\frac{k^{n+1}}{\knl^{n+3}}}_{\rm linear}
+\underbrace{\frac{k^{2n+4}}{\knl^{2n+6}}}_{\rm 1-loop}
+\underbrace{\frac{k}{\knl^3}+\frac{k^3}{\knl^5}+\frac{k\kp^2}{\knl^5}  }_{\rm stochastic}\Bigg]\\
&=\underbrace{\frac{\kp}{\knl}-\frac{k^{n+3}_\parallel}{\knl^{n+3}}}_{\rm linear}
+\underbrace{\frac{\kp}{\knl}-\frac{k^{2n+6}_\parallel}{\knl^{2n+6}}}_{\rm 1-loop}
+\underbrace{\frac{\kp}{\knl}-\frac{k^3_\parallel}{\knl^3}+\frac{\kp^3}{\knl^3} - \frac{\kp^5}{\knl^5} }_{\rm stochastic} \,.
\end{split} 
\ee
First, we see that each term in this expansion, even the deterministic ones, {produces} 
stochastic-type contributions $\propto \kp/\knl$ from the upper limit of the integration.
They contribute to the calculable part of the constant shot noise contribution in 1D.
The calculable parts of the higher-derivative stochastic contributions 
in 3D scale as $(\kp/\knl)^3$ and $(\kp/\knl)^5$, and hence are clearly suppressed 
in the limit $\kp/\knl \to 0$. The deterministic parts of the linear and one-loop 
corrections dominate over the stochastic contributions, but not so strongly as they did 
in 3D. In particular, the proper one loop part scales as $(\kp/\knl)^1$, the same way as 
the shot noise $(\kp/\knl)$ terms, in contrast with $(\kp/\knl)$ against  
$(\kp/\knl)^{3}$ in the 3D case, see Eq.~\eqref{eq:scale1l}.
This explains why the stochastic counterterms $P_{\rm shot}$, $a_0$ and $a_2$
were not needed in the 3D fits, but appeared important 
for the 1D correlations.

To illustrate how the EFT model for the 1D power spectrum Eq.~\eqref{eq:p1deft}
{works} in practice, it is natural to evaluate this expression for the best-fit model
to the 3D power spectrum from the previous section at $\kmax=3~\hMpc$. 
The 3D power spectrum was actually fit only up to $\kmax\approx 2.5~\hMpc$
since there were no data points in the range $2.5<k/(\hMpc)<3$. 
Extrapolating our 3D best-fit into this range leads to 
an additional error, which affects the 1D modes
with $k_\parallel \simeq k_{\rm trust}$. 
We found, however, that this error can be largely canceled 
by introducing a higher order counterterm $\mathcal{C}_{{3}} \kp^6$. 
Thus, in addition to the usual 3D EFT counterterms, we need to introduce
4 free parameters in order to describe the 1D power 
spectrum at the percent precision.
We fit all these extra counterterms $\mathcal{C}_{{0,1,2,3}}$ to the 1D power
spectrum data extracted from the same simulation box. 
The best-fit 
counterterm piece is given by 
\be 
\label{eq:bf1d}
\begin{split}
\frac{P_{\rm 1D}^{\rm stoch}}{\Mpch}= & 0.085 + 1.8\cdot 10^{-3}[\Mpch]^2 k_\parallel^2 \\
& + 3.1\cdot 10^{-4}[\Mpch]^4 k_\parallel^4 -4.2\cdot 10^{-5}[\Mpch]^6 k_\parallel^6 ~\,.
\end{split}
\ee
Note that these best fit values are in perfect agreement with our estimates~\eqref{eq:Cest}.
The results are shown in Fig.~\ref{fig:1D}. We observe a perfect match between the theory and the data 
within the percent statistical scatter characteristic {of} the 1D power spectra 
of the Sherwood simulations, cf. Fig. 10 of~\cite{Givans:2022qgb}.
Note{,} however, that this good fit is obtained at the price of 
four extra parameters \eqref{eq:bf1d}. 
Given that one can probe only line-of-sight modes, 
a large number of nuisance parameters may represent 
a problem in the context of actual cosmological analyses of the 
1D Lyman-$\a$ flux power spectrum. 
We believe that the best way to 
proceed is to determine these parameters from high-fidelity Ly-$\a$
simulations and use these measurements as priors when analyzing the actual data. 
Similar ideas have been recently discussed in the context of galaxy clustering, 
see e.g.~\cite{Barreira:2021ukk,Lazeyras:2021dar,Cabass:2022wjy,Cabass:2022epm}.
We leave this line of research for future work.

\section{Discussion and conclusions}
\label{sec:disc}

We have developed a one-loop EFT model for the 3D and 1D power
spectra of the Lyman alpha forest. As we discussed in Sec.~\ref{sec:eft},
this model is identical to that presented by Desjacques et al.~\cite{Desjacques:2018pfv}
in the context of line-of-sight dependent selection effects in galaxy bias. 
We showed that 
the relevant loop integrals can be efficiently evaluated with the FFTLog method, 
which allows for an opportunity to carry out 
global full-shape analyses of the Lyman alpha data that include
variations of cosmological parameters.
This will enable one to explore the information content of the Lyman alpha 
forest along the lines of the recent EFT-based full-shape analyses of the 
galaxy power spectra~\cite{Ivanov:2019pdj,DAmico:2019fhj}.
This extension of our work is 
currently underway. 

From the theoretical point of view, it is not immediately obvious that
the line-of-sight dependent selection effect bias model 
of Desjacques et al.~\cite{Desjacques:2018pfv}
would 
exactly match a perturbative expansion of the exponential map of the 
optical depth field in the context of the Lyman alpha forest transmission.
The fact that the two should be in direct relationship
was first pointed out by Chen et al.~\cite{Chen:2021rnb}. 
We have carried out an explicit derivation
that shows how the full EFT model emerges from the exponential map
of the optical depth. From the technical point of view, 
the EFT expansion gets  
generated by the renormalization of the contact operators that stem 
from the exponential map. 
To underline the importance of renormalization, we also carried out
a naive expansion of the exponential map as one does in SPT. 
In that case an incomplete set of line-of-sight dependent operators is generated. 
Their bias coefficients are fixed by the optical depth contrast biases
and the background optical depth value. 
We call this expansion the tree-level Tau model. 
We explicitly show this model is 
inconsistent as the loop corrections break constraints between the bias coefficients. 
Our calculation suggests 
that even if we ignored the counterterms to start with, 
the loop corrections would generate all 
line-of-sight dependent operators that are missing in the tree-level Tau model.\footnote{Although this 
exercise is of purely academic interest, the cleanest way to explicitly 
check this would be a calculation 
of the one-loop bispectrum of the Lyman alpha forest, which can be done 
along the lines of the recent one-loop redshift space 
galaxy bispectrum calculations~\cite{Philcox:2022frc,DAmico:2022ukl}.}

We have also derived the complete set of one-loop renormalization conditions, presented in Appendix~\ref{app:renorm}. They show that our stochastic and higher-derivative operators are sufficient to renormalize the 
UV-sensitive loop corrections. We also show that a line-of-sight dependent tracer requires a stochastic counterterm $\propto k^2\mu^4$, which is absent for galaxies.

On the data-related side, we have carried out a precision comparison 
of the one-loop EFT model and the highly accurate 3D Lyman alpha power spectra 
extracted from the Sherwood hydrodynamical simulations. 
We have found that the EFT model can successfully describe 
the broadband shape of 3D and 1D power spectra, with $\lesssim 1\%$ residuals 
at $\kmax=3~\hMpc$ for $z=2.8$. Note that at face value, 
these residuals are smaller
than the ones of the phenomenological models used to fit the same {data} in Ref.~\cite{Givans:2022qgb}. 
The results of this work, however, cannot be directly compared
to ours, as in contrast to us, 
Ref.~\cite{Givans:2022qgb} assumed an additional $5\%$ noise floor in their covariance.

As far as the measured values of the Lyman alpha biases are 
concerned, we have found strong deviations from the naive tree-level Tau model. 
First, we have observed a breakdown of the tight relationships between quadratic and linear 
bias parameters.
In addition, we have
detected quadratic and cubic line-of-sight operators
missing in the tree-level Tau model. 
We warn, however, that our measurements could be 
affected by degeneracies between the EFT
nuisance parameters, covariance matrix assumptions,
and simulations' systematics.
It would be interesting to carry out a more 
detailed search for line-of-sight-dependent biases,
especially at the bispectrum level.
This line of research is also motivated by measurements
of the Lyman alpha 3-point correlations in simulations~\cite{Tie:2019tpi}
and the real data~\cite{Zaldarriaga:2000rg,Fang:2003wg}.
We leave this for future work.

Finally, we have also described how to consistently extend the EFT framework 
to the 1D power spectrum. Thus, our calculations can be 
applied to the one dimensional flux power spectrum data from BOSS~\cite{Chabanier:2018rga}.

Going forward, it would be important to explicitly implement IR resummation 
and use it to 
quantify the systematics of the BAO measurements.
In addition, it would be interesting to study cross-correlations of 
galaxies and quasars with the Lyman alpha forest, similar to Refs.~\cite{Font-Ribera:2013fha,Givans:2022qgb}.
Another important line of work is to convert our formalism 
into configuration space, either using the FFTLog technique similar to~\cite{Ivanov:2019pdj}, or developing the Lagrangian space EFT, along the lines 
of~\cite{Porto:2013qua,Vlah:2015sea,Vlah:2016bcl,Vlah:2018ygt,Chen:2020fxs,Chen:2020zjt}.
Eventually, we will have to account for realistic effects 
present in the actual data, such as the photoionization fluctuations
and patchy reionization. We leave all these research directions for future exploration.

\paragraph{Acknowledgments.} 
We would like to thank Stephen Chen, Giovanni Cabass and Matias Zaldarriaga for useful discussions. 
We are grateful to Diego Blas, 
Stephen Chen,
Vincent Desjacques, 
Andreu Font-Ribera,
Mathias Garny, 
Donghui Jeong, 
Fabian Schmidt, and Martin White
for their valuable comments on the draft.

\appendix

\section{Master Integrals}
\label{app:master}

In this Appendix we present master integrals that are necessary to compute the 
selection-dependent one-loop corrections. We will use the notation that ($\nu_{12}\equiv \nu_1+\nu_2$)
\be
J[\nu_1,\nu_2]\equiv \frac{1}{8\pi^{3/2}}\frac{\Gamma\left(\frac{3}{2}-\nu_1\right)
\Gamma\left(\frac{3}{2}-\nu_2\right)
\Gamma\left(\nu_{12}-\frac{3}{2}\right)
}{\Gamma(\nu_1)\Gamma(\nu_2)\Gamma(3-\nu_{12})}\,.
\ee
We have:
\begin{align*}
&A_5 = \frac{15}{256} \Bigg( J[-5 + \nu_1, \nu_2] - 5 J[-4 + \nu_1, -1 + \nu_2] - 3 J[-4 + \nu_1, \nu_2] \\
&+ 10 J[-3 + \nu_1, -2 + \nu_2] + 4 J[-3 + \nu_1, -1 + \nu_2] + 2 J[-3 + \nu_1, \nu_2] \\
&- 10 J[-2 + \nu_1, -3 + \nu_2] + 6 J[-2 + \nu_1, -2 + \nu_2] \\
&+ 2 J[-2 + \nu_1, -1 + \nu_2] + 2 J[-2 + \nu_1, \nu_2] + 5 J[-1 + \nu_1, -4 + \nu_2] \\
&- 12 J[-1 + \nu_1, -3 + \nu_2] + 6 J[-1 + \nu_1, -2 + \nu_2] + 4 J[-1 + \nu_1, -1 + \nu_2] \\
& - 3 J[-1 + \nu_1, \nu_2] - J[\nu_1, -5 + \nu_2] + 5 J[\nu_1, -4 + \nu_2] - 10 J[\nu_1, -3 + \nu_2] \\
&+ 10 J[\nu_1, -2 + \nu_2] - 5 J[\nu_1, -1 + \nu_2] + J[\nu_1, \nu_2] \Bigg)
\end{align*}

\begin{align*}
B_5 &= -\frac{35}{128} J[-5 + \nu_1, \nu_2] + \frac{175}{128} J[-4 + \nu_1, -1 + \nu_2] + \frac{25}{128} J[-4 + \nu_1, \nu_2] \\
&\quad - \frac{175}{64} J[-3 + \nu_1, -2 + \nu_2] + \frac{25}{32} J[-3 + \nu_1, -1 + \nu_2] + \frac{5}{64} J[-3 + \nu_1, \nu_2] \\
&\quad + \frac{175}{64} J[-2 + \nu_1, -3 + \nu_2] - \frac{225}{64} J[-2 + \nu_1, -2 + \nu_2] + \frac{45}{64} J[-2 + \nu_1, -1 + \nu_2] \\
&\quad + \frac{5}{64} J[-2 + \nu_1, \nu_2] - \frac{175}{128} J[-1 + \nu_1, -4 + \nu_2] + \frac{125}{32} J[-1 + \nu_1, -3 + \nu_2] \\
&\quad - \frac{225}{64} J[-1 + \nu_1, -2 + \nu_2] + \frac{25}{32} J[-1 + \nu_1, -1 + \nu_2] + \frac{25}{128} J[-1 + \nu_1, \nu_2] \\
&\quad + \frac{35}{128} J[\nu_1, -5 + \nu_2] - \frac{175}{128} J[\nu_1, -4 + \nu_2] + \frac{175}{64} J[\nu_1, -3 + \nu_2] \\
&\quad - \frac{175}{64} J[\nu_1, -2 + \nu_2] + \frac{175}{128} J[\nu_1, -1 + \nu_2] - \frac{35}{128} J[\nu_1, \nu_2],
\end{align*}

\begin{align*}
C_5 &= \frac{63}{256} J[-5 + \nu_1, \nu_2] - \frac{315}{256} J[-4 + \nu_1, -1 + \nu_2] + \frac{35}{256} J[-4 + \nu_1, \nu_2] \\
&\quad + \frac{315}{128} J[-3 + \nu_1, -2 + \nu_2] - \frac{105}{64} J[-3 + \nu_1, -1 + \nu_2] + \frac{15}{128} J[-3 + \nu_1, \nu_2] \\
&\quad - \frac{315}{128} J[-2 + \nu_1, -3 + \nu_2] + \frac{525}{128} J[-2 + \nu_1, -2 + \nu_2] - \frac{225}{128} J[-2 + \nu_1, -1 + \nu_2] \\
&\quad + \frac{15}{128} J[-2 + \nu_1, \nu_2] + \frac{315}{256} J[-1 + \nu_1, -4 + \nu_2] - \frac{245}{64} J[-1 + \nu_1, -3 + \nu_2] \\
&\quad + \frac{525}{128} J[-1 + \nu_1, -2 + \nu_2] - \frac{105}{64} J[-1 + \nu_1, -1 + \nu_2] + \frac{35}{256} J[-1 + \nu_1, \nu_2] \\
&\quad - \frac{63}{256} J[\nu_1, -5 + \nu_2] + \frac{315}{256} J[\nu_1, -4 + \nu_2] - \frac{315}{128} J[\nu_1, -3 + \nu_2] \\
&\quad + \frac{315}{128} J[\nu_1, -2 + \nu_2] - \frac{315}{256} J[\nu_1, -1 + \nu_2] + \frac{63}{256} J[\nu_1, \nu_2].
\end{align*}

\begin{align*}
& A_6 = -\frac{1}{1024} \times 5 \Big( J[-6 + \nu_1, \nu_2] - 6 J[-5 + \nu_1, -1 + \nu_2] - 6 J[-5 + \nu_1, \nu_2] \\
& + 15 J[-4 + \nu_1, -2 + \nu_2] + 18 J[-4 + \nu_1, -1 + \nu_2] + 15 J[-4 + \nu_1, \nu_2] \\
& - 20 J[-3 + \nu_1, -3 + \nu_2] - 12 J[-3 + \nu_1, -2 + \nu_2] - 12 J[-3 + \nu_1, -1 + \nu_2] \\
& - 20 J[-3 + \nu_1, \nu_2] + 15 J[-2 + \nu_1, -4 + \nu_2] - 12 J[-2 + \nu_1, -3 + \nu_2] \\
& - 6 J[-2 + \nu_1, -2 + \nu_2] - 12 J[-2 + \nu_1, -1 + \nu_2] + 15 J[-2 + \nu_1, \nu_2] \\
& - 6 J[-1 + \nu_1, -5 + \nu_2] + 18 J[-1 + \nu_1, -4 + \nu_2] - 12 J[-1 + \nu_1, -3 + \nu_2] \\
& - 12 J[-1 + \nu_1, -2 + \nu_2] + 18 J[-1 + \nu_1, -1 + \nu_2] - 6 J[-1 + \nu_1, \nu_2] \\
& + J[\nu_1, -6 + \nu_2] - 6 J[\nu_1, -5 + \nu_2] + 15 J[\nu_1, -4 + \nu_2] \\
& - 20 J[\nu_1, -3 + \nu_2] + 15 J[\nu_1, -2 + \nu_2] - 6 J[\nu_1, -1 + \nu_2] + J[\nu_1, \nu_2] \Big)
\end{align*}

\begin{align*}
B_6 = \frac{15}{1024}\Big(&7\,J[-6+\nu_1,\nu_2] -42\,J[-5+\nu_1,-1+\nu_2] \\
&\qquad -18\,J[-5+\nu_1,\nu_2] +105\,J[-4+\nu_1,-2+\nu_2] \\
&\qquad +30\,J[-4+\nu_1,-1+\nu_2] +9\,J[-4+\nu_1,\nu_2] \\
&\qquad -140\,J[-3+\nu_1,-3+\nu_2] +60\,J[-3+\nu_1,-2+\nu_2] \\
&\qquad +12\,J[-3+\nu_1,-1+\nu_2] +4\,J[-3+\nu_1,\nu_2] \\
&\qquad +105\,J[-2+\nu_1,-4+\nu_2] -180\,J[-2+\nu_1,-3+\nu_2] \\
&\qquad +54\,J[-2+\nu_1,-2+\nu_2] +12\,J[-2+\nu_1,-1+\nu_2] \\
&\qquad +9\,J[-2+\nu_1,\nu_2] -42\,J[-1+\nu_1,-5+\nu_2] \\
&\qquad +150\,J[-1+\nu_1,-4+\nu_2] -180\,J[-1+\nu_1,-3+\nu_2] \\
&\qquad +60\,J[-1+\nu_1,-2+\nu_2] +30\,J[-1+\nu_1,-1+\nu_2] \\
&\qquad -18\,J[-1+\nu_1,\nu_2] +7\,J[\nu_1,-6+\nu_2] \\
&\qquad -42\,J[\nu_1,-5+\nu_2] +105\,J[\nu_1,-4+\nu_2] \\
&\qquad -140\,J[\nu_1,-3+\nu_2] +105\,J[\nu_1,-2+\nu_2] \\
&\qquad -42\,J[\nu_1,-1+\nu_2] +7\,J[\nu_1,\nu_2]\Big)
\end{align*}

\begin{align*}
& C_6 = -\frac{315 J[-6 + \nu_1, \nu_2]}{1024} + \frac{945}{512} J[-5 + \nu_1, -1 + \nu_2] + \frac{105}{512} J[-5 + \nu_1, \nu_2] \\
& - \frac{4725 J[-4 + \nu_1, -2 + \nu_2]}{1024} + \frac{525}{512} J[-4 + \nu_1, -1 + \nu_2] + \frac{75 J[-4 + \nu_1, \nu_2]}{1024} \\
& + \frac{1575}{256} J[-3 + \nu_1, -3 + \nu_2] - \frac{1575}{256} J[-3 + \nu_1, -2 + \nu_2] + \frac{225}{256} J[-3 + \nu_1, -1 + \nu_2] \\
& + \frac{15}{256} J[-3 + \nu_1, \nu_2] - \frac{4725 J[-2 + \nu_1, -4 + \nu_2]}{1024} + \frac{2625}{256} J[-2 + \nu_1, -3 + \nu_2] \\
& - \frac{3375}{512} J[-2 + \nu_1, -2 + \nu_2] + \frac{225}{256} J[-2 + \nu_1, -1 + \nu_2] + \frac{75 J[-2 + \nu_1, \nu_2]}{1024} \\
& + \frac{945}{512} J[-1 + \nu_1, -5 + \nu_2] - \frac{3675}{512} J[-1 + \nu_1, -4 + \nu_2] + \frac{2625}{256} J[-1 + \nu_1, -3 + \nu_2] \\
& - \frac{1575}{256} J[-1 + \nu_1, -2 + \nu_2] + \frac{525}{512} J[-1 + \nu_1, -1 + \nu_2] + \frac{105}{512} J[-1 + \nu_1, \nu_2] \\
& - \frac{315 J[\nu_1, -6 + \nu_2]}{1024} + \frac{945}{512} J[\nu_1, -5 + \nu_2] - \frac{4725 J[\nu_1, -4 + \nu_2]}{1024} \\
& + \frac{1575}{256} J[\nu_1, -3 + \nu_2] - \frac{4725 J[\nu_1, -2 + \nu_2]}{1024} + \frac{945}{512} J[\nu_1, -1 + \nu_2] \\
& - \frac{315 J[\nu_1, \nu_2]}{1024};
\end{align*}

\begin{align*}
& D_6 = \frac{231 J[-6 + \nu_1, \nu_2]}{1024} - \frac{693}{512} J[-5 + \nu_1, -1 + \nu_2] + \frac{63}{512} J[-5 + \nu_1, \nu_2] \\
& + \frac{3465 J[-4 + \nu_1, -2 + \nu_2]}{1024} - \frac{945}{512} J[-4 + \nu_1, -1 + \nu_2] + \frac{105 J[-4 + \nu_1, \nu_2]}{1024} \\
& - \frac{1155}{256} J[-3 + \nu_1, -3 + \nu_2] + \frac{1575}{256} J[-3 + \nu_1, -2 + \nu_2] - \frac{525}{256} J[-3 + \nu_1, -1 + \nu_2] \\
& + \frac{25}{256} J[-3 + \nu_1, \nu_2] + \frac{3465 J[-2 + \nu_1, -4 + \nu_2]}{1024} - \frac{2205}{256} J[-2 + \nu_1, -3 + \nu_2] \\
& + \frac{3675}{512} J[-2 + \nu_1, -2 + \nu_2] - \frac{525}{256} J[-2 + \nu_1, -1 + \nu_2] + \frac{105 J[-2 + \nu_1, \nu_2]}{1024} \\
& - \frac{693}{512} J[-1 + \nu_1, -5 + \nu_2] + \frac{2835}{512} J[-1 + \nu_1, -4 + \nu_2] - \frac{2205}{256} J[-1 + \nu_1, -3 + \nu_2] \\
& + \frac{1575}{256} J[-1 + \nu_1, -2 + \nu_2] - \frac{945}{512} J[-1 + \nu_1, -1 + \nu_2] + \frac{63}{512} J[-1 + \nu_1, \nu_2] \\
& + \frac{231 J[\nu_1, -6 + \nu_2]}{1024} - \frac{693}{512} J[\nu_1, -5 + \nu_2] + \frac{3465 J[\nu_1, -4 + \nu_2]}{1024} \\
& - \frac{1155}{256} J[\nu_1, -3 + \nu_2] + \frac{3465 J[\nu_1, -2 + \nu_2]}{1024} - \frac{693}{512} J[\nu_1, -1 + \nu_2] + \frac{231 J[\nu_1, \nu_2]}{1024};
\end{align*}

\begin{align*}
A_7 = -\frac{35}{2048}\Big(&\,J[-7+\nu_1,\nu_2] -7\,J[-6+\nu_1,-1+\nu_2] \\
&\qquad -5\,J[-6+\nu_1,\nu_2] +21\,J[-5+\nu_1,-2+\nu_2] \\
&\qquad +18\,J[-5+\nu_1,-1+\nu_2] +9\,J[-5+\nu_1,\nu_2] \\
&\qquad -35\,J[-4+\nu_1,-3+\nu_2] -15\,J[-4+\nu_1,-2+\nu_2] \\
&\qquad -9\,J[-4+\nu_1,-1+\nu_2] -5\,J[-4+\nu_1,\nu_2] \\
&\qquad +35\,J[-3+\nu_1,-4+\nu_2] -20\,J[-3+\nu_1,-3+\nu_2] \\
&\qquad -6\,J[-3+\nu_1,-2+\nu_2] -4\,J[-3+\nu_1,-1+\nu_2] \\
&\qquad -5\,J[-3+\nu_1,\nu_2] -21\,J[-2+\nu_1,-5+\nu_2] \\
&\qquad +45\,J[-2+\nu_1,-4+\nu_2] -18\,J[-2+\nu_1,-3+\nu_2] \\
&\qquad -6\,J[-2+\nu_1,-2+\nu_2] -9\,J[-2+\nu_1,-1+\nu_2] \\
&\qquad +9\,J[-2+\nu_1,\nu_2] +7\,J[-1+\nu_1,-6+\nu_2] \\
&\qquad -30\,J[-1+\nu_1,-5+\nu_2] +45\,J[-1+\nu_1,-4+\nu_2] \\
&\qquad -20\,J[-1+\nu_1,-3+\nu_2] -15\,J[-1+\nu_1,-2+\nu_2] \\
&\qquad +18\,J[-1+\nu_1,-1+\nu_2] -5\,J[-1+\nu_1,\nu_2] \\
&\qquad -\,J[\nu_1,-7+\nu_2] +7\,J[\nu_1,-6+\nu_2] \\
&\qquad -21\,J[\nu_1,-5+\nu_2] +35\,J[\nu_1,-4+\nu_2] \\
&\qquad -35\,J[\nu_1,-3+\nu_2] +21\,J[\nu_1,-2+\nu_2] \\
&\qquad -7\,J[\nu_1,-1+\nu_2] +\,J[\nu_1,\nu_2]\Big)
\end{align*}

\begin{align*}
& B_7  = {\frac{105}{2048}}  \bigg( 3 J[-7 + \nu_1, \nu_2] - 21 J[-6 + \nu_1, -1 + \nu_2] - 7 J[-6 + \nu_1, \nu_2] \\
& + 63 J[-5 + \nu_1, -2 + \nu_2] + 14 J[-5 + \nu_1, -1 + \nu_2] + 3 J[-5 + \nu_1, \nu_2] \\
& - 105 J[-4 + \nu_1, -3 + \nu_2] + 35 J[-4 + \nu_1, -2 + \nu_2] + 5 J[-4 + \nu_1, -1 + \nu_2] \\
& + J[-4 + \nu_1, \nu_2] + 105 J[-3 + \nu_1, -4 + \nu_2] - 140 J[-3 + \nu_1, -3 + \nu_2] \\
& + 30 J[-3 + \nu_1, -2 + \nu_2] + 4 J[-3 + \nu_1, -1 + \nu_2] + J[-3 + \nu_1, \nu_2] \\
& - 63 J[-2 + \nu_1, -5 + \nu_2] + 175 J[-2 + \nu_1, -4 + \nu_2] - 150 J[-2 + \nu_1, -3 + \nu_2] \\
& + 30 J[-2 + \nu_1, -2 + \nu_2] + 5 J[-2 + \nu_1, -1 + \nu_2] + 3 J[-2 + \nu_1, \nu_2] \\
& + 21 J[-1 + \nu_1, -6 + \nu_2] - 98 J[-1 + \nu_1, -5 + \nu_2] + 175 J[-1 + \nu_1, -4 + \nu_2] \\
& - 140 J[-1 + \nu_1, -3 + \nu_2] + 35 J[-1 + \nu_1, -2 + \nu_2] + 14 J[-1 + \nu_1, -1 + \nu_2] \\
& - 7 J[-1 + \nu_1, \nu_2] - 3 J[\nu_1, -7 + \nu_2] + 21 J[\nu_1, -6 + \nu_2] \\
& - 63 J[\nu_1, -5 + \nu_2] + 105 J[\nu_1, -4 + \nu_2] - 105 J[\nu_1, -3 + \nu_2] \\
& + 63 J[\nu_1, -2 + \nu_2] - 21 J[\nu_1, -1 + \nu_2] + 3 J[\nu_1, \nu_2] \bigg);
\end{align*}

{\begin{align*}
C_7 = -\frac{21}{2048}\Big(&33\,J[-7+\nu_1,\nu_2] -231\,J[-6+\nu_1,-1+\nu_2] \\
&\qquad -21\,J[-6+\nu_1,\nu_2] +693\,J[-5+\nu_1,-2+\nu_2] \\
&\qquad -126\,J[-5+\nu_1,-1+\nu_2] -7\,J[-5+\nu_1,\nu_2] \\
&\qquad -1155\,J[-4+\nu_1,-3+\nu_2] +945\,J[-4+\nu_1,-2+\nu_2] \\
&\qquad -105\,J[-4+\nu_1,-1+\nu_2] -5\,J[-4+\nu_1,\nu_2] \\
&\qquad +1155\,J[-3+\nu_1,-4+\nu_2] -2100\,J[-3+\nu_1,-3+\nu_2] \\
&\qquad +1050\,J[-3+\nu_1,-2+\nu_2] -100\,J[-3+\nu_1,-1+\nu_2] \\
&\qquad -5\,J[-3+\nu_1,\nu_2] -693\,J[-2+\nu_1,-5+\nu_2] \\
&\qquad +2205\,J[-2+\nu_1,-4+\nu_2] -2450\,J[-2+\nu_1,-3+\nu_2] \\
&\qquad +1050\,J[-2+\nu_1,-2+\nu_2] -105\,J[-2+\nu_1,-1+\nu_2] \\
&\qquad -7\,J[-2+\nu_1,\nu_2] +231\,J[-1+\nu_1,-6+\nu_2] \\
&\qquad -1134\,J[-1+\nu_1,-5+\nu_2] +2205\,J[-1+\nu_1,-4+\nu_2] \\
&\qquad -2100\,J[-1+\nu_1,-3+\nu_2] +945\,J[-1+\nu_1,-2+\nu_2] \\
&\qquad -126\,J[-1+\nu_1,-1+\nu_2] -21\,J[-1+\nu_1,\nu_2] \\
&\qquad -33\,J[\nu_1,-7+\nu_2] +231\,J[\nu_1,-6+\nu_2] \\
&\qquad -693\,J[\nu_1,-5+\nu_2] +1155\,J[\nu_1,-4+\nu_2] \\
&\qquad -1155\,J[\nu_1,-3+\nu_2] +693\,J[\nu_1,-2+\nu_2] \\
&\qquad -231\,J[\nu_1,-1+\nu_2] +33\,J[\nu_1,\nu_2]\Big)
\end{align*}}

\begin{align*}
& D_7  = \frac{1}{2048}  \bigg( 429 J[-7 + \nu_1, \nu_2] - 3003 J[-6 + \nu_1, -1 + \nu_2] + 231 J[-6 + \nu_1, \nu_2] \\
& + 9009 J[-5 + \nu_1, -2 + \nu_2] - 4158 J[-5 + \nu_1, -1 + \nu_2] + 189 J[-5 + \nu_1, \nu_2] \\
& - 15015 J[-4 + \nu_1, -3 + \nu_2] + 17325 J[-4 + \nu_1, -2 + \nu_2] - 4725 J[-4 + \nu_1, -1 + \nu_2] \\
& + 175 J[-4 + \nu_1, \nu_2] + 15015 J[-3 + \nu_1, -4 + \nu_2] - 32340 J[-3 + \nu_1, -3 + \nu_2] \\
& + 22050 J[-3 + \nu_1, -2 + \nu_2] - 4900 J[-3 + \nu_1, -1 + \nu_2] + 175 J[-3 + \nu_1, \nu_2] \\
& - 9009 J[-2 + \nu_1, -5 + \nu_2] + 31185 J[-2 + \nu_1, -4 + \nu_2] - 39690 J[-2 + \nu_1, -3 + \nu_2] \\
& + 22050 J[-2 + \nu_1, -2 + \nu_2] - 4725 J[-2 + \nu_1, -1 + \nu_2] + 189 J[-2 + \nu_1, \nu_2] \\
& + 3003 J[-1 + \nu_1, -6 + \nu_2] - 15246 J[-1 + \nu_1, -5 + \nu_2] + 31185 J[-1 + \nu_1, -4 + \nu_2] \\
& - 32340 J[-1 + \nu_1, -3 + \nu_2] + 17325 J[-1 + \nu_1, -2 + \nu_2] - 4158 J[-1 + \nu_1, -1 + \nu_2] \\
& + 231 J[-1 + \nu_1, \nu_2] - 429 J[\nu_1, -7 + \nu_2] + 3003 J[\nu_1, -6 + \nu_2] \\
& - 9009 J[\nu_1, -5 + \nu_2] + 15015 J[\nu_1, -4 + \nu_2] - 15015 J[\nu_1, -3 + \nu_2] \\
& + 9009 J[\nu_1, -2 + \nu_2] - 3003 J[\nu_1, -1 + \nu_2] + 429 J[\nu_1, \nu_2] \bigg);
\end{align*}

{\begin{align*}
A_8 = \frac{35}{32768}\Big(&\,J[-8+\nu_1,\nu_2] -8\,J[-7+\nu_1,-1+\nu_2] \\
&\qquad -8\,J[-7+\nu_1,\nu_2] +28\,J[-6+\nu_1,-2+\nu_2] \\
&\qquad +40\,J[-6+\nu_1,-1+\nu_2] +28\,J[-6+\nu_1,\nu_2] \\
&\qquad -56\,J[-5+\nu_1,-3+\nu_2] -72\,J[-5+\nu_1,-2+\nu_2] \\
&\qquad -72\,J[-5+\nu_1,-1+\nu_2] -56\,J[-5+\nu_1,\nu_2] \\
&\qquad +70\,J[-4+\nu_1,-4+\nu_2] +40\,J[-4+\nu_1,-3+\nu_2] \\
&\qquad +36\,J[-4+\nu_1,-2+\nu_2] +40\,J[-4+\nu_1,-1+\nu_2] \\
&\qquad +70\,J[-4+\nu_1,\nu_2] -56\,J[-3+\nu_1,-5+\nu_2] \\
&\qquad +40\,J[-3+\nu_1,-4+\nu_2] +16\,J[-3+\nu_1,-3+\nu_2] \\
&\qquad +16\,J[-3+\nu_1,-2+\nu_2] +40\,J[-3+\nu_1,-1+\nu_2] \\
&\qquad -56\,J[-3+\nu_1,\nu_2] +28\,J[-2+\nu_1,-6+\nu_2] \\
&\qquad -72\,J[-2+\nu_1,-5+\nu_2] +36\,J[-2+\nu_1,-4+\nu_2] \\
&\qquad +16\,J[-2+\nu_1,-3+\nu_2] +36\,J[-2+\nu_1,-2+\nu_2] \\
&\qquad -72\,J[-2+\nu_1,-1+\nu_2] +28\,J[-2+\nu_1,\nu_2] \\
&\qquad -8\,J[-1+\nu_1,-7+\nu_2] +40\,J[-1+\nu_1,-6+\nu_2] \\
&\qquad -72\,J[-1+\nu_1,-5+\nu_2] +40\,J[-1+\nu_1,-4+\nu_2] \\
&\qquad +40\,J[-1+\nu_1,-3+\nu_2] -72\,J[-1+\nu_1,-2+\nu_2] \\
&\qquad +40\,J[-1+\nu_1,-1+\nu_2] -8\,J[-1+\nu_1,\nu_2] \\
&\qquad +\,J[\nu_1,-8+\nu_2] -8\,J[\nu_1,-7+\nu_2] \\
&\qquad +28\,J[\nu_1,-6+\nu_2] -56\,J[\nu_1,-5+\nu_2] \\
&\qquad +70\,J[\nu_1,-4+\nu_2] -56\,J[\nu_1,-3+\nu_2] \\
&\qquad +28\,J[\nu_1,-2+\nu_2] -8\,J[\nu_1,-1+\nu_2] \\
&\qquad +\,J[\nu_1,\nu_2]\Big)
\end{align*}}

{\begin{align*}
B_8 = -\frac{35}{8192}\Big(&9\,J[-8+\nu_1,\nu_2] -72\,J[-7+\nu_1,-1+\nu_2] \\
&\qquad -40\,J[-7+\nu_1,\nu_2] +252\,J[-6+\nu_1,-2+\nu_2] \\
&\qquad +168\,J[-6+\nu_1,-1+\nu_2] +60\,J[-6+\nu_1,\nu_2] \\
&\qquad -504\,J[-5+\nu_1,-3+\nu_2] -168\,J[-5+\nu_1,-2+\nu_2] \\
&\qquad -72\,J[-5+\nu_1,-1+\nu_2] -24\,J[-5+\nu_1,\nu_2] \\
&\qquad +630\,J[-4+\nu_1,-4+\nu_2] -280\,J[-4+\nu_1,-3+\nu_2] \\
&\qquad -60\,J[-4+\nu_1,-2+\nu_2] -24\,J[-4+\nu_1,-1+\nu_2] \\
&\qquad -10\,J[-4+\nu_1,\nu_2] -504\,J[-3+\nu_1,-5+\nu_2] \\
&\qquad +840\,J[-3+\nu_1,-4+\nu_2] -240\,J[-3+\nu_1,-3+\nu_2] \\
&\qquad -48\,J[-3+\nu_1,-2+\nu_2] -24\,J[-3+\nu_1,-1+\nu_2] \\
&\qquad -24\,J[-3+\nu_1,\nu_2] +252\,J[-2+\nu_1,-6+\nu_2] \\
&\qquad -840\,J[-2+\nu_1,-5+\nu_2] +900\,J[-2+\nu_1,-4+\nu_2] \\
&\qquad -240\,J[-2+\nu_1,-3+\nu_2] -60\,J[-2+\nu_1,-2+\nu_2] \\
&\qquad -72\,J[-2+\nu_1,-1+\nu_2] +60\,J[-2+\nu_1,\nu_2] \\
&\qquad -72\,J[-1+\nu_1,-7+\nu_2] +392\,J[-1+\nu_1,-6+\nu_2] \\
&\qquad -840\,J[-1+\nu_1,-5+\nu_2] +840\,J[-1+\nu_1,-4+\nu_2] \\
&\qquad -280\,J[-1+\nu_1,-3+\nu_2] -168\,J[-1+\nu_1,-2+\nu_2] \\
&\qquad +168\,J[-1+\nu_1,-1+\nu_2] -40\,J[-1+\nu_1,\nu_2] \\
&\qquad +9\,J[\nu_1,-8+\nu_2] -72\,J[\nu_1,-7+\nu_2] \\
&\qquad +252\,J[\nu_1,-6+\nu_2] -504\,J[\nu_1,-5+\nu_2] \\
&\qquad +630\,J[\nu_1,-4+\nu_2] -504\,J[\nu_1,-3+\nu_2] \\
&\qquad +252\,J[\nu_1,-2+\nu_2] -72\,J[\nu_1,-1+\nu_2] \\
&\qquad +9\,J[\nu_1,\nu_2]\Big)
\end{align*}}

{\begin{align*}
C_8 = \frac{105}{16384}\Big(&33\,J[-8+\nu_1,\nu_2] -264\,J[-7+\nu_1,-1+\nu_2] \\
&\qquad -72\,J[-7+\nu_1,\nu_2] +924\,J[-6+\nu_1,-2+\nu_2] \\
&\qquad +168\,J[-6+\nu_1,-1+\nu_2] +28\,J[-6+\nu_1,\nu_2] \\
&\qquad -1848\,J[-5+\nu_1,-3+\nu_2] +504\,J[-5+\nu_1,-2+\nu_2] \\
&\qquad +56\,J[-5+\nu_1,-1+\nu_2] +8\,J[-5+\nu_1,\nu_2] \\
&\qquad +2310\,J[-4+\nu_1,-4+\nu_2] -2520\,J[-4+\nu_1,-3+\nu_2] \\
&\qquad +420\,J[-4+\nu_1,-2+\nu_2] +40\,J[-4+\nu_1,-1+\nu_2] \\
&\qquad +6\,J[-4+\nu_1,\nu_2] -1848\,J[-3+\nu_1,-5+\nu_2] \\
&\qquad +4200\,J[-3+\nu_1,-4+\nu_2] -2800\,J[-3+\nu_1,-3+\nu_2] \\
&\qquad +400\,J[-3+\nu_1,-2+\nu_2] +40\,J[-3+\nu_1,-1+\nu_2] \\
&\qquad +8\,J[-3+\nu_1,\nu_2] +924\,J[-2+\nu_1,-6+\nu_2] \\
&\qquad -3528\,J[-2+\nu_1,-5+\nu_2] +4900\,J[-2+\nu_1,-4+\nu_2] \\
&\qquad -2800\,J[-2+\nu_1,-3+\nu_2] +420\,J[-2+\nu_1,-2+\nu_2] \\
&\qquad +56\,J[-2+\nu_1,-1+\nu_2] +28\,J[-2+\nu_1,\nu_2] \\
&\qquad -264\,J[-1+\nu_1,-7+\nu_2] +1512\,J[-1+\nu_1,-6+\nu_2] \\
&\qquad -3528\,J[-1+\nu_1,-5+\nu_2] +4200\,J[-1+\nu_1,-4+\nu_2] \\
&\qquad -2520\,J[-1+\nu_1,-3+\nu_2] +504\,J[-1+\nu_1,-2+\nu_2] \\
&\qquad +168\,J[-1+\nu_1,-1+\nu_2] -72\,J[-1+\nu_1,\nu_2] \\
&\qquad +33\,J[\nu_1,-8+\nu_2] -264\,J[\nu_1,-7+\nu_2] \\
&\qquad +924\,J[\nu_1,-6+\nu_2] -1848\,J[\nu_1,-5+\nu_2] \\
&\qquad +2310\,J[\nu_1,-4+\nu_2] -1848\,J[\nu_1,-3+\nu_2] \\
&\qquad +924\,J[\nu_1,-2+\nu_2] -264\,J[\nu_1,-1+\nu_2] \\
&\qquad +33\,J[\nu_1,\nu_2]\Big)
\end{align*}}

{\begin{align*}
D_8 = -\frac{7}{8192}\Big(&429\,J[-8+\nu_1,\nu_2] -3432\,J[-7+\nu_1,-1+\nu_2] \\
&\qquad -264\,J[-7+\nu_1,\nu_2] +12012\,J[-6+\nu_1,-2+\nu_2] \\
&\qquad -1848\,J[-6+\nu_1,-1+\nu_2] -84\,J[-6+\nu_1,\nu_2] \\
&\qquad -24024\,J[-5+\nu_1,-3+\nu_2] +16632\,J[-5+\nu_1,-2+\nu_2] \\
&\qquad -1512\,J[-5+\nu_1,-1+\nu_2] -56\,J[-5+\nu_1,\nu_2] \\
&\qquad +30030\,J[-4+\nu_1,-4+\nu_2] -46200\,J[-4+\nu_1,-3+\nu_2] \\
&\qquad +18900\,J[-4+\nu_1,-2+\nu_2] -1400\,J[-4+\nu_1,-1+\nu_2] \\
&\qquad -50\,J[-4+\nu_1,\nu_2] -24024\,J[-3+\nu_1,-5+\nu_2] \\
&\qquad +64680\,J[-3+\nu_1,-4+\nu_2] -58800\,J[-3+\nu_1,-3+\nu_2] \\
&\qquad +19600\,J[-3+\nu_1,-2+\nu_2] -1400\,J[-3+\nu_1,-1+\nu_2] \\
&\qquad -56\,J[-3+\nu_1,\nu_2] +12012\,J[-2+\nu_1,-6+\nu_2] \\
&\qquad -49896\,J[-2+\nu_1,-5+\nu_2] +79380\,J[-2+\nu_1,-4+\nu_2] \\
&\qquad -58800\,J[-2+\nu_1,-3+\nu_2] +18900\,J[-2+\nu_1,-2+\nu_2] \\
&\qquad -1512\,J[-2+\nu_1,-1+\nu_2] -84\,J[-2+\nu_1,\nu_2] \\
&\qquad -3432\,J[-1+\nu_1,-7+\nu_2] +20328\,J[-1+\nu_1,-6+\nu_2] \\
&\qquad -49896\,J[-1+\nu_1,-5+\nu_2] +64680\,J[-1+\nu_1,-4+\nu_2] \\
&\qquad -46200\,J[-1+\nu_1,-3+\nu_2] +16632\,J[-1+\nu_1,-2+\nu_2] \\
&\qquad -1848\,J[-1+\nu_1,-1+\nu_2] -264\,J[-1+\nu_1,\nu_2] \\
&\qquad +429\,J[\nu_1,-8+\nu_2] -3432\,J[\nu_1,-7+\nu_2] \\
&\qquad +12012\,J[\nu_1,-6+\nu_2] -24024\,J[\nu_1,-5+\nu_2] \\
&\qquad +30030\,J[\nu_1,-4+\nu_2] -24024\,J[\nu_1,-3+\nu_2] \\
&\qquad +12012\,J[\nu_1,-2+\nu_2] -3432\,J[\nu_1,-1+\nu_2] \\
&\qquad +429\,J[\nu_1,\nu_2]\Big)
\end{align*}}

\begin{align*}
& E_8 = \frac{1}{32768} (6435 J[-8 + \nu_1, \nu_2] - 51480 J[-7 + \nu_1, -1 + \nu_2] \\
&+ 3432 J[-7 + \nu_1, \nu_2] + 180180 J[-6 + \nu_1, -2 + \nu_2] - 72072 J[-6 + \nu_1, -1 + \nu_2] \\
&+ 2772 J[-6 + \nu_1, \nu_2] - 360360 J[-5 + \nu_1, -3 + \nu_2] + 360360 J[-5 + \nu_1, -2 + \nu_2] \\
&- 83160 J[-5 + \nu_1, -1 + \nu_2] + 2520 J[-5 + \nu_1, \nu_2] + 450450 J[-4 + \nu_1, -4 + \nu_2] \\
&- 840840 J[-4 + \nu_1, -3 + \nu_2] + 485100 J[-4 + \nu_1, -2 + \nu_2] - 88200 J[-4 + \nu_1, -1 + \nu_2] \\
&+ 2450 J[-4 + \nu_1, \nu_2] - 360360 J[-3 + \nu_1, -5 + \nu_2] + 1081080 J[-3 + \nu_1, -4 + \nu_2] \\
&- 1164240 J[-3 + \nu_1, -3 + \nu_2] + 529200 J[-3 + \nu_1, -2 + \nu_2] - 88200 J[-3 + \nu_1, -1 + \nu_2] \\
&+ 2520 J[-3 + \nu_1, \nu_2] + 180180 J[-2 + \nu_1, -6 + \nu_2] - 792792 J[-2 + \nu_1, -5 + \nu_2] \\
&+ 1372140 J[-2 + \nu_1, -4 + \nu_2] - 1164240 J[-2 + \nu_1, -3 + \nu_2] + 485100 J[-2 + \nu_1, -2 + \nu_2] \\
&- 83160 J[-2 + \nu_1, -1 + \nu_2] + 2772 J[-2 + \nu_1, \nu_2] - 51480 J[-1 + \nu_1, -7 + \nu_2] \\
&+ 312312 J[-1 + \nu_1, -6 + \nu_2] - 792792 J[-1 + \nu_1, -5 + \nu_2] + 1081080 J[-1 + \nu_1, -4 + \nu_2] \\
&- 840840 J[-1 + \nu_1, -3 + \nu_2] + 360360 J[-1 + \nu_1, -2 + \nu_2] - 72072 J[-1 + \nu_1, -1 + \nu_2] \\
&+ 3432 J[-1 + \nu_1, \nu_2] + 6435 J[\nu_1, -8 + \nu_2] - 51480 J[\nu_1, -7 + \nu_2] \\
&+ 180180 J[\nu_1, -6 + \nu_2] - 360360 J[\nu_1, -5 + \nu_2] + 450450 J[\nu_1, -4 + \nu_2] \\
&- 360360 J[\nu_1, -3 + \nu_2] + 180180 J[\nu_1, -2 + \nu_2] - 51480 J[\nu_1, -1 + \nu_2] \\
&+ 6435 J[\nu_1, \nu_2]).
\end{align*}

\section{Calculation of 13-contributions in the Tau model}
\label{app:p13uv}

{We use $\delta_F=F/\langle F\rangle-1$ and $\langle\delta_\tau\rangle=0$, i.e.\ mean-subtracted composite operators
in the bias expansion of $\delta_\tau$. The UV part of the 13-type contribution then reads
\be
\label{eq:p13_2}
\delta P_{13}\Big|_{\rm UV}=2K_1(\k)P_{\rm lin}(k)\,V_{\rm UV}\,,\qquad
\frac{V_{\rm UV}}{\sigma^2_\Lambda}=-\tau_0\left(\frac{34}{21}b_2^\tau+\frac{1}{2}b_3^\tau-\frac43b^\tau_{\delta\mathcal G_2}\right)+2\tau_0^2\,W(\mu)\,,
\ee
with $K_1=-\tau_0(b^\tau_1+f\mu^2)$ at tree level and
\be
W(\mu)\equiv\frac{1}{\sigma^2_\Lambda}\int_\q P_{\rm lin}(q)\,Z_1(\q)\,Z_2(\k,\q)\Big|_{\rm UV}\,.
\ee
The first term in $V_{\rm UV}$ comes from $-\tau_0 Z_3$. The cubic term $\sim\tau_0^3\delta_\tau^3$ does not contribute because its
$P_{13}$ term gets canceled by the $\langle F\rangle$  normalization.
To compute $W$ one expands the integrand for $q\gg k$. An explicit computation gives
\be
\begin{split}
W=\,&\frac{17}{21}(b_1^\tau)^2+\frac12b_1^\tau b_2^\tau-\frac23b^\tau_1b^\tau_{\mathcal G_2}
+f\left[\frac16(b^\tau_1)^2+\frac{13}{35}b^\tau_1+\frac16b^\tau_2-\frac4{15}b^\tau_{\mathcal G_2}\right]+f^2\left[\frac{1}{10}b^\tau_1+\frac{18}{245}\right]\\
&+\mu^2f\left[\frac12(b^\tau_1)^2+\frac12fb^\tau_1+\frac{11}{35}b^\tau_1+\frac{2}{15}b^\tau_{\mathcal G_2}+\frac15f^2+\frac{37}{245}f\right].
\end{split}
\ee
Writing $2W=\mathcal A+\mu^2f\mathcal B$, Eq.~\eqref{eq:p13_2} gives Eqs.~(\ref{eq:renormP13UV})--(\ref{eq:bren}) with
\be
\label{eq:AB}
\begin{split}
\mathcal A&=\frac{34}{21}(b_1^\tau)^2+b_1^\tau b_2^\tau-\frac43b^\tau_1b^\tau_{\mathcal G_2}
+f\left[\frac13(b^\tau_1)^2+\frac{26}{35}b^\tau_1+\frac13b^\tau_2-\frac{8}{15}b^\tau_{\mathcal G_2}\right]+f^2\left[\frac{1}{5}b^\tau_1+\frac{36}{245}\right],\\
\mathcal B&=(b_1^\tau)^2+fb_1^\tau+\frac{22}{35}b_1^\tau+\frac{4}{15}b^\tau_{\mathcal G_2}+\frac25f^2+\frac{74}{245}f\,.
\end{split}
\ee
The $b^\tau$ inside $\mathcal A$ and $\mathcal B$ may be taken bare or renormalized, as the difference is of order $\sigma^4_\Lambda$.
In terms of the mean flux, $\tau_{\rm eff}\equiv-\ln\langle F\rangle$, the relation for $b_\eta^{\rm ren}$ in Eq.~\eqref{eq:bren} reads
$b_\eta^{\rm ren}=\tau_{\rm eff}-\tau_{\rm eff}^2\sigma^2_\Lambda\left[\frac12(b^\tau_1)^2+\frac23fb^\tau_1+\frac{22}{35}b^\tau_1+\frac4{15}b^\tau_{\mathcal G_2}+\frac3{10}f^2+\frac{74}{245}f\right]$,
so $b_\eta=-\ln\langle F\rangle$ is also broken at order $\tau_{\rm eff}^2\sigma^2_\Lambda$.
The same result follows from the general EFT conditions of Appendix~\ref{app:renorm} evaluated at the
Tau-model values of the bare bias coefficients.

The Tau model also requires the higher-derivative and stochastic counterterms of Appendix~\ref{app:renorm}.
For instance, the $k^2$ coefficient $T(\mu)$ of Eq.~\eqref{eq:p13uvC} is
\be
\begin{split}
T_{\rm Tau}=\,&\tau_0\left[\tfrac{61}{630}b_1^\tau+\tfrac{32}{21}b^\tau_{\mathcal G_2}+\tfrac{64}{105}b^\tau_{\Gamma_3}\right]-\tau_0^2\left[\tfrac{24}{245}b_1^\tau f+\tfrac{4}{49}f^2\right]\\
&+\mu^2\left\{\tau_0f\left[\frac{3}{10}+\frac{8f}{35}-\frac{b_1^\tau}{105}+\frac{b_1^\tau f}{6}\right]+\tau_0^2f\left[\frac{72}{245}b_1^\tau+\frac{16}{245}f\right]\right\}\\
&+\mu^4\left\{\tau_0f^2\left[\frac{23}{105}+\frac f6\right]-\frac{4}{735}\tau_0^2f^2\right\},
\end{split}
\ee
and the $\mu^4$ coefficient of the $P_{22}$ UV limit, Eq.~\eqref{eq:A4}, is also non-zero,
\be
A^{(4)}_{\rm Tau}=\frac{8}{245}f^2\tau_0^3(7b_1^\tau+3f)\,.
\ee
\par}

\section{One-loop renormalization conditions}
\label{app:renorm}

In this Appendix we collect the renormalization conditions of the one-loop power spectrum
of the full EFT model, Eqs.~\eqref{eq:F_full2}--\eqref{eq:stochP}.
All bias coefficients are the observed (redshift-space) ones of Eqs.~\eqref{eq:F_full2}--\eqref{eq:K3full},
and all composite operators are mean-subtracted.
The UV-sensitive integrals are
\be
\begin{split}
&\sigma^2_\Lambda=\int^\Lambda_\q P_{\rm lin}(q)\,,\qquad
s_v=\int_\q\frac{P_{\rm lin}(q)}{q^2}=3\sigma_v^2\,,\\
&J_0=\int_\q P^2_{\rm lin}(q)\,,\qquad
J_1=\int_\q\frac{P^2_{\rm lin}(q)}{q^2}\,,\qquad
J_4=\int_\q \left(\frac{dP_{\rm lin}}{dq}\right)^2\,.
\end{split}
\ee
In dimensional regularization, and hence in our FFTLog evaluation, the $\sigma^2_\Lambda$ terms vanish
and the fitted parameters are the renormalized ones. The conditions below relate them to bare parameters
defined with a cutoff $\Lambda$.

\subsection{$P_{13}$}
For $q\gg k$ the angle-averaged 13-type integrand of each operator $\mathcal O$ behaves as
\be
\label{eq:hardlimit}
3\big\langle K_3(\k,\q,-\q)\big\rangle_{\hat{q}}\supset b_{\mathcal O}\left[\Sigma_{\mathcal O}(\mu)+\frac{k^2}{q^2}T_{\mathcal O}(\mu)+O\!\left(\frac{k^4}{q^4}\right)\right],
\ee
with the coefficients listed in Table~\ref{tab:renorm13} for the full set of EFT cubic operators.~\footnote{The cubic structure $(KKK)_\parallel\equiv\hat{z}^iK_{ij}K_{jl}K_{lm}\hat{z}^m$ is not an independent operator. Since $K_{ij}$ is traceless,
the Cayley--Hamilton theorem gives $K^3=\frac12{\rm tr}(K^2)\,K+\det K$, which in terms of $\Pi^{[1]}_{ij}=\d_i\d_j\Phi$ reads
\be
\label{eq:CH}
(KKK)_\parallel=-\frac13\mathcal{G}_3-\frac{1}{27}\delta^3+\frac12\Pi^{[1]}_\parallel\mathcal{G}_2+\frac13\Pi^{[1]}_\parallel\delta^2\,,
\ee
with $\mathcal{G}_3=-\frac12\big[2\,{\rm tr}(\Pi^{[1]})^3+\delta^3-3\,{\rm tr}(\Pi^{[1]})^2\,\delta\big]=-3\det\Pi^{[1]}$.
At the order considered here $\Pi^{[1]}_\parallel=-\eta/f$, so that
$(KKK)_\parallel=-\frac13\mathcal{G}_3-\frac{1}{27}\delta^3-\frac{1}{2f}\eta\,\mathcal{G}_2-\frac{1}{3f}\eta\,\delta^2$
is a combination of operators in Table~\ref{tab:renorm13}.} Terms that grow with $q/k$, and odd powers of $k/q$,
vanish after the angular average. Hence
\be
\label{eq:p13uvC}
P_{13}\Big|_{\rm UV}=2K_1(\k)P_{\rm lin}(k)\left[\Sigma(\mu)\sigma^2_\Lambda+k^2T(\mu)s_v\right],
\ee
with $\Sigma=\sum_{\mathcal O}b_{\mathcal O}\Sigma_{\mathcal O}$ and $T=\sum_{\mathcal O}b_{\mathcal O}T_{\mathcal O}$,
and the renormalization conditions are
\be
\label{eq:renorm13}
b_1^R=b_1+\Sigma^{(0)}\sigma^2_\Lambda\,,\quad
b_\eta^R=b_\eta-\frac{\Sigma^{(2)}}{f}\,\sigma^2_\Lambda\,,\quad
\gamma_n^R=\gamma_n-T^{(n)}s_v\ \ (n=0,2,4)\,,
\ee
where $X^{(n)}$ is the coefficient of $\mu^n$ and $\gamma_n$ are the counterterms of Eq.~\eqref{eq:hd}.
$\Sigma$ contains no $\mu^4$ term and $T$ no $\mu^6$ term, so the one-loop $P_{13}$ is renormalizable with
the counterterms of Eqs.~\eqref{eq:F_full} and~\eqref{eq:hd}.
The nine cubic operators in the lower part of Table~\ref{tab:renorm13} are not included in Eq.~\eqref{eq:F_full}:
their $P_{13}$ contribution is a pure $\sigma^2_\Lambda$ constant, which only renormalizes $b_1$ and $b_\eta$.

In the galaxy limit ($b_\eta=-1$, $b_{\delta\eta}=-b_1$, $b_{\eta^2}=1$, no selection operators)
Eq.~\eqref{eq:renorm13} gives $b_1^R=b_1+(\tfrac{34}{21}b_2+\tfrac12b_3-\tfrac43b_{\delta\mathcal G_2})\sigma^2_\Lambda$
and $b_\eta^R=-1$, i.e.\ $f$ is not renormalized. Without the mean subtraction of the composite
operators, $b_\eta^R$ would be shifted by
the tadpole of the RSD mapping
$-\sigma^2_\Lambda[\tfrac{b_2}{2}-\tfrac f3(b_1+b_{\delta\eta})+\tfrac{f^2}{5}(b_\eta+b_{\eta^2})+\tfrac29b_{(KK)_\parallel}+\tfrac13b_{\Pi^{[2]}_\parallel}]$.

Finally, the higher-derivative counterterms needed to renormalize $P_{13}$
are precisely the ones given in Eq.~\eqref{eq:hd}. They contribute $-2K_1(\k)(\gamma_0+\gamma_2\mu^2+\gamma_4\mu^4)k^2P_{\rm lin}(k)$ to the one-loop
power spectrum.

All in all, the deterministic operators 
derived in our work are sufficient for a 
renormalization of the $P_{13}$ integrals.
The set is somewhat restrictive.
In particular, there are no $P_{13}$ integrand (i.e. $\big\langle K_3(\k,\q,-\q)\big\rangle_{\hat{q}}$) divergences
that carry the coefficients proportional to $\mu^4\sigma^2_\Lambda$ or $k^2\mu^6 s_v$. 
In agreement with this observation,
symmetries do not allow $\mu^4\delta$ 
and $\mu^6 k^2\delta$ counterterms.

\newpage

{\footnotesize
\renewcommand{\arraystretch}{1.45}
\begin{longtable}{|l|>{\raggedright\arraybackslash}p{0.27\textwidth}|>{\raggedright\arraybackslash}p{0.40\textwidth}|}
\hline
coefficient & $\Sigma_{\mathcal O}(\mu)$ & $T_{\mathcal O}(\mu)$\\ \hline\endhead
$b_1$ & $-\frac{26}{35}f-\frac{22}{35}f\mu^{2}$ & $-\frac{61}{630}+\frac{24}{245}f-\left(\frac{209}{735}f+\frac{1}{6}f^{2}\right)\mu^{2}$ \\
$b_2$ & $\frac{34}{21}$ & $0$ \\
$b_{\mathcal{G}_2}$ & $0$ & $-\frac{32}{21}$ \\
$b_{\Gamma_3}$ & $0$ & $-\frac{64}{105}$ \\
$b_\eta$ & $\frac{72}{245}f^{2}+\frac{148}{245}f^{2}\mu^{2}$ & $-\frac{8}{49}f^{2}+\left(\frac{3}{10}f+\frac{88}{245}f^{2}\right)\mu^{2}+\left(\frac{51}{245}f^{2}+\frac{1}{6}f^{3}\right)\mu^{4}$ \\
$b_{\delta\eta}$ & $-\frac{26}{35}f-\frac{22}{35}f\mu^{2}$ & $\frac{24}{245}f-\frac{72}{245}f\mu^{2}$ \\
$b_{\eta^2}$ & $\frac{72}{245}f^{2}+\frac{148}{245}f^{2}\mu^{2}$ & $-\frac{8}{49}f^{2}+\frac{32}{245}f^{2}\mu^{2}-\frac{8}{735}f^{2}\mu^{4}$ \\
$b_{(KK)_\parallel}$ & $\frac{622}{945}+\frac{58}{315}\mu^{2}$ & $-\frac{64}{147}-\frac{32}{147}\mu^{2}$ \\
$b_{\Pi^{[2]}_\parallel}$ & $\frac{26}{35}-\frac{12}{49}f+\left(\frac{22}{35}-\frac{419}{735}f\right)\mu^{2}$ & $-\frac{64}{147}+\frac{20}{147}f+\left(\frac{2432}{2205}-\frac{80}{147}f\right)\mu^{2}+\frac{100}{441}f\mu^{4}$ \\
$b_{\delta\Pi^{[2]}_\parallel}$ & $\frac{5}{7}+\frac{10}{21}\mu^{2}$ & $-\frac{8}{49}+\frac{24}{49}\mu^{2}$ \\
$b_{\eta\Pi^{[2]}_\parallel}$ & $-\frac{12}{49}f-\frac{419}{735}f\mu^{2}$ & $\frac{20}{147}f-\frac{16}{147}f\mu^{2}+\frac{4}{441}f\mu^{4}$ \\
$b_{(K\Pi^{[2]})_\parallel}$ & $\frac{22}{105}+\frac{142}{315}\mu^{2}$ & $-\frac{32}{147}-\frac{16}{147}\mu^{2}$ \\
$b_{\Pi^{[3]}_\parallel}$ & $\frac{59}{105}+\frac{58}{105}\mu^{2}$ & $-\frac{52}{147}+\frac{2116}{2205}\mu^{2}$ \\
\hline
$b_3$\ \ ($\delta^3/6$) & $\frac{1}{2}$ & $0$ \\
$b_{\delta\mathcal{G}_2}$\ \ ($\delta\mathcal{G}_2$) & $-\frac{4}{3}$ & $0$ \\
$b_{\mathcal{G}_3}$\ \ ($\mathcal{G}_3$) & $0$ & $0$ \\
$b_{\delta^2\eta}$\ \ ($\delta^2\eta$) & $-\frac{2}{3}f-f\mu^{2}$ & $0$ \\
$b_{\delta\eta^2}$\ \ ($\delta\eta^2$) & $\frac{1}{5}f^{2}+\frac{2}{3}f^{2}\mu^{2}$ & $0$ \\
$b_{\eta^3}$\ \ ($\eta^3$) & $-\frac{3}{5}f^{3}\mu^{2}$ & $0$ \\
$b_{\eta\mathcal{G}_2}$\ \ ($\eta\,\mathcal{G}_2$) & $\frac{8}{15}f-\frac{4}{15}f\mu^{2}$ & $0$ \\
$b_{\delta(KK)_\parallel}$\ \ ($\delta\,(KK)_\parallel$) & $\frac{2}{9}$ & $0$ \\
$b_{\eta(KK)_\parallel}$\ \ ($\eta\,(KK)_\parallel$) & $\frac{8}{135}f-\frac{2}{5}f\mu^{2}$ & $0$ \\
\hline
\caption{UV coefficients of the 13-type integrands per unit bias coefficient, Eq.~(\ref{eq:hardlimit}).
Upper part: the operators of Eqs.~\eqref{eq:F_full}--\eqref{eq:F_full2}, observed coefficients.
Lower part: cubic operators that only renormalize $b_1$ and $b_\eta$ (rest-frame coefficients; in brackets the operator).
The identities~\eqref{eq:degen} hold for $\Sigma_{\mathcal O}$, with the constants on their right-hand sides, and the corresponding combinations of $T_{\mathcal O}$ vanish.
\label{tab:renorm13}}
\end{longtable}}

\subsection{$P_{22}$ and stochastic counterterms}
With the symmetric routing $\q=\k/2+\p$, $\k-\q=\k/2-\p$, the integrand is even in $\k$, and for $p\gg k$
\be
\label{eq:p22uv}
P_{22}\big|_{\rm UV}=S\,J_0+k^2\left[A(\mu)J_1+B(\mu)J_4\right],
\ee
up to total derivatives. Note that $S$ does not depend on $\mu$, while $A$, $B$ are polynomials of degree four and two in $\mu$, respectively.
The constant term is fixed by the kernel at zero total momentum,
$K_2(\p,-\p)=\frac{b_2}{2}+\frac{b_{(KK)_\parallel}}{9}+\kappa_2\mu_p^2+\kappa_4\mu_p^4$, with
$\kappa_2=-f(b_1+b_{\delta\eta})+\frac13b_{(KK)_\parallel}+b_{\Pi^{[2]}_\parallel}$ and $\kappa_4=f^2(b_\eta+b_{\eta^2})$:
\be
\begin{split}
&S=2\big\langle K_2(\p,-\p)^2\big\rangle_{\hat{p}}=2\left(\mathcal K_0^2+\frac{\mathcal K_2^2}{5}+\frac{\mathcal K_4^2}{9}\right),\\
&\mathcal K_0=\frac{b_2}{2}+\frac{b_{(KK)_\parallel}}{9}+\frac{\kappa_2}{3}+\frac{\kappa_4}{5}\,,\quad
\mathcal K_2=\frac{2\kappa_2}{3}+\frac{4\kappa_4}{7}\,,\quad
\mathcal K_4=\frac{8\kappa_4}{35}\,.
\end{split}
\ee
The term $\propto \mu^4$ only enters the $A$ coefficient:
\be
\label{eq:A4}
A^{(4)}=\frac{4}{35}\left(3fb_\eta-5b_{\Pi^{[2]}_\parallel}\right)\mathcal K_2\,,\qquad B^{(4)}=0\,.
\ee
The remaining coefficients are listed in Table~\ref{tab:renorm22}.
Eq.~\eqref{eq:p22uv} is absorbed by the stochastic counterterms of Eq.~\eqref{eq:stochP},
\be
P_{\rm shot}^R=P_{\rm shot}+S\,J_0\,,\qquad
\frac{a_n^R}{k_{\rm NL}^2}=\frac{a_n}{k_{\rm NL}^2}+A^{(n)}J_1+B^{(n)}J_4\quad(n=0,2,4)\,.
\ee
In the galaxy limit $S=b_2^2/2$ and $A$ has no $\mu^4$ term, in agreement with Eq.~\eqref{eq:stochP} for galaxies.
With a finite cutoff, the boundary terms of the integration by parts that leads to Eq.~\eqref{eq:p22uv} affect the shifts of
$a_0$ and $a_2$ but not those of $P_{\rm shot}$ and $a_4$, since the derivative terms only contribute up to $\mu^2$. For $\Lambda$CDM, $J_0$, $J_1$ and $J_4$
are dominated by IR modes, so that the contributions of modes above $k_{\rm NL}$ to all these shifts are numerically small.

{\footnotesize
\renewcommand{\arraystretch}{1.45}
\begin{longtable}{|l|l|l|}
\hline
pair & $A(\mu)$ & $B(\mu)$\\ \hline\endhead
$b_1^2$ & $-\frac{2}{21}f-\frac{4}{35}f^{2}+\left(\frac{8}{21}f+\frac{22}{105}f^{2}\right)\mu^{2}$ & $-\frac{1}{35}f^{2}-\frac{4}{35}f^{2}\mu^{2}$ \\
$b_1\,b_2$ & $-\frac{1}{21}+\frac{1}{15}f+\frac{2}{15}f\mu^{2}$ & $\frac{1}{15}f+\frac{2}{15}f\mu^{2}$ \\
$b_1\,b_{\mathcal{G}_2}$ & $\frac{16}{15}f-\frac{8}{15}f\mu^{2}$ & $0$ \\
$b_1\,b_\eta$ & $\frac{22}{245}f^{2}+\frac{22}{63}f^{3}-\left(\frac{394}{735}f^{2}+\frac{74}{105}f^{3}\right)\mu^{2}-\frac{8}{35}f^{2}\mu^{4}$ & $\frac{2}{63}f^{3}+\frac{4}{21}f^{3}\mu^{2}$ \\
$b_1\,b_{\delta\eta}$ & $-\frac{2}{21}f-\frac{8}{35}f^{2}+\left(\frac{8}{21}f+\frac{2}{7}f^{2}\right)\mu^{2}$ & $-\frac{2}{35}f^{2}-\frac{8}{35}f^{2}\mu^{2}$ \\
$b_1\,b_{\eta^2}$ & $\frac{22}{245}f^{2}+\frac{22}{63}f^{3}-\left(\frac{16}{49}f^{2}+\frac{32}{105}f^{3}\right)\mu^{2}$ & $\frac{2}{63}f^{3}+\frac{4}{21}f^{3}\mu^{2}$ \\
$b_1\,b_{(KK)_\parallel}$ & $\frac{4}{189}+\frac{464}{945}f-\left(\frac{8}{63}+\frac{188}{945}f\right)\mu^{2}$ & $\frac{32}{945}f+\frac{20}{189}f\mu^{2}$ \\
$b_1\,b_{\Pi^{[2]}_\parallel}$ & $\frac{2}{21}+\frac{22}{35}f-\left(\frac{8}{21}+\frac{124}{105}f\right)\mu^{2}+\frac{8}{21}f\mu^{4}$ & $\frac{2}{35}f+\frac{8}{35}f\mu^{2}$ \\
$b_2^2$ & $0$ & $-\frac{1}{12}$ \\
$b_2\,b_{\mathcal{G}_2}$ & $-\frac{4}{3}$ & $0$ \\
$b_2\,b_\eta$ & $-\frac{6}{35}f^{2}+\left(\frac{3}{7}f+\frac{11}{35}f^{2}\right)\mu^{2}$ & $-\frac{1}{35}f^{2}-\frac{4}{35}f^{2}\mu^{2}$ \\
$b_2\,b_{\delta\eta}$ & $\frac{1}{15}f-\frac{1}{5}f\mu^{2}$ & $\frac{1}{15}f+\frac{2}{15}f\mu^{2}$ \\
$b_2\,b_{\eta^2}$ & $-\frac{6}{35}f^{2}-\frac{2}{105}f^{2}\mu^{2}$ & $-\frac{1}{35}f^{2}-\frac{4}{35}f^{2}\mu^{2}$ \\
$b_2\,b_{(KK)_\parallel}$ & $-\frac{16}{45}-\frac{4}{15}\mu^{2}$ & $-\frac{8}{135}-\frac{2}{45}\mu^{2}$ \\
$b_2\,b_{\Pi^{[2]}_\parallel}$ & $-\frac{2}{5}+\frac{86}{105}\mu^{2}$ & $-\frac{1}{15}-\frac{2}{15}\mu^{2}$ \\
$b_{\mathcal{G}_2}\,b_\eta$ & $-\frac{24}{35}f^{2}+\frac{16}{35}f^{2}\mu^{2}$ & $0$ \\
$b_{\mathcal{G}_2}\,b_{\delta\eta}$ & $\frac{16}{15}f-\frac{8}{15}f\mu^{2}$ & $0$ \\
$b_{\mathcal{G}_2}\,b_{\eta^2}$ & $-\frac{24}{35}f^{2}+\frac{16}{35}f^{2}\mu^{2}$ & $0$ \\
$b_{\mathcal{G}_2}\,b_{(KK)_\parallel}$ & $-\frac{88}{135}+\frac{8}{45}\mu^{2}$ & $0$ \\
$b_{\mathcal{G}_2}\,b_{\Pi^{[2]}_\parallel}$ & $-\frac{16}{15}+\frac{8}{15}\mu^{2}$ & $0$ \\
$b_\eta^2$ & $-\frac{20}{99}f^{4}+\left(\frac{26}{245}f^{3}+\frac{38}{99}f^{4}\right)\mu^{2}+\frac{48}{245}f^{3}\mu^{4}$ & $-\frac{1}{99}f^{4}-\frac{8}{99}f^{4}\mu^{2}$ \\
$b_\eta\,b_{\delta\eta}$ & $\frac{22}{63}f^{3}-\left(\frac{22}{105}f^{2}+\frac{8}{15}f^{3}\right)\mu^{2}-\frac{8}{35}f^{2}\mu^{4}$ & $\frac{2}{63}f^{3}+\frac{4}{21}f^{3}\mu^{2}$ \\
$b_\eta\,b_{\eta^2}$ & $-\frac{40}{99}f^{4}+\left(\frac{26}{245}f^{3}+\frac{334}{693}f^{4}\right)\mu^{2}+\frac{48}{245}f^{3}\mu^{4}$ & $-\frac{2}{99}f^{4}-\frac{16}{99}f^{4}\mu^{2}$ \\
$b_\eta\,b_{(KK)_\parallel}$ & $-\frac{416}{945}f^{2}+\left(\frac{52}{315}f+\frac{44}{105}f^{2}\right)\mu^{2}+\frac{8}{105}f\mu^{4}$ & $-\frac{16}{945}f^{2}-\frac{4}{45}f^{2}\mu^{2}$ \\
$b_\eta\,b_{\Pi^{[2]}_\parallel}$ & $-\frac{40}{63}f^{2}+\left(\frac{22}{105}f+\frac{878}{735}f^{2}\right)\mu^{2}+\left(\frac{8}{35}f-\frac{16}{49}f^{2}\right)\mu^{4}$ & $-\frac{2}{63}f^{2}-\frac{4}{21}f^{2}\mu^{2}$ \\
$b_{\delta\eta}^2$ & $-\frac{4}{35}f^{2}+\frac{8}{105}f^{2}\mu^{2}$ & $-\frac{1}{35}f^{2}-\frac{4}{35}f^{2}\mu^{2}$ \\
$b_{\delta\eta}\,b_{\eta^2}$ & $\frac{22}{63}f^{3}-\frac{2}{15}f^{3}\mu^{2}$ & $\frac{2}{63}f^{3}+\frac{4}{21}f^{3}\mu^{2}$ \\
$b_{\delta\eta}\,b_{(KK)_\parallel}$ & $\frac{464}{945}f-\frac{8}{35}f\mu^{2}$ & $\frac{32}{945}f+\frac{20}{189}f\mu^{2}$ \\
$b_{\delta\eta}\,b_{\Pi^{[2]}_\parallel}$ & $\frac{22}{35}f-\frac{22}{21}f\mu^{2}+\frac{8}{21}f\mu^{4}$ & $\frac{2}{35}f+\frac{8}{35}f\mu^{2}$ \\
$b_{\eta^2}^2$ & $-\frac{20}{99}f^{4}+\frac{68}{693}f^{4}\mu^{2}$ & $-\frac{1}{99}f^{4}-\frac{8}{99}f^{4}\mu^{2}$ \\
$b_{\eta^2}\,b_{(KK)_\parallel}$ & $-\frac{416}{945}f^{2}+\frac{40}{189}f^{2}\mu^{2}$ & $-\frac{16}{945}f^{2}-\frac{4}{45}f^{2}\mu^{2}$ \\
$b_{\eta^2}\,b_{\Pi^{[2]}_\parallel}$ & $-\frac{40}{63}f^{2}+\frac{584}{735}f^{2}\mu^{2}-\frac{16}{49}f^{2}\mu^{4}$ & $-\frac{2}{63}f^{2}-\frac{4}{21}f^{2}\mu^{2}$ \\
$b_{(KK)_\parallel}^2$ & $-\frac{638}{2835}-\frac{2}{315}\mu^{2}$ & $-\frac{104}{8505}-\frac{64}{2835}\mu^{2}$ \\
$b_{(KK)_\parallel}\,b_{\Pi^{[2]}_\parallel}$ & $-\frac{44}{63}+\frac{628}{945}\mu^{2}-\frac{8}{63}\mu^{4}$ & $-\frac{32}{945}-\frac{20}{189}\mu^{2}$ \\
$b_{\Pi^{[2]}_\parallel}^2$ & $-\frac{18}{35}+\frac{34}{35}\mu^{2}-\frac{8}{21}\mu^{4}$ & $-\frac{1}{35}-\frac{4}{35}\mu^{2}$ \\
\hline
\caption{Coefficients of the bias pairs in $A(\mu)$ and $B(\mu)$ of Eq.~\eqref{eq:p22uv} (observed coefficients).
Pairs not listed do not contribute at this order.
\label{tab:renorm22}}
\end{longtable}}

\subsection{The origin of the stochastic $k^2\mu^4$ term}
\label{app:a4}

Rest-frame noise of the tracer is local, so its spectrum is analytic in $\k$; with $SO(2)$ symmetry and
$z\to-z$ parity it reads $N_0+N_2k^2+N_2'k_z^2+O(k^4)$, which contributes to $P_{\rm shot}$, $a_0$ and $a_2$ only.
A $k^2\mu^4=k_z^4/k^2$ term is formally non-analytic (non-polynomial) in $k$. 
The usual tensor-like 
noise structures in the EFT expansion of the Lyman-$\alpha$ stochasticity, like 
$\hat z^i\epsilon_{i},~\hat z^i\hat z^j\epsilon_{ij}$, cannot generate such non-locality:
it requires the projector $\hat{k}_i\hat{k}_j$, which long-range fields
introduce when they respond to noise. By the equivalence principle and mass and momentum conservation,
matter responds to short-scale noise only through the stochastic part of the stress tensor, $\Delta\tau_{ab}$,
i.e.\ through the force per unit mass $\mathcal F_a=-ik_b\Delta\tau_{ab}/\bar\rho$. The linear velocity response is
\be
v_a(\k,\tau)=\int^\tau d\tau'\left[G_T\,\delta_{al}+(G_L-G_T)\,\hat{k}_a\hat{k}_l\right](\tau,\tau')\,\mathcal F_l(\tau')\,,
\ee
where $G_L$ ($G_T$) is the Green's function of the longitudinal (vortical) velocity. They differ because gravity acts
only on the longitudinal mode. Denoting by $G_\delta$ the density response
and by $\Delta\tau_{\hat{k}\hat{k}}=\hat{k}_a\hat{k}_b\Delta\tau_{ab}$, $\Delta\tau_{\hat{z}\hat{k}}=\hat{z}_a\hat{k}_b\Delta\tau_{ab}$
the projections of the noise, the following selection operators acquire the stochastic parts (subscript ``st'') 
\be
\begin{split}
\eta_{\rm st}&=\frac{k^2}{\HH\bar\rho}\int^\tau d\tau'\left[G_T\,\mu\,\Delta\tau_{\hat{z}\hat{k}}+(G_L-G_T)\,\mu^2\Delta\tau_{\hat{k}\hat{k}}\right],\\
\Pi^{[1]}_{\parallel,{\rm st}}&=\d_z^2\Phi_{\rm st}=\frac{k^2\mu^2}{\bar\rho}\int^\tau d\tau'\,G_\delta\,\Delta\tau_{\hat{k}\hat{k}}\,,\qquad
\Pi^{[2]}_{\parallel,{\rm st}}=\frac{k^2\mu^2}{\bar\rho}\int^\tau d\tau'\,G^{[2]}_\delta\,\Delta\tau_{\hat{k}\hat{k}}\,,
\end{split}
\ee
with $G_\delta^{[2]}=-G_L/(f\HH)-G_\delta$. 
Importantly, $\Pi^{[2]}_{\parallel}$ is 2nd order in 
deterministic fields, but it is first order in the noise fields.

The non-analytic behaviour stems from terms in which the factor $\mu^2$ multiplies $\Delta\tau_{\hat{k}\hat{k}}$.
The cross-correlation with the tracer noise allowed by $SO(2)$ is
$\langle\epsilon(\tau)\Delta\tau_{ab}(\tau')\rangle'=C_{\epsilon\tau}\delta_{ab}+C'_{\epsilon\tau}\hat{z}_a\hat{z}_b$, so that
$\langle\epsilon\Delta\tau_{\hat{k}\hat{k}}\rangle'=C_{\epsilon\tau}+C'_{\epsilon\tau}\mu^2$, and the $k^2\mu^4$ coefficient is
\be
\label{eq:a4mech}
\frac{a_4}{k_{\rm NL}^2}=\frac{2}{\bar\rho}\int^\tau d\tau'\,C'_{\epsilon\tau}(\tau,\tau')\left[\frac{b_\eta}{\HH}(G_L-G_T)+b_{\Pi^{[1]}_\parallel}G_\delta+b_{\Pi^{[2]}_\parallel}G^{[2]}_\delta\right](\tau,\tau')\,.
\ee
$a_4\neq 0$ requires both an anisotropic noise cross-correlation, $C'_{\epsilon\tau}\neq0$, and a non-local response,
$G_L\neq G_T$ or a tidal operator.
In the basis of Eq.~\eqref{eq:F_full}, $\Pi^{[1]}_\parallel$ is degenerate with $\eta$
for the deterministic growing mode, which allows us to set $b_{\Pi^{[1]}_\parallel}=0$. This choice of basis does not lead to a loss of generality even when the noise contributions 
are considered because 
$a_4\neq0$ can be generated from the $\eta$ and $\Pi^{[2]}_\parallel$ channels.
For galaxies, rest-frame isotropy implies $C'_{\epsilon\tau}=0$ and therefore $a_4=0$,
even though the RSD mapping contributes $b_\eta=-1$. For the Lyman alpha forest $C'_{\epsilon\tau}$ is generically non-zero.

Three comments are in order. First, if the noise cross-correlation is instantaneous, $C'_{\epsilon\tau}\propto\delta_D(\tau-\tau')$,
the $\eta$ and $\Pi^{[1]}_\parallel$ channels of Eq.~\eqref{eq:a4mech} vanish, since $(G_L-G_T)(\tau,\tau)=G_\delta(\tau,\tau)=0$,
while the $\Pi^{[2]}_\parallel$ channel survives because $G^{[2]}_\delta(\tau,\tau)=-1/(f\HH)\neq0$ since $\Pi^{[2]}_\parallel$ involves the
velocity divergence. 
$\Pi^{[2]}$ is in general produced by an expansion
in convective time derivatives that capture non-trivial time evolution.
Hence, all three channels suggest that
 the appearance 
of the $k^2\mu^4$ noise can be interpreted
as a non-locality in time effect, specific
to the line-of-sight dependent tracers. 
Second, galaxy samples with
line-of-sight dependent selection break rest-frame isotropy and can have $a_4\neq0$. Third, in the 1D power spectrum
the $a_4$ term produces a non-analytic contribution $\propto k_\parallel^4\ln k_\parallel$ (Section~\ref{sec:1d}).

\section{Fits to the simulations for z=3.2}
\label{app:z3p2}

\begin{figure}
    \centering
    \includegraphics[width=\textwidth]{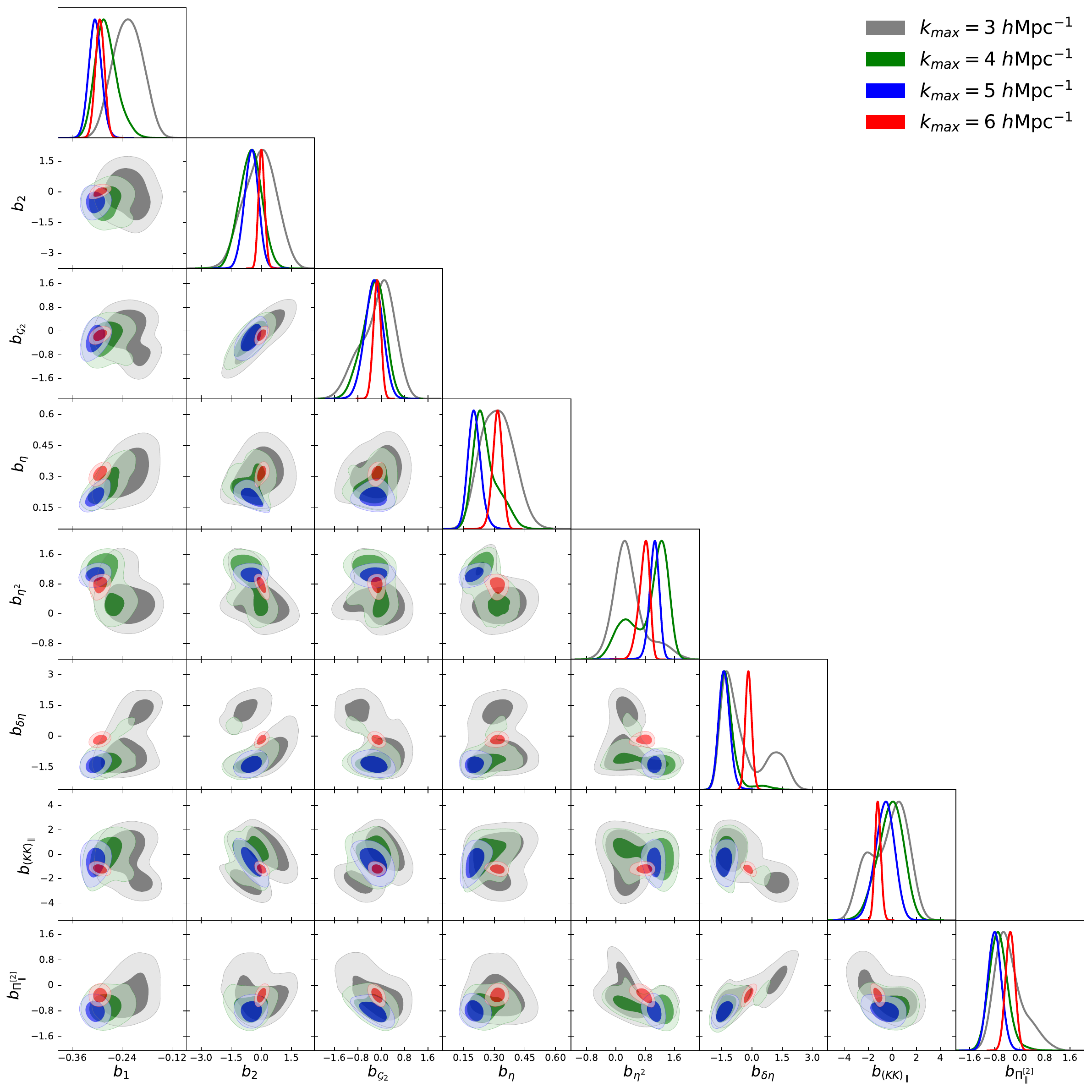}
    \caption{Triangle plot and marginalized projections for bias parameters
    of the EFT model for the Lyman alpha flux power spectrum of the Sherwood 
    simulations at $z=3.2$. We show results for four choices of $\kmax$: 3, 4, 5 and 6 $\hMpc$ (gray, green, blue, red, respectively).}
    \label{fig:bias_eftz32}
\end{figure}

\begin{table}[!htb]
\centering
\begin{tabular}{|l|c|c|c|c|}
 \hline
Param & best-fit & mean$\pm\sigma$ & 95\% lower & 95\% upper \\ \hline
$b_{1 }$ &$-0.3049$ & $-0.3051_{-0.016}^{+0.015}$ & $-0.3362$ & $-0.2738$ \\
$b_\eta$ &$0.2037$ & $0.2026_{-0.034}^{+0.029}$ & $0.1402$ & $0.2679$ \\
$b_{2 }$ &$-0.6221$ & $-0.5054_{-0.34}^{+0.36}$ & $-1.206$ & $0.1749$ \\
$b_{\mathcal{G}_2 }$ &$-0.3887$ & $-0.2627_{-0.29}^{+0.32}$ & $-0.8758$ & $0.3404$ \\
$b_{\eta^2}$ &$1.052$ & $1.044_{-0.12}^{+0.15}$ & $0.764$ & $1.331$ \\
$b_{\delta\eta}$ &$-1.461$ & $-1.376_{-0.28}^{+0.26}$ & $-1.918$ & $-0.8311$ \\
$b_{(KK)_\parallel}$ &$-0.5014$ & $-0.6172_{-0.73}^{+0.82}$ & $-2.16$ & $0.8908$ \\
$b_{\Pi^{[2]}_\parallel}$ &$-0.81$ & $-0.8054_{-0.22}^{+0.21}$ & $-1.243$ & $-0.3755$ \\
\hline
$10^3c_0/[\Mpch]^2$ &$7.91$ & $7.91_{-0.90}^{+0.90}$  & $ 6.12$ & $9.69$ \\
$10^3c_2/[\Mpch]^2$ &$4.43$ & $4.43_{-0.57}^{+0.57}$  & $3.29$ & $5.57$\\
$10^3c_4/[\Mpch]^2$ &$-12.4$ & $-12.4_{-0.57}^{+0.57}$ & $-13.5$ & $-11.3$  \\
$b_{\Pi^{[3]}_\parallel}$ &$2.36$ & $2.36_{-0.09}^{+0.09}$ & $2.18$ & $2.55$  \\
$b_{\delta\Pi^{[2]}_\parallel}$ &$-5.15$ & $-5.15_{-0.11}^{+0.11}$ & $-5.36$ & $-4.94$ \\
$b_{(K\Pi^{[2]})_\parallel}$ &$-3.16$ & $-3.16_{-0.16}^{+0.16}$ & $-3.48$ & $-2.83$ \\
$b_{\eta\Pi^{[2]}_\parallel}$ &$0.13$ & $0.13_{-0.19}^{+0.19}$ & $-0.25$ & $0.52$ \\
 \hline
 \end{tabular} 
 \caption{Same as Table~\ref{tab:kmax08}, but for $z=3.2$ and $\kmax=5~\hMpc$ data.
 \label{tab:kmax1}}
\end{table}

In this Section we present results for the Sherwood simulation
data analysis at $z=3.2$. This redshift is somewhat higher than $z=2.8$
used in the main text, so one may expect that the non-linear corrections 
would be suppressed and the fit would {extend to larger} $\kmax$.
This is indeed the case. In Fig.~\ref{fig:bias_eftz32} we show the results
for four choices of $\kmax$: 3, 4, 5 and 6 $\hMpc$. We see that the 
posteriors for $\kmax=3,4,5~\hMpc$
are fully consistent with each other. 
The posteriors for $\kmax=3,4$ are highly non-Gaussian, and even feature 
a multi-island structure for certain parameters.
This is likely due to a large number of parameters in the fit, 
which the data at a given $\kmax$ cannot constrain.

The $\kmax=5~\hMpc$
posterior has significant overlap with those of the $\kmax=3,4$ cases.
In contrast, 
the $\kmax=5~\hMpc$ and $\kmax=6~\hMpc$ posteriors 
are in noticeable tension with each other in the $b_{\delta \eta}-b_\eta$
plane. This suggests that the $\kmax=6~\hMpc$ results are biased. 
 Thus, we select 
$\kmax=5~\hMpc$ as a baseline for $z=3.2$. As {anticipated}, this is larger 
than our baseline choice $\kmax=3~\hMpc$ for $z=2.8$. 
The 1d marginalized posteriors for this case are shown in Table~\ref{tab:kmax1}.

\bibliography{short}
\bibliographystyle{JHEP}

\end{document}